\documentclass[12pt]{article}
\usepackage[margin=1in]{geometry}
\usepackage{amsmath,amssymb,amsthm}
\usepackage[authoryear]{natbib}
\usepackage{titletoc}
\usepackage{setspace}
\usepackage[dvipsnames]{xcolor}
\usepackage{rotating}
\usepackage{lscape}
\usepackage{float}
\usepackage{graphicx}
\usepackage{subcaption}
\usepackage{tikz}
\usepackage{booktabs}
\usepackage{enumerate}
\usepackage{verbatim}
\usepackage{tabularx}
\usetikzlibrary{arrows,calc,trees}
\tikzset{
  treenode/.style = {},
  root/.style     = {},
  env/.style      = {},
  dummy/.style    = {}
}
\usepackage{titletoc}

\def\spacingset#1{\renewcommand{\baselinestretch}%
{#1}\small\normalsize} \spacingset{1}

\makeatletter
\g@addto@macro\normalsize{%
  \setlength\abovedisplayskip{4pt}
  \setlength\belowdisplayskip{4pt}
  \setlength\abovedisplayshortskip{4pt}
  \setlength\belowdisplayshortskip{4pt}
}
\makeatother

\usepackage{hyperref}

\definecolor{darkblue}{rgb}{0, 0, 0.5}
\hypersetup{
     colorlinks = true,
     citecolor = Maroon,
     urlcolor = darkblue,
     linkcolor = darkblue
}
\usepackage{titlesec}

\titlespacing*{\section}{0pt}{3pt}{2pt}
\titlespacing*{\subsection}{0pt}{3pt}{2pt}
\titlespacing*{\subsubsection}{0pt}{3pt}{2pt}

 \theoremstyle{plain}
 \newtheorem{lem}{Lemma}[section]
 \newtheorem{prop}{Proposition}[section]
 \newtheorem{cor}{Corollary}[section]
 
 \theoremstyle{definition}
 \newtheorem{defn}{Definition}[section]
  \newtheorem{rem}{Remark}[section]
 \newtheorem{ass}{Assumption}[section]
  \newtheorem{exmp}{Example}[section]

\newcommand{\sgn}{\operatorname{sgn}}

\spacingset{1.5}

\begin{document}

\title{A model of multiple hypothesis testing\thanks{Earlier versions of this paper were circulated under the title: ``(When) should you adjust inferences for multiple hypothesis testing?'' (with and without the parentheses).  
First version on arXiv: 27 April 2021. Viviano and W\"uthrich contributed equally to this work. Email: {\tt dviviano@fas.harvard.edu, kasparwu@umich.edu, pniehaus@ucsd.edu}\vspace{1.5ex}}}  
\author{Davide Viviano\\Harvard University \and Kaspar W\"{u}thrich\\University of Michigan \and Paul Niehaus\\UC San Diego}
  \date{\today}

\maketitle
\thispagestyle{empty}

\begin{abstract}

Multiple hypothesis testing practices vary widely, without consensus on which are appropriate when. This paper provides an economic foundation for these practices designed to capture leading examples, such as regulatory approval on the basis of clinical trials. MHT adjustments are appropriate in our framework to the extent that research costs are invariant to the number of hypotheses. Control of average size, as for example via a Bonferroni correction, emerges in the limit case where all costs are fixed; in the opposite limit, where costs vary in proportion to the hypothesis count, no correction is needed. We illustrate implications by calculating explicit critical values using data on actual costs in the drug approval process and in program evaluation research; these suggest that some MHT adjustment is warranted in these applications, but not as much as implied by standard practice.

\bigskip 
\noindent \textbf{Keywords:} Bonferroni, family-wise error rate, multiple subgroups, multiple treatments, research costs

\bigskip 
\noindent \textbf{JEL Codes:} C12

\end{abstract}

\clearpage
\setcounter{page}{1}

\section{Introduction}

Hypothesis testing plays a prominent role in evidence-based decision-making. Typically, researchers report results from more than one test, and there has recently been increasing interest in and debate over whether the testing procedures they employ should reflect this in some way---that is, whether to apply some form of multiple hypothesis testing (MHT) adjustment. As a concrete example, consider pharmaceutical companies reporting the results of clinical trials to regulators when seeking approval to market new drugs: the U.S. regulator (the Food and Drug Administration, FDA) recently released guidelines calling for MHT adjustments on the grounds that omitting them could ``increase the chance of false conclusions regarding the effects of the drug'' \citep{fda2022guidance}. Analogous concerns arise in other settings, including experimental program evaluation in economics. As a result, a number of procedures for MHT adjustment have been proposed, and their statistical properties are well-understood \citep[see, e.g.,][for an overview]{romano2010hypothesis}. 

What is less clear is whether and when these procedures are economically desirable. That is, under what conditions does MHT adjustment lead to better decision-making from the point of view of the actor designing the process? The answer is far from obvious. It is certainly true, for example, that without MHT adjustments, the chance of making at least one type I error increases with the number of tests. But this is analogous to the truism that the more decisions one makes, the more likely one is to make at least one mistake. It is indisputable, but sheds no light on the pertinent questions, which are whether and how the rule for making individual decisions should change with the total number being made. 

This paper provides a framework for analyzing such questions. We focus, in particular, on whether and when MHT adjustments arise as a solution to incentive misalignment between a researcher and a mechanism designer. Our interest in this case reflects two primary considerations. The first is substantive: incentives are clearly an issue in real-world cases of interest (e.g., in the drug approval process, which we will use as a running example). The instinctive concern many seem to have is that without MHT adjustments, the researcher would have an undue incentive to test many hypotheses in the hopes of getting lucky. We would like to formalize and scrutinize that intuition. And the second is pragmatic: to have a theory of MHT adjustments, we must have a theory that rationalizes hypothesis testing at standard levels in the first place, which (as we discuss below) is hard to do convincingly in a non-strategic setting \citep[e.g.,][]{tetenov2012statistical,tetenov2016economic}.

Specifically, we study a model in which a benevolent social planner chooses norms with respect to MHT adjustments, taking into account the way this shapes researchers' incentives. The model embeds two core ideas. First, social welfare is affected by the summary recommendations (in particular, hypothesis tests) contained in research studies, and the planner also cares about the more generic benefits to society and to the researcher of conducting research per se.\footnote{We describe the case where hypothesis \emph{rejections} affect welfare; under a straightforward reinterpretation the framework can also accommodate situations in which ``precise null'' results do so.} Second, while this makes the research a public good, the costs of producing it are borne privately by the researcher. She decides whether or not to incur these costs and conduct a (pre-specified) experiment based on the private returns to doing so. The planner must, therefore, balance the goals of (i) limiting the possibility of harm due to mistaken conclusions and (ii) motivating the production of research. We represent these preferences with a utility function that includes both ambiguity-averse and expected-utility components \citep[in the spirit of, for example,][]{gilboa1989maxmin,banerjee2020theory}, which (we show) turn out to have intuitive connections with the statistical concepts of size control and power. We focus on cases where multiplicity takes the form of testing multiple \emph{treatments} or estimating effects within multiple \emph{sub-populations};\footnote{These forms of multiplicity are common in practice. For example, the majority of the clinical trials reviewed in \citet[][Table 1]{pocock2002subgroup} tested for effects in more than one subgroup. In economics, 27 of 124 field experiments published in ``top-5'' journals between 2007 and 2017 feature factorial designs with more than one treatment \citep{muralidharan2025factorial}.} multiple \emph{outcomes} are an economically distinct case covered in an earlier version of the paper \citep{VWN2025MHTv8}.

We start by characterizing optimal hypothesis testing protocols. We show that the class of optimal protocols is the class of maximin optimal and unbiased protocols, where maximin optimality is closely connected to size control while unbiasedness requires the power of the protocols to exceed their size. We then prove that separate $t$-tests, which are ubiquitous in applied work, are maximin optimal and unbiased, and we provide an explicit characterization of the optimal critical value in terms of the researcher's costs.\footnote{We focus on one-sided $t$-tests in the main text and consider two-sided tests in Appendix \ref{app: two-sided tests}.}

We next characterize the role of multiplicity, drawing two broad conclusions. First, it is generically optimal to adjust testing thresholds (i.e., critical values) for the number of hypotheses. A loose intuition is as follows. The worst states of the world are those in which the status quo of no treatment is best; in these states, a research study has only a downside, and it is desirable to keep the benefits from experimentation low enough that the researcher chooses not to experiment. If the hypothesis testing protocol were invariant to the number of hypotheses being tested, then for sufficiently many hypotheses, this condition would be violated: the researcher's expected payoff from false positives alone would be high enough to warrant experimentation. Some adjustment for hypothesis count may thus be needed. This logic aligns fairly well with the lay intuition that researchers should not be allowed to test many hypotheses and then ``get credit'' for false discoveries. Interestingly, the same logic immediately implies that critical values should adjust for other factors that influence cost such as the sample size, though these have not attracted the same degree of attention.

Second (and as this suggests), economic fundamentals---in particular, the research cost function---determine exactly how much adjustment is required. When hypotheses are equally important, for example, the optimal critical values for the separate $t$-tests are given by
\begin{equation}
    \label{eqn:t_threshold_intro}
    t(J,\Sigma) = \Phi^{-1}\left(1 - \frac{C(J, \Sigma)}{b |J|} \right),
\end{equation}
where $J$ is the set of hypotheses tested (with $|J|$ denoting its cardinality), $\Sigma$ captures features of the experimental design such as the sample size, $C(J, \Sigma)$ the cost of the experiment, and $b$ the benefit to the researcher of rejecting a null. When research costs are fixed, so that $C$ is invariant to $J$, this implies a Bonferroni correction.\footnote{Including, for example, in subgroup analysis contexts where experimentation costs are sunk.} When costs scale in proportion to the number of hypotheses, on the other hand, \emph{no} MHT adjustment is required. Intuitively, the researcher has no undue incentive to test many hypotheses in this scenario because doing so is costly. This cost-based perspective also helps to clarify confusion about the boundaries of MHT adjustment and whether researchers should adjust for multiple testing \textit{across} different studies. It suggests that MHT adjustments may be appropriate when there are cost complementarities across studies but not otherwise.\footnote{The broader principle is that optimal MHT adjustments depend on how exactly hypotheses interact. Our base model emphasizes interactions in the research cost; Appendix \ref{sec:additional_interactions_main} considers interactions of other kinds, through non-linearities in the researcher's payoff and through interactions in the planner's objective.}

\begin{figure}[t]
    \begin{center}
    \caption{Multiple hypothesis testing adjustment in ``top-5'' experimental papers}     \label{fig:publishing_trends}
    \vspace{-4mm}
    \includegraphics[scale = 0.4]{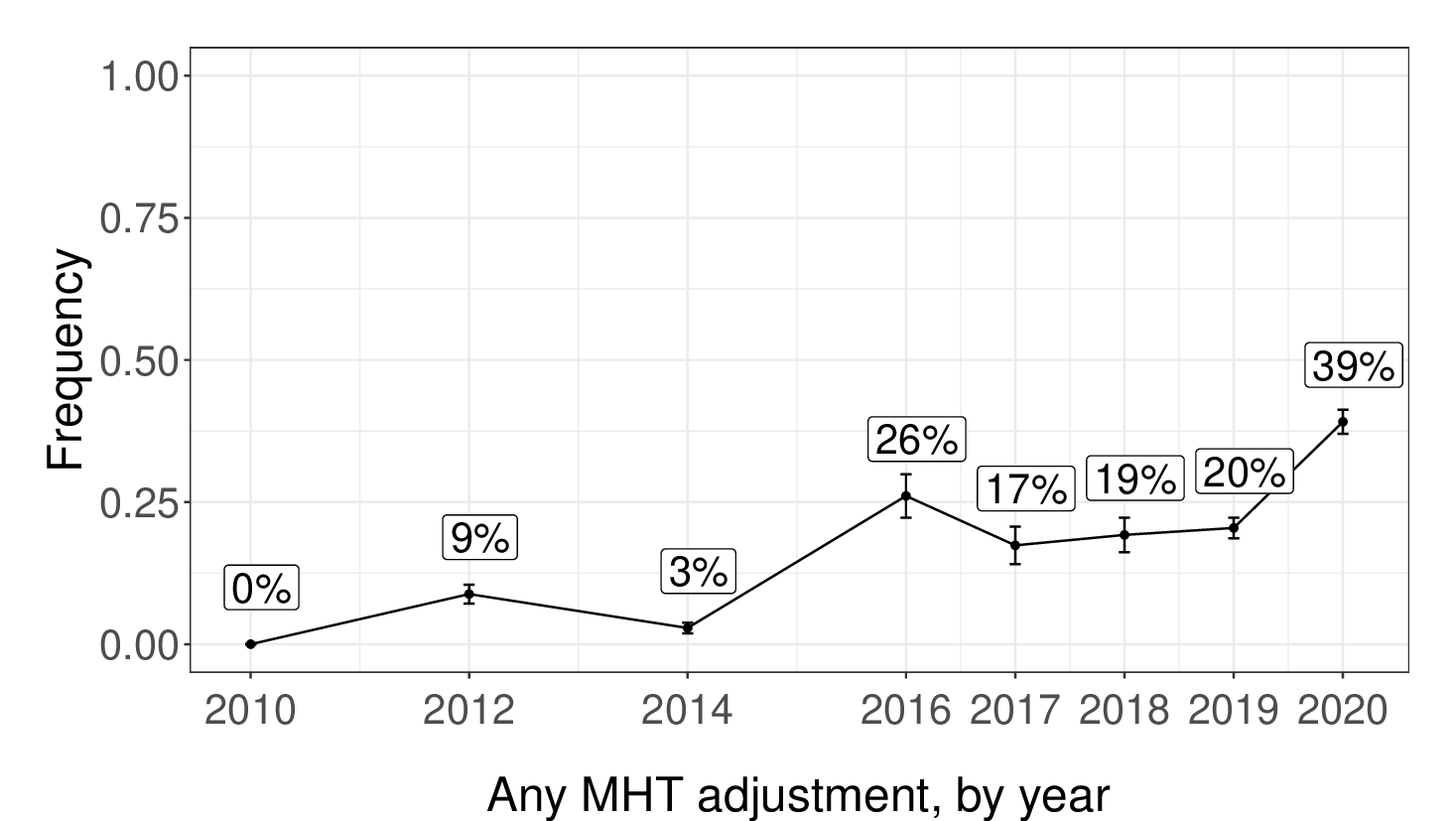}
     \includegraphics[scale = 0.4]{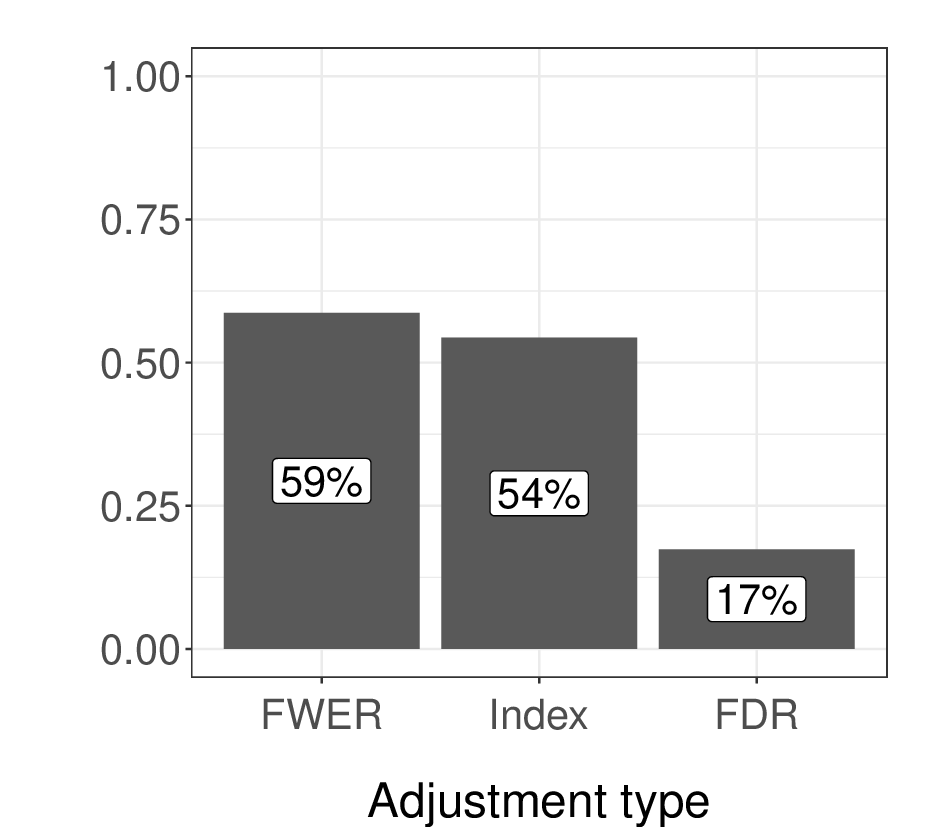}
     
     \end{center}
     \vspace{-1em}
     {\footnotesize\emph{Notes:} The left-hand panel reports the share of experimental papers (both field and lab experiments) that conduct at least one MHT adjustment, including both indexing and control of compound error rates, by year of publication. (Note that almost all experimental studies have more than one hypothesis). The right-hand panel reports the frequency of each adjustment type, pooling across years. Adjustment types are not mutually exclusive. FWER: Family-Wise Error Rate control. Index: Methods based on creating an index using multiple outcomes. FDR: False Discovery Rate control. Authors' calculations based on a review of publications in the \emph{American Economic Review} (excluding Papers and Proceedings), \emph{Econometrica}, the \emph{Journal of Political Economy}, the \emph{Quarterly Journal of Economics}, and the \emph{Review of Economic Studies}. }

\end{figure}

To illustrate the quantitative implications of the model, we apply it to our running example, regulatory approval by the FDA. Applying the formulae implied by the model to published data on the cost structure of clinical trials, we calculate adjusted critical values that are neither as liberal as unadjusted testing, nor as conservative as those implied by some of the procedures in current use. If the appropriate level in the single-hypothesis case is 5\%, for example, then the optimal level according to our formulae is 3.2\% with two tests, 2.6\% with three tests, and tends to 1.4\% as $|J|\rightarrow\infty$. By comparison, the level implied by Sidak's correction \citep{vsidak1968multivariate} for controlling the Family-Wise Error Rate (FWER) under independence (and, up to rounding, also Bonferroni), is 2.5\% for two tests, 1.7\% for three tests, and tends to zero as $|J|\rightarrow\infty$.\footnote{The Sidak correction is a natural benchmark because it is exact for controlling the FWER with independent tests, and FWER control is common in practice (see Figure \ref{fig:publishing_trends}).} These results suggest both that some adjustments are warranted but also that standard practices may be overly-conservative. Moreover, because costs also scale with the sample size, optimal adjustments must be less conservative for larger samples in order to induce researchers to incur the correspondingly larger costs. 
 
It is also natural to wonder about applicability to economic research. The share of experimental papers published in ``top 5'' journals that used some form of MHT adjustment grew rapidly, from 0\% in 2010 to 39\% in 2020, so that there is now wide variability in whether (and how) these papers adjust (see Figure \ref{fig:publishing_trends}) and little consensus on what the norms should be. In Appendix \ref{app:jpal}, we develop an additional empirical application to experimental program evaluation, using a unique dataset on the costs of projects submitted to the Abdul Latif Jameel Poverty Action Lab (J-PAL) from 2009 to 2021 which we assembled for this purpose. In this application, we find that the estimated adjustments implied by our model are less conservative than FWER control using Sidak's correction, but only slightly so. This is because the relationship between costs and number of treatment arms, while significant, is relatively weak in this setting, with a cross-sectional elasticity of approximately 15\%.  

Our paper draws inspiration from other work using economic models to select statistical procedures, in which researchers' preferences and incentives drive the analysis. This includes work on scientific communication \citep[e.g.,][]{andrews2021model,frankel2022findings}, several aspects of which have been studied in more recent papers including \citet{bates2022principal,bates2023incentive} and \citet{kasy2023optimal}.\footnote{See also \citet{chassang2012selective,banerjee2017decision,spiess2018optimal,henry2019research,banerjee2020theory,williams2021preregistration,mccloskey2022incentive,yoder2022designing}.}  None of these papers analyze multiple hypothesis testing, however. The most closely related paper is the insightful work by \citet{tetenov2016economic}, who shows that $t$-tests are maximin optimal and uniformly most powerful in the single-hypothesis case. Our extension to a multiple-hypothesis setting requires us to deal with two major challenges. First, the notions of maximin optimality and the corresponding theoretical results are more complex because the effects of different treatments may have opposite signs. Second, within the (large) class of maximin optimal protocols, none uniformly dominates all others, requiring us to develop new notions of optimality suitable to the context.

Our paper also relates to an extensive literature at the intersection between decision theory and hypothesis testing, dating back to \cite{wald1950statistical} and \citet{robbins1951asymptotically}. Previous work has motivated notions of compound error control in single-agent non-strategic environments; see in particular \citet{kline2022systemic,kline2024discrimination} for recent examples in economics based on a Bayesian interpretation of the false discovery rate (FDR), as well as \citet{storey2003positive}, \cite{lehmann2005testing}, \citet{ efron2008simultaneous}, and \cite{hirano2020handbook} for further examples.\footnote{The literature on statistical treatment choice has similarly focused for the most part on non-strategic planner problems. See \citet{manski2004} and \citet{tetenov2012statistical} as well as \citet{hirano2009asymptotics,KitagawaTetenov_EMCA2018,athey2017efficient} for recent contributions.} We complement this literature (as well as the statistical literature discussed below) by developing a model that explicitly incorporates the incentives and constraints of the researchers. Relative to the decision-theoretic approach, this has two main advantages. First, it lets us characterize \emph{when} MHT adjustments are appropriate---and also when they are \emph{not}---as a function of measurable features of the research process. Second, it allows us to justify and discriminate between different notions of compound error (e.g., average error rate or the FWER) in the same framework based on these same economic fundamentals. 

Finally, we aim to provide guidance for navigating the extensive statistical literature on MHT. This literature provides procedures for controlling particular notions of compound error,\footnote{See \citet{efron2008microarrays} and \citet{romano2010hypothesis} for overviews. 
} but few statistical optimality results \citep[e.g.,][]{spjotvoll1972optimality, lehmann2005optimality,romano2011consonance}, and none in which MHT procedures address an incentive problem. We maximize a different (social planner's) objective, subject to incentive compatibility constraints. We also draw on \citet{list2019multiple}'s helpful distinction between different types of multiplicity, and show how these lead to different optimal testing procedures.

\section{Model}
\label{sec:model}

We study MHT in a game between a social planner who chooses statistical procedures and a representative experimental researcher with private incentives. In our running example, we can think of the planner as a regulator (e.g., the FDA) who defines testing protocols for studies submitted  in support of applications for the approval of new drugs, and the researcher as a pharmaceutical company running a pre-specified clinical trial of a new drug hoping to obtain such approval. Multiple testing issues arise whenever research informs multiple decisions. We focus on settings with multiple \emph{treatments} (e.g., multiple drugs) or different \emph{subpopulations} (e.g., multiple demographic groups for which a drug may be approved); for brevity we will refer throughout to treatments, taking this to refer to multiplicity of both types. 

To say something coherent about MHT, a framework must be able to rationalize conventional (single) hypothesis testing in the first place. This is known to be a challenging problem, requiring non-trivial restrictions on the research process \citep[see, e.g., Section 1 in][for a discussion]{tetenov2016economic}---in particular, strong asymmetries to match the inherently asymmetric nature of null hypothesis testing. For example, \citet{tetenov2012statistical} shows that justifying testing at conventional levels in a single agent model with minimax regret requires extreme degrees of asymmetry: statistical tests at the 5\% (1\%) level correspond to decision-makers placing 102 (970) times more weight on type I than type II regret. Here the asymmetry necessary for rationalizing hypothesis testing will arise naturally from the planner's desire to prevent the implementation of harmful treatments.

\subsection{A game between a researcher and a social planner}

We consider a two-stage game between the planner and the researcher. In the first stage, the planner prescribes and commits to a hypothesis testing protocol, restricting how the researcher can report findings. In the second stage, given this protocol, the researcher decides whether or not to run one of several possible experiments by comparing the private benefits of experimentation to the private costs. Importantly, these private benefits may differ from the planner's objective. Unless noted otherwise, we will assume that the researcher's preferences are common knowledge and that she is not allowed to mis-characterize them. 

Hypothesis testing protocols take as input the data from the experiment and output multiple binary findings indicating whether the treatments were found to be effective. These findings, in turn, affect social welfare: we will interpret a finding as equivalent to the planner's decision to implement the corresponding treatment. Since the planner selects the hypothesis testing protocol, this is equivalent to the planner pre-committing to a decision rule and the researcher truthfully reporting the decisions it implies, given the observed data.

\subsubsection{The researcher's problem}

The researcher takes the hypothesis testing protocol as given and decides, before observing data, whether and how to experiment. We first describe the experiment and hypothesis testing protocol and then introduce the researcher's optimization problem.

\smallskip

\noindent \textbf{Experiment.}
Let $\mathcal{J}$ denote the finite set of all combinations of non-exclusive treatments that can be tested in an experiment (the ``power set''), with $\emptyset \in \mathcal{J}$ denoting no experimentation. The parameter vector $\theta \in \Theta$, where $\Theta$ is a compact parameter space, captures the effects corresponding to all possible combinations of treatments in $\mathcal{J}$.

An experiment consists of a set of treatments $J \in \mathcal{J}$ and a design $\Sigma \in \mathcal{S}(J)$, which are chosen by the researcher. Here $\mathcal{S}(J)$ is the set of all possible designs given $J$. If the researcher experiments, she draws a vector of statistics $X$ from a distribution $F_{\theta, J, \Sigma}$, indexed by $J$, $\theta$, and $\Sigma$.
The design $\Sigma$ summarizes all the relevant features of the distribution of $X$ the researcher can choose ex-ante, such as the sample size of the experiment. The researcher pre-specifies and reports $J$ and $\Sigma$ before running the experiment, so that they become common knowledge. $J$ and $\Sigma$ may depend on the researcher's prior knowledge and private incentives, but not on the realized statistics $X$. We thus abstract from issues of $p$-hacking and selective reporting. This case is relevant for considering decision-making at the FDA, for example, which requires pre-registration.\footnote{Specifically, the summary of the Final Rule for Clinical Trials Registration and Results Information Submission (42 CFR \S 11) on ClinicalTrials.gov states that ``[r]egistration is required for studies that meet the definition of an `applicable clinical trial' (ACT) and either were initiated after September 27, 2007, or initiated on or before that date and were still ongoing as of December 26, 2007'' \citep{final_rule}. Registration must specify, among other things, the intervention(s), primary outcomes, and intended enrollment and study design \cite[][42 CFR \S 11.28]{42cfr1128}.}

\begin{rem}[Mutually exclusive treatments]
We focus on settings where the treatments may not be mutually exclusive. This is relevant in the regulatory approval process context, for example,  when there are multiple subgroups and the pharmaceutical companies receive separate approvals for each group, or when there are different drugs that can be provided to the same set of individuals.  That said, our framework also accommodates settings with mutually exclusive treatments, such as competing drugs for treating the same condition. First, if the researcher can report multiple findings and each treatment will be implemented with an (exogenous) probability and these probabilities sum to one, the results in Sections \ref{sec:multiple_treatments} and \ref{sec:t-tests} apply due to the linearity of the planner's objective defined below. Second, if the researcher is only allowed to report one finding, the resulting model is isomorphic to the one discussed at the end of Appendix \ref{sec:additional_interactions_main}.  \qed
\end{rem}

\noindent \textbf{Hypothesis testing protocols.} As described above, the researcher first chooses and pre-specifies an experiment $(J,\Sigma)$. She then runs the experiment, as a result of which the vector of statistics $X$ is realized and becomes publicly available. The results from the experiment are reported in the form of a vector of non-exclusive \emph{findings} or \emph{recommendations},
\begin{equation} \label{eqn:protocol}
r(X;J,\Sigma) = \left(r_{1}(X;J, \Sigma), \dots, r_{|J|}(X;J, \Sigma)\right)^\top\in \{0,1\}^{|J|},  
\end{equation}  
where $r_{j}(X;J, \Sigma) = 1$ if and only if the treatment corresponding to $J_{j}$ is found to be effective, with $J_{j}$ denoting the $j^{th}$ entry of $J$. If $J = \emptyset$, no findings are reported, $r(X,\emptyset,\Sigma)=0$. If there are no findings ($r(X,J,\Sigma)=0$), the status quo prevails.
We will refer to $r$ as a 
\emph{hypothesis testing protocol}. 

To simplify notation when describing the researcher's payoff and welfare below, it is useful to introduce the selector function $\delta(r(X;J, \Sigma); J) \in \{0,1\}^{2^{|J|} - 1}$. Each entry of $\delta(r(X;J, \Sigma); J)= \left(\delta_1(r(X;J, \Sigma);J),\dots,\delta_{2^{|J|} - 1}(r(X;J, \Sigma);J)\right)$, corresponds to one of the $2^{|J|} - 1$ possible combinations of the $J$ treatments. Specifically, for $k=1,\dots,2^{|J|} - 1$, $\delta_k(r(X;J, \Sigma);J)=1$ if the treatment combination $k$ is found to be effective and $\delta_k(r(X;J, \Sigma);J)=0$ otherwise. We let $\delta(r;\emptyset) = 0$ for all $r$. By definition of $\delta$, we have that
\begin{equation*}
\sum_{k=1}^{2^{|J|} - 1}
\delta_k(r(X;J, \Sigma);J) \in \{0,1\}.
\end{equation*}
Here $\delta(\cdot;J)$ only takes into account combinations of the treatments in the set $J$, ignoring combinations not studied in the experiment. Example \ref{exmp:1} provides an illustration of $\delta$. 

\smallskip

\begin{exmp} \label{exmp:1}
The researcher uses linear regression to estimate the effects of treatments $J \in \mathcal{J}$ on an outcome $Y$ based on an experiment with $n$ units. Let $D_{i,j}=1$ if unit $i$ received treatment $j$ and $D_{i,j}=0$ otherwise. Suppose that $J = \{1,2\}$ and that 
\begin{equation}  
    Y_i = \mu + \theta_1 D_{i,1} + \theta_2 D_{i,2} +\varepsilon_i,~ \varepsilon_i \overset{i.i.d.}\sim \mathcal{N}(0, \eta^2),\label{eq:regression_model}
\end{equation}
where $\mu$, $\theta_1$, and $\theta_2$ are unknown parameters and $\eta^2$ is known.
Let
$X \sim \mathcal{N}(\theta, \Sigma)$, where $X$ is the OLS estimator of $(\theta_1,\theta_2)$ and where, specializing notation, in this example the design $\Sigma$ denotes the covariance matrix of $X$. An example of a hypothesis testing protocol is separate (one-sided) $t$-testing,
$
    r(X;\{1,2\}, \Sigma)=(1\{X_1/\sqrt{\Sigma_{1,1}}\ge t \},1\{X_2/\sqrt{\Sigma_{2,2}}\ge t\})^{\top}
$.

In this example, the power set $\mathcal{J}$ is $\mathcal{J} = \{\emptyset, \{1\}, \{2\}, \{1,2\}\}$, and $J$ is an element of $\mathcal{J}$. The selector function $\delta$ is defined as follows (suppressing the dependence of $r(X;J,\Sigma)$ on $\Sigma$): $\delta(r(X;\{1\});\{1\}) = r(X;\{1\})$, $\delta(r(X;\{2\});\{2\}) = r(X;\{2\})$, 
\begin{eqnarray*} 
&&\delta(r(X;\{1,2\});\{1, 2\})=\Big(\delta_1(r(X;\{1,2\});\{1, 2\}),\delta_2(r(X;\{1,2\});\{1, 2\}),\delta_3(r(X;\{1,2\});\{1, 2\})\Big)\\
&&=\Big(r_1(X;\{1,2\})(1 - r_2(X;\{1,2\}), r_2(X;\{1,2\})(1 - r_1(X;\{1,2\}), r_1(X;\{1,2\})r_2(X;\{1,2\})\Big).
\end{eqnarray*}
That is, the first (second) entry of $\delta(r(X;\{1,2\});\{1, 2\})$ is equal to one if treatment 1 (treatment 2) is found to be effective but not treatment 2 (treatment 1), and the last entry is equal to one if both treatments are found to be effective. \qed
\end{exmp}

\noindent \textbf{The researcher's objective.} 
For each $(J,\Sigma)$, the researcher takes as given the corresponding hypothesis testing protocol $r(\cdot;J,\Sigma)$, which is chosen by the planner in the first stage. For simplicity, we assume that the researcher knows $\theta$, but our main results continue to hold when the researcher is imperfectly informed and has a prior about $\theta$ (see Section \ref{sec:uncertain}). 

We consider settings where the researcher's and the planner's objective are misaligned. We model misalignment using researcher's utility of the form 
$p(J)^\top \delta(r(X;J,\Sigma)) - c_\theta(J,\Sigma)$ where $p_k(J)\geq 0$, $k=1,\dots, 2^{|J|} - 1$, is the benefit from getting approval for the $k$th combination of treatments, conditional on the set of treatments $J$ being tested, and $c_{\theta}(J, \Sigma)$ captures the costs of research, which can be a function of $(J,\Sigma,\theta)$.\footnote{For instance, we could write $c_{\theta}(J,\Sigma) = C(J,\Sigma) - b(J,\theta)$ for some function $b(\theta, J)$ of  $(J,\theta)$ that is the part of benefits that depend on $\theta$ ($b(\theta,J)$ can be an implicit function of $p$) and the costs $C(J,\Sigma)$ as a function of the design. Practically speaking however this requires being able to measure $b(J, \theta)$ in addition to the costs.}  
Taking expectations over $X$ yields the following class of expected researcher payoff functions,
\begin{equation} \label{eqn:B_R_general}
\small 
\begin{aligned} 
\beta_r(\theta, J,\Sigma) = \underbrace{\sum_{k=1}^{2^{|J|} - 1} p_k(J)  \int \delta_k(r(x;J,\Sigma), J)  dF_{\theta,J,\Sigma}(x)}_{\text{expected benefits}} - \underbrace{c_\theta(J,\Sigma)}_{\text{costs}}, \quad c_{\theta}(J,\Sigma) \ge 0 \quad \forall (\theta,J,\Sigma). 
\end{aligned} 
\end{equation}
Thus, $\beta_r(\theta, J,\Sigma)$ captures the net benefits of experimenting. We let $c_\theta(J,\Sigma) \ge 0$ for all $(J,\Sigma,\theta)$ and $c_\theta(\emptyset, \Sigma) = 0$ for all $(\theta,\Sigma)$, so that the researcher's net benefits are normalized to zero if no experiment is conducted. Misalignment arises because the researcher's payoff is different from the planner's objective. In the following, whenever we consider settings where the costs do not depend on $\theta$, we will write $c_{\theta}(J,\Sigma) \equiv C(J,\Sigma)$ for a function $C(J,\Sigma)$ that does not depend on $\theta$.

The researcher chooses which treatments to study and how to design the experiment so as to maximize her net benefits. Formally, the researcher's problem is
\begin{equation} \label{eqn:optimality_sigma}
(J_{r,\theta}^*, \Sigma_{r,\theta}^*) \in  \mathrm{arg} \max_{J \in \mathcal{J},\Sigma \in \mathcal{S}(J)} \beta_r(\theta, J, \Sigma),  
\end{equation}
where $J_{r,\theta}^* = \emptyset$ corresponds to no experimentation. To state the theoretical results, we also define the experiment the researcher would choose when forced to run an experiment,
\begin{equation} \label{eqn:optimality_sigma_v2}
\Big(J_{r,\theta}^{+}, \Sigma_{r,\theta}^{+}\Big) = \mathrm{arg} \max_{J \in \mathcal{J} \setminus \emptyset, \Sigma \in \mathcal{S}(J)}  \beta_r(\theta, J, \Sigma).
\end{equation} 
We impose a standard tie-breaking rule: whenever the researcher is indifferent regarding whether to experiment (i.e., when $\beta_r(\theta, J_{r,\theta}^*, \Sigma_{r,\theta}^*) = 0$), she experiments if the planner's utility (defined below) is weakly positive. Let $e_r^\ast(\theta)$ indicate whether the researcher experiments, $e_r^\ast(\theta)=1\left\{J_{r,\theta}^* \neq \emptyset\right\}$. 

\smallskip

\noindent \textbf{Example \ref{exmp:1} continued.} \label{exmp:null_space}
    Suppose that $\mathcal{J}=\left\{\emptyset, \{1\}, \{2\}, \{1,2\}\right\}$. Let $p(\{1\}) \in \mathbb{R}_+$, $p(\{2\}) \in \mathbb{R}_+$, and $p(\{1,2\}) \in \mathbb{R}_+^3$ denote the researcher's (vector of) benefits from getting approval, conditional on the set of treatments $J$ being tested. In this example, the researcher's payoff for $J = \{1,2\}$ equals 
    $$
    \begin{aligned} 
    \beta_r(\theta,\{1,2\},\Sigma) =  & P(r_1(X;\{1,2\},\Sigma) = 1, r_2(X) = 0|\theta) p_1(\{1,2\})  \\ & +  P(r_2(X;\{1,2\},\Sigma) = 1, r_1(X;\{1,2\},\Sigma) = 0|\theta) p_2(\{1,2\}) \\ 
    &+ P(r_1(X;\{1,2\},\Sigma) = 1, r_2(X;\{1,2\},\Sigma) = 1|\theta) p_3(\{1,2\}) - C(\{1,2\},\Sigma).
    \end{aligned} 
    $$
\qed

\subsubsection{The planner's problem}

The social planner chooses a hypothesis testing protocol $r \in \mathcal{R}$ to maximize her utility, where $\mathcal{R}$ is the class of all (pointwise measurable) protocols. That is,  $\mathcal{R}$ contains all protocols typically found in practice, including (and not limited to) standard $t$-tests.  The planner's utility will depend on the welfare effects of implementing the recommended treatments as well as a measure of the more generic benefits to society and to the researcher of conducting research per se.

\smallskip

\noindent \textbf{Welfare.}
Welfare depends on whether the researcher experiments and on her findings if she does. To define welfare, for $k=1,\dots,2^{|J|}-1$, let $u_k(\theta; J)$ denote the effect on welfare that would result from implementing the combination of treatments $k$.  

Given $r(X;J, \Sigma)$, the overall welfare is $u(\theta; J)^\top \delta(r(X;J, \Sigma);J)$. We normalize $u(\theta;\emptyset) = 0$ for all $\theta \in \Theta$. That is, welfare is equal to zero under the status quo when no experimentation occurs. 
For a given experiment $(J,\Sigma)$, the expected welfare is
\begin{equation} \label{eqn:aab} 
v_r(\theta, J,\Sigma) = 
  \int \delta(r(x;J,\Sigma); J)^\top u(\theta; J)dF_{\theta, J,\Sigma}(x).
\end{equation}
If the researcher does not experiment, $v_r(\theta, \emptyset, \Sigma) = 0$ for all $\theta \in \Theta$.

\smallskip 

\noindent \textbf{Example \ref{exmp:1} continued.} \label{exmp:null_space}
    Suppose that $\mathcal{J}=\left\{\emptyset, \{1\}, \{2\}, \{1,2\}\right\}$ and denote by $\theta_1$ and $\theta_2$, the welfare from implementing treatment 1 and 2, respectively. In this case, $u(\theta;\emptyset)=0$, $u(\theta;\{1\})=\theta_1$, and $u(\theta;\{2\})=\theta_2$. If there are no interaction effects, so that the welfare effect from implementing both treatments is equal to the sum of effects from implementing each of them separately, then 
\begin{equation} \label{eqn:exmp_additive}
u(\theta;\{1,2\}) = (\theta_1,\theta_2,\theta_1 + \theta_2).
\end{equation}  
\qed
\medskip

\noindent \textbf{The planner's objective.} We consider a planner who wishes to increase welfare while also limiting the possibility of harm due to mistaken conclusions, and to encourage research. Specifically, the planner chooses $r$ to maximize
\begin{equation}
\label{eqn:plannerobjective}
    U(r; \lambda, \pi)=\min_{\theta \in \Theta} v_r(\theta,J_{r,\theta}^*, \Sigma_{r,\theta}^*) + \lambda \int e_r^\ast(\theta) \pi(\theta) d\theta, 
\end{equation}
where $\lambda\ge 0$ and $\pi(\theta) \ge 0$ for all $\theta \in \Theta$. The first component, which depends on which treatments are actually implemented, captures the desire to raise welfare while limiting harm using a standard ambiguity-averse (maximin) formulation. The second component depends on whether or not the researcher experiments. (If $\pi$ is a probability density, that second component is equal to the probability of experimentation since $\int e_r^\ast(\theta) \pi(\theta) d\theta=\int 1\{J_{r,\theta}^\ast\ne \emptyset\} \pi(\theta) d\theta$). It can be interpreted as capturing the benefits of scientific research per se, and as internalizing some aspects of the researcher's utility. In Appendix \ref{sec:discussion_objective}, we formalize the latter interpretation, showing that protocols that are optimal under $U$ remain approximately optimal if the second component is replaced by  $\int\beta_r^\ast(\theta)\pi(\theta)d\theta$, the expected researcher utility (if $\pi$ is a density).
The parameter $\lambda$ allows us to trade-off each of these components. We show below that under suitable assumptions on $\pi$, the two components of $U$ have intuitive connections to the statistical concepts of size control and power. Moreover, $U$ admits optimal protocols that do not depend on $\lambda$ and $\pi$. This is important because the relative importance of the two components may be difficult to determine and choosing high-dimensional weights $\pi$ is difficult and often somewhat arbitrary. Working with $U$ thus provides a cogent rationale for testing protocols that control size, have non-trivial power, and do not depend on $\lambda$ and $\pi$.

In the regulatory approval example, the structure of the planner objective $U$ is motivated by regulators such as the FDA being tasked by legislators with several distinct objectives \citep[e.g.,][]{fda_mission}. Each component of Equation \eqref{eqn:plannerobjective} relates to a distinct objective. The first captures the desire to avoid implementing harmful treatments, as for example under the ``do no harm'' principle (since, we show, our framework naturally rationalizes one-sided hypothesis testing). The second captures the broader value of scientific research, which need not be directly related to the immediate regulatory decision being made.\footnote{As the international guidelines for clinical trials state, for example, ``the rationale and design of confirmatory trials nearly always rests on earlier clinical work carried out in a series of exploratory studies'' \citep{lewis1999statistical}. More broadly, the results of one study may lead to new conceptual insights or scientific hypotheses which are valuable independent of any immediate clinical application.} 
The relative importance of these two objectives is generally not specified, however, which motivates focusing on protocols that are optimal for all $\lambda\ge 0$.

The weighting of ambiguity-averse and expected-utility components in the planner objective $U$ echoes a long tradition in economic theory \citep[e.g.,][]{gilboa1989maxmin,banerjee2020theory}. The planner objective $U$ differs from (but, as we discuss below, approximates) the objectives in \citet{gilboa1989maxmin} and \citet{banerjee2020theory}, which, in our notation, correspond to
\begin{equation}
\label{eqn:plannerobjective_prime}
 U'(r;\lambda, w)  =\min_{\theta \in \Theta} v_r(\theta,J_{r,\theta}^*, \Sigma_{r,\theta}^*) + \lambda \int v_r^\ast(\theta, J_{r,\theta}^*, \Sigma_{r,\theta}^*) w(\theta) d\theta, 
\end{equation}
for weights $w(\theta)$. The second component of $U'$ captures the welfare from implementing the treatments, whereas the second component of $U$ captures a preference for experimentation. The objective $U'$ has a decision-theoretic interpretation \citep{gilboa1989maxmin} and is related to Huber's $\varepsilon$-contamination model \citep[see][]{banerjee2020theory}. However, Appendix \ref{sec:discussion_objective} shows that exact solutions under $U'$ do not necessarily guarantee size control (as the solution depends on $\lambda$ and $w$). By contrast, working with the planner objective $U$ allows us to justify notions of size control and power and to obtain optimal protocols that do not depend on $w$ and $\lambda$, while retaining an approximate decision-theoretic justification.

\section{Optimal hypothesis testing protocols} 
\label{sec:multiple_treatments}

In this section, we characterize optimal hypothesis testing protocols without imposing additional functional form restrictions on the researcher's payoff or the planner's utility. 

\subsection{Null space, alternative space, and notions of optimality}
\noindent \textbf{Null space and alternative space.}
For a given set of treatments $J$, define the (global) \emph{null space}, the set of parameters for which the welfare effect of implementing any combination of treatments is negative, as
\begin{equation} \label{eqn:null_space}
\Theta_0(J) = \Big\{\theta: u_k(\theta; J) < 0 \text{ for all } k \in \{1, \dots, 2^{|J|} - 1\} \Big\} \subseteq \Theta.
\end{equation}
Similarly, define the null space given the treatments chosen by the researcher ex-ante after excluding the option not to experiment, $J_{r,\theta}^+$, as
$$
\Theta_0^*(r) = \left\{\theta: u_k(\theta, J_{r,\theta}^+) < 0 \text{ for all } k \in \{1, \dots, 2^{|J_{r,\theta}^+|}\} \right\}. 
$$ 

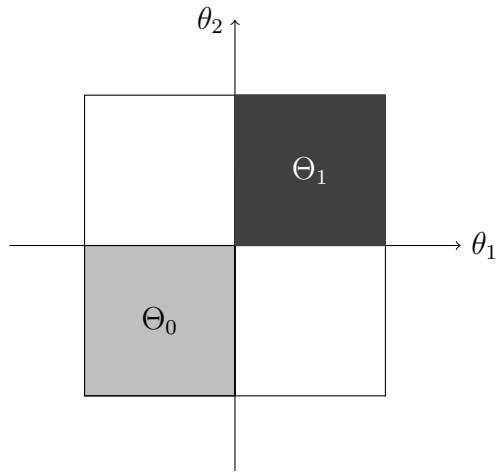
\begin{figure}[tp]
 \begin{center}
 \caption{Graphical illustration} \label{fig:illustrative}
    \begin{tikzpicture}
   
\coordinate (1) at (-4,3);
\coordinate (2) at (-2,3);
\coordinate (3) at (-2,5);
\coordinate (4) at (-4,5);
\coordinate (5) at ($(1)!.5!(2)$); 
\coordinate (6) at ($(2)!.5!(3)$);
\coordinate (7) at ($(3)!.5!(4)$);
\coordinate (8) at ($(1)!.5!(4)$);
\coordinate (9) at ($(1)!.5!(3)$);

\coordinate (10) at (-4,3);
\coordinate (11) at (-2,3);
\coordinate (12) at (-2,1);
\coordinate (13) at (-4,1);
\coordinate (14) at ($(10)!.5!(11)$); 
\coordinate (15) at ($(11)!.5!(12)$);
\coordinate (16) at ($(12)!.5!(13)$);
\coordinate (17) at ($(10)!.5!(13)$);
\coordinate (18) at ($(10)!.5!(14)$);

\coordinate (21) at (-2,3);
\coordinate (22) at (0,3);
\coordinate (23) at (0,5);
\coordinate (24) at (-2, 5);
\coordinate (25) at ($(21)!.5!(22)$); 
\coordinate (26) at ($(22)!.5!(23)$);
\coordinate (27) at ($(23)!.5!(24)$);
\coordinate (28) at ($(21)!.5!(24)$);
\coordinate (29) at ($(21)!.5!(23)$);

\coordinate (30) at (-2,3);
\coordinate (31) at (0,3);
\coordinate (32) at (0,1);
\coordinate (33) at (-2,1);
\coordinate (34) at ($(30)!.5!(31)$); 
\coordinate (35) at ($(31)!.5!(32)$);
\coordinate (36) at ($(32)!.5!(33)$);
\coordinate (37) at ($(30)!.5!(33)$);
\coordinate (38) at ($(30)!.5!(34)$);

 \draw [fill=lightgray] (-4,3) rectangle (-2,1); 

 \node[circle] (a) at (-3, 2) {$\Theta_0$};
\node[circle] (a) at (-1, 2) {};
\node[circle] (a) at (-3, 4) {};
\node[circle] (a) at (-1, 4) {};
 
\draw (1)--(2)--(3)--(4)-- cycle;

\draw (21)--(22)--(23)--(24)-- cycle;
\draw (10)--(11)--(12)--(13)-- cycle;
 \draw (30)--(31)--(32)--(33)-- cycle;
  
 \draw[->] (-5,3) -- (1,3) node[right] {$\theta_1$};
    \draw[->] (-2,0) -- (-2,6) node[left] {$\theta_2$};
    \draw [fill=,darkgray] (-2,5) rectangle (0,4); 
  \draw [fill=,darkgray,darkgray] (-2,5) rectangle (0,3);

   \node[circle] (a) at (-1, 4) {\textcolor{white}{$\Theta_1$}};
 
    \end{tikzpicture}
   \end{center}
 {\footnotesize\emph{Notes:} Graphical illustration of the null space $\Theta_0(J)=\left\{ \theta \in \Theta: \theta_1 < 0 \text{ and }  \theta_2 < 0 \right\}$ and the alternative space $\Theta_1(J)=\left\{ \theta \in \Theta: \theta_1 \ge 0 \text{ and }  \theta_2 \ge 0 \right\}$ for $J =\{1,2\}$, with additive welfare as in \eqref{eqn:exmp_additive}.
 }

\end{figure}

Moreover, define the (global) \textit{alternative space}, the set of parameters for which welfare is always positive, as 
\begin{equation} 
\Theta_1(J) = \Big\{\theta: u_k(\theta; J) \ge 0 \text{ for all } k \in \{1, \dots, 2^{|J|} - 1\}\Big\} \subseteq \Theta. 
\end{equation}
A graphical illustration of $\Theta_0(J)$ and $\Theta_1(J)$ is provided in Figure \ref{fig:illustrative}. We impose the following assumption.
\begin{ass}[Non-emptiness of null and alternative space] \label{ass:non_emptyness} Let $\bigcap_{J \in \mathcal{J} \setminus \emptyset} \Theta_0(J)\neq \emptyset$ and $\bigcap_{J \in \mathcal{J} \setminus \emptyset} \Theta_1(J) \neq \emptyset$ (and therefore $\Theta_0(J), \Theta_1(J) \neq \emptyset$ for all $J \neq \emptyset$).
\end{ass} 
Assumption \ref{ass:non_emptyness} states that for all combinations of treatments $J$, welfare is strictly negative for some values of $\theta$ and weakly positive for some other values of $\theta$.\footnote{This precludes interventions that everyone believes are sure to do good or those sure to cause harm, which are not of interest and would be precluded by, for example, rules regarding research ethics.}

Finally, denote by $\bar{\Theta}_1$ the set of parameters for which welfare is weakly positive for each choice of treatments $J$,
$
\bar{\Theta}_1 = \bigcap_{J \in \mathcal{J} \setminus \emptyset} \Theta_1(J).
$ 

\smallskip

\noindent \textbf{Notions of optimality.} The main notion of optimality we consider is uniform global optimality. We say that a protocol $r^*$ is \emph{uniformly globally optimal} if 
\begin{equation}
r^* \in \arg \max_{r\in \mathcal{R}} \left\{\min_{\theta \in \Theta} v_r(\theta, J_{r,\theta}^*, \Sigma_{r,\theta}^*) + \lambda \int_\Theta e_r^\ast(\theta') \pi(\theta') d\theta'\right\}\label{eq:planner_problem_subjective_utility}, \quad \forall \lambda \ge 0, \pi \in \Pi, 
\end{equation} 
for a given set $\Pi$. Uniformly globally optimal protocols do not depend on $\lambda$ and $\pi$, which is important in practice, as argued above.
In Appendix \ref{sec:discussion_objective}, we discuss the relationship between uniform global optimality and alternative notions of optimality. 

We also introduce two additional definitions that will be helpful for characterizing uniformly globally optimal protocols: maximin optimality and unbiasedness. We say that $r^*$ is \emph{maximin optimal} if it maximizes the planner's objective \eqref{eqn:plannerobjective} for $\lambda = 0$, that is,
$$
r^* \in \arg \max_{r \in \mathcal{R}} \min_{\theta \in \Theta} v_r(\theta, J_{r,\theta}^*, \Sigma_{r,\theta}^*). 
$$ 
We say that $r^*$ is \textit{unbiased} if
\begin{equation} \label{eqn:unbiased}
\beta_{r^*}^*(\theta) \ge 0\text{ for all } \theta \in \bar{\Theta}_1, \text{ where } \beta_{r^*}^*(\theta) = \max_{J \in \mathcal{J} \setminus \emptyset, \Sigma \in \mathcal{S}(J)} \beta_{r^*}(\theta,J,\Sigma).
 \end{equation} 
Here $\beta_r^*(\theta)$ is the largest net benefit the researcher can achieve when conducting an experiment. 
We use the term ``unbiased'' because, as we show below, this definition has a close connection to the definition of unbiased tests in the hypothesis testing literature, where a test is called unbiased if its power exceeds its size \citep{lehmann2005testing}.

\subsection{Characterizing optimal protocols}

Here, we characterize uniformly globally optimal and maximin optimal protocols. We impose the following assumption on the planner's weights $\pi$.
\begin{ass}[Weights over $\bar{\Theta}_1$] \label{ass:alternative_space} Suppose that $\pi \in \Pi$, where $\Pi$ is the class of functions $\pi$ satisfying $\pi(\theta) \ge 0$ for all $\theta \in \Theta$ and $\int_{\Theta} \pi(\theta) d\theta = \int_{\bar{\Theta}_1} \pi(\theta) d\theta > 0$. 
 \end{ass}
 
Assumption \ref{ass:alternative_space} restricts the support of the planner's weights to the alternative space $\bar{\Theta}_1$.   In other words, the planner desires to promote experimentation if she expects that treatments will generate a positive welfare effect and are therefore worth exploring. She derives no benefit (but also no harm), on the other hand, from exploration of parts of the parameter space in which some treatment effects are negative. 

Using classical hypothesis testing terminology, Assumption \ref{ass:alternative_space} allows for arbitrary alternative hypotheses over the positive orthant, including those in Section 4 of \citet{romano2011consonance} and Chapter 9.2 of \citet{lehmann2005testing}. Economically speaking, one can think of this assumption as ensuring that the components of the planner's utility function \eqref{eqn:plannerobjective} cleanly separate the two motives we wish to capture: avoiding harm and pursuing benefit. Doing so has the benefit that the components of utility will then map directly into the classical statistical concepts of size and power. As we discuss in more detail in Remark \ref{rem:remark_alternative}, Assumption \ref{ass:alternative_space} is necessary to justify testing protocols that control size, and we thus see Assumption \ref{ass:alternative_space} as natural when size control is a desideratum.

The following proposition characterizes uniformly globally optimal protocols.

\begin{prop}[Necessary and sufficient conditions for uniform global optimality]\label{cor:unbiasedness}  Suppose that Assumptions \ref{ass:non_emptyness} and \ref{ass:alternative_space} hold and that a maximin optimal and unbiased protocol exists. Then a protocol is uniformly globally optimal if and only if it is maximin optimal and unbiased.
\end{prop}

\begin{proof} See Appendix \ref{proof:cor:unbiasedness}.
\end{proof}

Proposition \ref{cor:unbiasedness} shows that uniformly globally optimal protocols are maximin optimal and unbiased. It assumes that a maximin and unbiased protocol exists. While the existence of such protocols is not guaranteed in general, we will show in Section \ref{sec:t-tests} that such protocols exist in leading cases. In the remainder of this section, we study maximin optimality and unbiasedness in more detail and show that these properties are related to notions of size control and power of testing protocols. 

The next proposition provides necessary and sufficient conditions for maximin optimality. It generalizes Proposition 1 in \citet{tetenov2016economic} (discussed in  Appendix \ref{sec:review_single_hypothesis}) to a setting in which $|J| > 1$ and where researchers can choose the experimental design and hypotheses to test. 


\begin{prop}[Necessary and sufficient conditions for maximin optimality] \label{prop:maximin2} Let Assumptions \ref{ass:non_emptyness} hold. A protocol $r^*$ is maximin optimal if and only if
 \begin{equation} \label{eqn:constraint} 
\begin{aligned} 
\beta_{r^*}(\theta, J_{r^*,\theta}^*, \Sigma_{r^*,\theta}^*) & \le 0 \quad & \forall \theta \in \Theta_0^*(r^*)  \\  v_{r^*}(\theta, J_{r^*,\theta}^*, \Sigma_{r^*,\theta}^*) & \ge 0 \quad & \forall \theta \in \Theta \setminus \Theta_0^*(r^*).  
\end{aligned} 
\end{equation} 
\end{prop} 
\begin{proof}
See Appendix \ref{app:prop:maximin}. 
\end{proof}
Proposition \ref{prop:maximin2} shows that maximin optimality is equivalent to two conditions. First, as in the case with a single hypothesis (i.e., $|J|=1$), maximin protocols deter experimentation over the null space $\Theta_0^*(r^*)$, where each configuration of treatments reduces welfare. This captures a notion of size control, as we discuss below.  Second, welfare must be non-negative for $\theta\in\Theta\setminus \Theta_0^*(r^*)$. This condition requires that if some treatments reduce welfare there must be others that compensate for them; it always holds in the single-hypothesis case but is non-trivial in the MHT case.

Building on Proposition \ref{cor:unbiasedness} and the sufficient conditions in Proposition \ref{prop:maximin2}, the following corollary provides sufficient conditions for uniform global optimality stated in terms of the global null and alternative spaces.

\begin{cor}[Sufficient condition for uniform global optimality]\label{cor:sufficient_maximin} Let Assumption \ref{ass:non_emptyness} hold. A protocol $r^*$ is maximin optimal if, for all $J \in \mathcal{J} \setminus \emptyset$ and $ \Sigma \in \mathcal{S}(J)$,
\begin{align}
\beta_{r^*}(\theta, J,\Sigma) \le 0 \quad & \forall \theta \in \Theta_0(J) \label{cor:sufficient_maximin_eq1} \\ 
v_{r^*}(\theta, J,\Sigma) \times 1\{\beta_{r^*}(\theta, J,\Sigma) > 0\}  \ge 0 \quad & \forall \theta \in \Theta \setminus \Theta_0(J) \label{cor:sufficient_maximin_eq2}. 
\end{align} 
If, in addition, Assumption \ref{ass:alternative_space} holds and $r^*$ satisfies, for all $J \in \mathcal{J} \setminus \emptyset, \Sigma \in \mathcal{S}(J)$,
\begin{align} 
\beta_{r^*}(\theta, J,\Sigma) & \ge 0 \quad & \forall \theta \in \Theta_1(J), \label{cor:sufficient_maximin_eq3}
\end{align}
then it is unbiased and therefore uniformly globally optimal. 
\end{cor}  
\begin{proof} See Appendix \ref{proof:cor:sufficient_maximin}. 
\end{proof}

Conditions \eqref{cor:sufficient_maximin_eq1} and \eqref{cor:sufficient_maximin_eq2} in Corollary \ref{cor:sufficient_maximin} mimic those in Proposition \ref{prop:maximin2}, but are required to hold for all potential choices of $J$ and $\Sigma$. These conditions may be easier to check than those in Proposition \ref{prop:maximin2} because they are stated in terms of $\Theta_0(J)$ rather than $\Theta_0^\ast(r^*)$, which is itself a function of the set of treatments pre-specified by the researcher in response to $r^\ast$. Note that in Condition \eqref{cor:sufficient_maximin_eq2}, we need weakly positive welfare only when the researcher finds it beneficial to experiment, since otherwise the researcher will not experiment and hence welfare will be zero. Condition \eqref{cor:sufficient_maximin_eq3} captures a stronger notion of unbiasedness that implies unbiasedness as defined in Equation \eqref{eqn:unbiased}. 

The theoretical results in this section establish close connections between uniform global optimality on the one hand, and size control and the power of testing protocols on the other. Specifically, maximin optimality captures a notion of size control, whereas unbiasedness captures a notion of power. 

\begin{exmp}[Average size control and unbiasedness with linear researcher benefits]\label{ex:average_size}
Suppose that $J = \{1,2\}$ and that welfare is additive as in Equation \eqref{eqn:exmp_additive}. Then the null space for $J  = \{1,2\}$ is $\Theta_0(\{1,2\}) = \left\{ \theta \in \Theta: \theta_1 < 0 \text{ and }  \theta_2 < 0 \right\}$ and the alternative space is $\Theta_1(\{1,2\})=\left\{ \theta \in \Theta: \theta_1 \ge 0 \text{ and }  \theta_2 \ge 0 \right\}$. Figure \ref{fig:illustrative} provides a graphical illustration. Suppose further that the researcher's payoff is 
$$
\beta_r(\theta;J,\Sigma) = b \sum_{j \in J} P(r_j(X;J,\Sigma) = 1|\theta) - C(J,\Sigma)
$$ 
for some constant $b > 0$. This payoff is a special case of the payoff in Equation \eqref{eqn:B_R_general}, where $p(\{1\}) = p(\{2\}) = b, p(\{1,2\}) = (b, b, 2b)$. Condition \eqref{cor:sufficient_maximin_eq1} in Corollary \ref{cor:sufficient_maximin} implies that 
\begin{equation} \label{eqn:illu_example_rast0}
 P(r^\ast_1(X;J,\Sigma) = 1|\theta_1, \theta_2) +  P(r_2^\ast(X;J,\Sigma) = 1|\theta_1, \theta_2) \le C(J, \Sigma)/b, \quad \forall \theta_1 < 0, \theta_2 < 0. 
\end{equation}
Equation \eqref{eqn:illu_example_rast0} is a size control requirement (i.e., a restriction on the probability of reporting false discoveries). Note that the right-hand side $C(J,\Sigma)/b$ is not a function of $r$. Here, $r^\ast$ satisfies Condition \eqref{cor:sufficient_maximin_eq3} in Corollary \ref{cor:sufficient_maximin} if and only if
\begin{equation} \label{eqn:illu_example_rast1}
 P(r^\ast_1(X;J,\Sigma) = 1|\theta_1, \theta_2) +  P(r_2^\ast(X;J,\Sigma) = 1|\theta_1, \theta_2) \ge C(\{1,2\},\Sigma)/b, ~\forall  \theta_1 \ge 0, \theta_2 \ge 0. 
\end{equation}
Equation \eqref{eqn:illu_example_rast1} can be interpreted as a power criterion: it requires the power of the protocol $r^\ast$ to (weakly) exceed the cost-to-benefit ratio of the experiment whenever $\theta \ge 0$. Taken together, Conditions \eqref{eqn:illu_example_rast0} and \eqref{eqn:illu_example_rast1} imply (under suitable continuity assumptions) that $ P(r^\ast_1(X;J,\Sigma) = 1|\theta_1, \theta_2) +  P(r_2^\ast(X;J,\Sigma) = 1|\theta_1, \theta_2) = C(J,\Sigma)/b$ for $\theta_1 = \theta_2 = 0$. In Section \ref{sec:t-tests}, we show that separate $t$-tests satisfy these optimality criteria.
\qed
\end{exmp}

\begin{exmp}[Weak FWER control with nonlinear research benefits] \label{exmp:fwer} Consider the setup in Example \ref{ex:average_size}, but suppose instead that the researcher's payoff is 
$$
\beta_r(\theta;J,\Sigma) = b P\left(\sum_{j \in J} r_j(X;J,\Sigma) \ge 1 | \theta\right) - C(J,\Sigma)
$$ 
for some constant $b > 0$, which is a special case of the payoff in Equation \eqref{eqn:B_R_general} with $p_j(J)=b$ for all $j$. Such a payoff might arise, for instance, if the firm conducting a trial needs at least one success in order to be solvent (or, when applied to academic publishing, if a researcher needs at least one significant result in order to publish). Condition \eqref{cor:sufficient_maximin_eq1} in Corollary \ref{cor:sufficient_maximin}  implies that 
\begin{equation}  
 P(\max_j r^\ast_j(X;J,\Sigma) = 1|\theta_1, \theta_2)  \le C(J, \Sigma)/b, \quad \forall \theta_1 < 0, \theta_2 < 0.
\end{equation}
That is, it implies weak control of the FWER (which coincides with the positive FDR for $\theta \in \Theta_0(J)$). See Appendix \ref{sec:additional_interactions_main} for additional details. \qed 
\end{exmp}

As discussed in Section \ref{sec:model}, rationalizing hypothesis testing (let alone multiple testing) is difficult in practice. The results and examples in this section show that one can write down a coherent economic objective function that rationalizes the standard statistical practice of choosing protocols that both control size and have non-trivial power. Specifically, optimal protocols must guarantee size control (encoded in the maximin optimality requirement) and also guarantee sufficient power against alternatives (encoded in the unbiasedness). Expressing these requirements in the form of an optimization problem has the benefit that it will allow us to then link the notions of size control and compound error rates directly to economic fundamentals, depending on the researcher's private costs and benefits and on how these scale with the number of hypotheses.

\begin{rem}[Assumption \ref{ass:alternative_space} and size control] \label{rem:remark_alternative}
Restrictions on $\Pi$ such as those in Assumption \ref{ass:alternative_space} are necessary to justify hypothesis testing protocols that control size (are maximin optimal).  Suppose instead that $\Pi$ contains all possible positive weights integrating to one (i.e., all prior distributions) on $\Theta \setminus \Theta_0(J)$ (the same reasoning applies if we simply consider all priors on $\Theta_0(J)$). Then there are no $r^*$ satisfying Equation \eqref{eq:planner_problem_subjective_utility} (for all $\lambda \ge 0, \pi \in \Pi$). To see this, fix a prior $\pi = 1\{\theta = (-1,0, \dots, 0)\}$ on $\Theta\setminus \Theta_0(J)$. Then for sufficiently large $\lambda$, assuming the researcher's payoff is strictly increasing in the number of findings, the protocol $r(X;J,\Sigma) = (1,  \dots, 1)$ dominates any maximin protocol, since when $\lambda$ is large enough the planner would prefer the researcher to experiment even when the resulting welfare effects are negative. In this example, the cost of approving harmful treatments is outweighed by the benefits of incentivizing experimentation for sufficiently large $\lambda$. Therefore, there exist combinations of $\pi$ and $\lambda$ such that the planner chooses protocols that do not control size and instead incentivizes experimentation regardless of the value of $\theta$. Restricting $\Pi$ to the strictly positive orthant guarantees that optimal protocols control size (are maximin optimal) for all $\lambda$ and motivates sufficiently powerful protocols only when $\theta$ is expected to have positive effects.   \qed 
\end{rem}

\begin{rem}[Deterrence of experimentation under the null] Proposition \ref{prop:maximin2} implies that experiments testing welfare-reducing treatments never occur in equilibrium, which might seem unrealistic. This result is a consequence of the simplifying assumption that the researcher has perfect information about $\theta$. Section \ref{sec:uncertain} extends our analysis to settings where the researchers has a prior about $\theta$, characterizing testing protocols that are maximin optimal with respect not just to point mass priors (equivalent to the problem in the main model) but with respect to arbitrary priors. In this scenario optimal protocols deter experimentation by a researcher who is \emph{certain} that some treatments are welfare-reducing, but may not deter testing by one who believes that treatment is very likely to be welfare-increasing but possibly harmful.  \qed 
\end{rem} 

\subsection{Robustness to researcher constraints in the design choice}

So far, we have assumed that after observing the protocol $r$, the researcher can choose any experiment $(J,\Sigma)$ with $J\in\mathcal{J}$ and $\Sigma\in S(J)$. In some applications, however, researchers may face
constraints that restrict the menu of treatments and designs they can implement. Because the planner may not know ex ante which experiments are feasible, we consider a refined notion of global optimality, referred to as \emph{design-robust global optimality}, that builds in robustness to design constraints.

\begin{defn}[Design-robust global optimality]\label{defn:robusta}
Consider a constrained environment in which the researcher can choose $(J,\Sigma) \in \mathcal{D}$ from a constrained set $\mathcal{D} \subseteq \{(J,\Sigma): J \in \mathcal{J}, \Sigma \in \mathcal{S}(J)\}$. For a given protocol $r$ and parameter $\theta$, let $\Big(J_{r,\theta}^*(\mathcal{D}), \Sigma_{r,\theta}^*(\mathcal{D})\Big) \in \mathrm{arg} \max_{(J',\Sigma') \in \mathcal{D} } \beta_{r}(\theta, J', \Sigma')$. 
We say that a protocol is \emph{design-robust globally
optimal} if
\[
r^{*}
\in
\arg\max_{r\in\mathcal{R}}
\left\{
\min_{\theta\in\Theta} v_r(\theta,J_{r,\theta}^*(\mathcal{D}), \Sigma_{r,\theta}^*(\mathcal{D})) 
+
\lambda \int_{\Theta} 1\{J_{r,\theta}^*(\mathcal{D}) \neq \emptyset \}\,\pi(\theta')\,d\theta'
\right\},
\]
for all $\lambda\ge 0$, $\pi\in\Pi$, and $\mathcal{D} \subseteq \{(J,\Sigma): J \in \mathcal{J}, \Sigma \in \mathcal{S}(J)\}$. 
\end{defn}

Definition~\ref{defn:robusta} requires protocols to be optimal for every potential set of experiments $(J,\Sigma)$ available to the researcher. Without the refinement of global optimality in Definition~\ref{defn:robusta}, the planner can afford to select protocols that are not unbiased (and hence may lead to low power) for designs that she expects the researcher not to choose. With it, she must consider the possibility that the researcher may be forced to choose any design, and thus must ensure that protocols are always unbiased.

Definition~\ref{defn:robusta} is appealing in settings where the planner has limited ex-ante knowledge of the researcher’s feasibility constraints, and thus wishes to mandate a protocol that performs well uniformly across all designs. 

The following proposition provides a characterization of design-robust globally optimal protocols.

\begin{prop}\label{cor:iff_robust}
Let Assumptions \ref{ass:non_emptyness} and \ref{ass:alternative_space} hold. Then $r^*$ is design-robust
globally optimal if and only if Equations
\eqref{cor:sufficient_maximin_eq1}, \eqref{cor:sufficient_maximin_eq2}, and \eqref{cor:sufficient_maximin_eq3} hold for all $J \in \mathcal{J} \setminus \emptyset$ and $ \Sigma \in \mathcal{S}(J)$, assuming such a protocol exists. 
\end{prop}

\begin{proof} See Appendix \ref{proof:cor:iff_robust}. 
\end{proof}
Proposition~\ref{cor:iff_robust} shows that the sufficient conditions for global optimality in Corollary \ref{cor:sufficient_maximin} become necessary once we strengthen the notion of global optimality to design-robust global optimality.

\subsection{Robustness to uncertainty about the researcher's payoff} 

So far, we have assumed that the planner knows the researcher's payoff. The following proposition shows that maximin optimality for protocols satisfying Equations \eqref{cor:sufficient_maximin_eq1} and \eqref{cor:sufficient_maximin_eq2} is preserved if the planner knows only an upper bound on the researcher's payoff. Uniform global optimality is preserved under additional restrictions on $\Pi$.  

\begin{prop}[Robustness to unknown researcher payoffs]  \label{cor:maximin_optimal} Let Assumptions \ref{ass:non_emptyness} and \ref{ass:alternative_space} hold. Then 
\begin{enumerate}[(i)]\setlength\itemsep{0pt}
\item Any protocol $r^*$ satisfying Equations \eqref{cor:sufficient_maximin_eq1} and \eqref{cor:sufficient_maximin_eq2} under payoff function $\beta_{r^*}(\theta, J, \Sigma)$ is also maximin optimal for any  $\beta_{r^*}'(\theta, J, \Sigma)$ such that $\beta_{r^*}'(\theta, J, \Sigma) \le \beta_{r^*}(\theta, J, \Sigma)$ for all $\theta \in \Theta, J \in \mathcal{J}, \Sigma \in \mathcal{S}(J)$. \label{cor:maximin_optimal_i}
\item Any protocol $r^*$ satisfying  the conditions in Corollary \ref{cor:sufficient_maximin} under payoff function $\beta_{r^*}(\theta, J, \Sigma)$ is uniformly globally optimal for any  $\beta_{r^*}'(\theta, J, \Sigma)$ such that $\beta_{r^*}'(\theta, J, \Sigma) \le \beta_{r^*}(\theta, J, \Sigma)$ for all $\theta \in \Theta, J \in \mathcal{J}, \Sigma \in \mathcal{S}(J)$ and any positive function $\pi \in \tilde{\Pi}$, where $\tilde{\Pi} =\{\pi: \int_{\tilde{\Theta}_1(r^*)} \pi(\theta) d\theta = \int_{\Theta} \pi(\theta) d\theta\}, \tilde{\Theta}_1(r^*) = \{\theta: \beta_{r^*}'(\theta, J_{r^*,\theta}^*, \Sigma_{r^*,\theta}^*) \ge 0\}$. \label{cor:maximin_optimal_ii}
\end{enumerate}
\end{prop} 

\begin{proof}[Proof of Proposition \ref{cor:maximin_optimal}] See Appendix \ref{proof:maximin_3}. 
\end{proof} 

Proposition \ref{cor:maximin_optimal}(\ref{cor:maximin_optimal_i}) demonstrates an important robustness property of our maximin optimality results in settings where the researcher's payoff function is unknown. Proposition \ref{cor:maximin_optimal}(\ref{cor:maximin_optimal_ii}) states that protocols that are uniformly globally optimal with respect to an upper bound $\beta_r(\theta,J,\Sigma)$ are also uniformly globally optimal for weights $\pi \in \tilde\Pi$. That is, uniform global optimality is preserved when considering a (weakly) smaller class of alternatives. For example, for the optimal separate $t$-tests in Section \ref{sec:t-tests}, the set of weights $\tilde{\Pi}$ is a subset of the set of weights on strictly positive treatment effects. Note that $\tilde{\Theta}_1(r^*)$ and $\tilde{\Pi}$ do not need to be known for the planner to implement the optimal protocol. 

Proposition \ref{cor:maximin_optimal} is particularly useful when applied to settings in which the planner's uncertainty about the researcher's payoff hinges on the researcher's costs. 
\begin{cor}[Unknown cost function] \label{cor:unknown_costs} Let the conditions in Proposition \ref{cor:maximin_optimal} hold, and suppose that the researcher's costs do not depend on $\theta$, so that $c_\theta(J,\Sigma)=C(J,\Sigma)$. Consider $\beta_r(\theta, J,\Sigma) - \beta_r'(\theta,J,\Sigma) = C'(J, \Sigma) - C(J, \Sigma)$, for some $C'(J,\Sigma) \ge C(J, \Sigma)$. Then any maximin and unbiased protocol $r^*$ under $\beta_{r}$ is also maximin under $\beta_{r}'$ and uniformly globally optimal for any $\pi \in \tilde{\Pi}$, with $\tilde{\Pi}$ defined in Proposition \ref{cor:maximin_optimal}.
\end{cor}
\begin{proof}The proof follows directly from Proposition \ref{cor:maximin_optimal}. \end{proof}
Corollary \ref{cor:unknown_costs} states that in settings with uncertainty over the true cost function $C'(J,\Sigma)$, the planner may use sensible lower bounds $C(J,\Sigma) \le C'(J,\Sigma)$. This result is important in empirical applications such as the ones we consider in Section \ref{sec:applications}.

\section{Optimal protocols under linearity and normality}

\label{sec:t-tests}
Which (if any) specific hypothesis testing protocols are uniformly globally optimal? The answer depends on the functional form of the researcher's payoff, the functional form of welfare, and the distribution of $X$. Here, we show that separate $t$-tests are optimal in settings with a linear researcher payoff function (Assumption \ref{ass:linear_rule}), a linear welfare function (Assumption  \ref{ass:additive2}), and a normally distributed vector of statistics $X$ (Assumption \ref{ass:partially_contractable}).

\subsection{Assumptions}

\noindent \textbf{Linearity assumptions.} Let $\omega$ denote a vector of weights and define $\bar{\omega}(J) = \sum_{j= 1}^{|J|} \omega_{J_j}$ for $J\in \mathcal{J}$. These weights will let us capture factors that affect the importance of the different treatments symmetrically from the point of view of both the researcher and the planner; they also nest the case in which all treatments are equally important ($\omega_{J_j} = 1$ $\forall j$). We consider the following assumption on the researcher's payoff.
\begin{ass}[Linear payoff] \label{ass:linear_rule} The researcher's payoff is (recall that the entry $r_j(X;J,\Sigma)$ corresponds to treatment $J_j$)
 \begin{equation} \label{eqn:jk} 
\beta_r(\theta, J, \Sigma) =   b \int \sum_{j = 1}^{|J|}  \omega_{J_j} r_j(x;J, \Sigma)  dF_{\theta, J, \Sigma}(x) - C(J,\Sigma),
\end{equation}
where $\omega_j \in (0,\infty)$ for all $j$,  $b > 0$, $b \bar{\omega}(J) > C(J,\Sigma) > 0$ for all $(J,\Sigma)$, and $C(J,\Sigma)$ does not depend on $\theta$. 
\end{ass} 
The payoff function \eqref{eqn:jk} in Assumption \ref{ass:linear_rule} is a special case of the general payoff function \eqref{eqn:B_R_general}, and the condition $b \bar{\omega}(J) > C(J,\Sigma)$ guarantees that the experiment $(J, \Sigma)$ is a relevant option; otherwise the researcher would never conduct this experiment, regardless of $r$.
Note that Assumption \ref{ass:linear_rule} rules out interactions between treatments in the researcher's utility, which we discuss in Appendix \ref{sec:additional_interactions_main}. In addition, we assume that the researcher costs do not depend on $\theta$ and write them as $ C(J,\Sigma)$. 

We consider the following linearity assumption on welfare. 
\begin{ass}[Linear welfare] \label{ass:additive2} For all $r \in \mathcal{R}$, 
$$\delta(r(x; J, \Sigma); J)^\top u(\theta; J)  = \sum_{j = 1}^{|J|} \omega_{J_j} r_j(x;J, \Sigma) u(\theta;J_j),$$ 
where $u(\theta;J_j)$ is the welfare from implementing treatment $J_j$.
\end{ass} 

Assumptions \ref{ass:linear_rule} and \ref{ass:additive2} capture a setting in which the researcher's and planner's objectives differ: the expected researcher's payoff depends on the (weighted) expected number of findings, whereas the welfare component of the planner's objective depends on the welfare effects that such findings generate. Before stating results, we provide an interpretation of these assumptions in the context of our leading example, the drug approval process. 

With \emph{multiple subgroups}, we interpret $r_j(X;J,\Sigma)$ as indicating whether the drug was found to be effective for subgroup $J_j$, which is of size $\omega_{J_j}$. The component $b \omega_{J_j}$ denotes the expected profits from selling the drug to subgroup $J_j$, where $b$ denotes the average per-sale profit. Assumption \ref{ass:linear_rule} then states that researchers care about the sum of the expected profits they can earn by selling the drug to each of the subpopulations for which its use is approved. Our specification of welfare in Assumption \ref{ass:additive2}, meanwhile, corresponds to the utilitarian welfare from approving the drug, where $u(\theta;J_j)$ denotes the per unit treatment effect on members of subgroup $J_j$. The economic import of the assumption is that there are no spillovers between different subgroups. 

With \emph{multiple treatments} the interpretation is similar, but here each treatment denotes a different drug in the same market. $\omega_{J_j}$ denotes the expected number of individuals that would purchase and use drug $J_j$ if approved, and $u(\theta;J_j)$ is the effect of drug $J_j$ on those individuals. As before, $b$ denotes the average per-sale profit. The economic import of Assumption \ref{ass:additive2} is that the sets of people who would use the different drugs are disjoint (or that the drugs do not exhibit interaction effects). Symmetry between the planner's and researcher's weights here implies that per-user profit and per-user consumer surplus are proportional across drugs. Without this property it is no longer necessarily the case that the separate $t$-tests defined in Equation \eqref{eq:optimal_t_test} below guarantee uniformly non-negative welfare (and therefore are maximin optimal); we do not characterize optimal testing protocols in that scenario, but note that this could be a fruitful direction for future work.

\smallskip
\noindent\textbf{Normality assumption.} We focus on the leading case where $X$ is normally distributed, which is motivated by standard normal approximations (e.g., Berry-Esseen bounds). Let $\theta_J$ denote the subvector of $\theta$ corresponding to treatments $J$. 

\begin{ass}[Normality and homogeneous variances] \label{ass:partially_contractable} For each experiment $(J,\Sigma)$, let $X\sim \mathcal{N}(\theta_J,\Sigma)$ and suppose that $u(\theta, j) = \theta_j$, with $\theta_j \in [-M,M]$ for a positive constant $M > 0$, so that $X_j$ is an unbiased normally distributed signal of the welfare effect of treatment $j$. The class of designs $\mathcal{S}(J)$ is such that $\Sigma_{i,i} = \Sigma_{j,j} \in (\underline{\gamma}, \bar{\gamma})$ for all $i,j \in \{1, \dots, |J|\}$ and some constants $0<\underline{\gamma} < \bar{\gamma} < \infty$. 
\end{ass}

Assumption \ref{ass:partially_contractable} imposes that the vector of statistics $X$ is normally distributed, centered around the vector of welfare effects $\theta_J$ and with finite sample covariance matrix $\Sigma$.\footnote{Because $\Sigma$ is the finite sample covariance matrix, it is proportional to the inverse of the square root of the sample size in the experiment.} The researcher can choose the covariance matrix, but we restrict the class of designs she can choose from to designs in which the variances (the diagonal entries of $\Sigma$) are positive and equal to each other. 
This implies that the researcher can choose the overall sample size of the experiment, for example, but is constrained in allocating sample across experimental arms. See Section \ref{sec:unknown_variances} for an extension to designs with heterogeneous variances. 


\subsection{Optimality of $t$-tests} \label{sec:linear}
\label{sec: most powerful tests}

The following proposition shows that separate $t$-tests are maximin optimal under Assumptions \ref{ass:linear_rule}, \ref{ass:additive2}, and \ref{ass:partially_contractable} and uniformly and design-robust globally optimal if, in addition, Assumption \ref{ass:alternative_space} holds.

\begin{prop}[Optimality of separate $t$-tests] \label{cor:threshold_main1}  Let Assumptions \ref{ass:linear_rule}, \ref{ass:additive2}, and \ref{ass:partially_contractable} hold. 
Consider the testing protocol $r^t(X; J, \Sigma)=\left(r^t_1(X;J, \Sigma), \dots, r^t_{|J|}(X;J, \Sigma)\right)^\top$, with 
\begin{equation}  \label{eq:optimal_t_test}
r_j^t(X;J, \Sigma) = 
1\left\{\frac{X_j}{\sqrt{\Sigma_{j,j}}} \ge t(J,\Sigma) \right\}, \forall j \in J, 
\end{equation}  
for a threshold $t(J,\Sigma) \in \mathbb{R}$. 
Then $r^t$ 
is maximin optimal if and only if $t(J,\Sigma) \ge \Phi^{-1}\left(1 - \frac{C(J, \Sigma)}{b \bar{\omega}(J)} \right)$ for all $(J,\Sigma)$. If, in addition, Assumption \ref{ass:alternative_space} holds, then $r^t$ is maximin optimal and unbiased (and thus uniformly globally optimal) if and only if it also satisfies $t(J,\Sigma) = \Phi^{-1}\left(1 - \frac{C(J, \Sigma)}{b \bar{\omega}(J)} \right)$ for at least some $(J,\Sigma)$ with $J \neq \emptyset$. Finally, $r^t$ is also design-robust globally optimal if and only if  $t(J,\Sigma) = \Phi^{-1}\left(1 - \frac{C(J, \Sigma)}{b \bar{\omega}(J)} \right)$ for all feasible $J \neq \emptyset$ and  $\Sigma$. 
\end{prop} 
\begin{proof}  
See Appendix \ref{app:cor:threshold}. 
\end{proof}

Proposition \ref{cor:threshold_main1} shows that separate one-sided $t$-tests with critical values $t\ge \Phi^{-1}\left(1 - \frac{C(J, \Sigma)}{b\bar{\omega}(J)} \right)$ are maximin optimal, where we write $t$ in lieu of $t(J,\Sigma)$ to simplify notation. The key technical step in the proof is to show that under this protocol welfare is non-negative even when parameters have different signs (Equation \eqref{cor:sufficient_maximin_eq2}). Proposition \ref{cor:threshold_main1} further shows that standard separate one-sided $t$-tests are maximin optimal and unbiased and thus uniformly globally optimal by Proposition \ref{cor:unbiasedness} and also design-robust globally optimal.\footnote{It may seem surprising that the result in Proposition \ref{cor:threshold_main1} holds for any $\lambda$. Intuitively, once we focus on $\Pi$ defined in Assumption \ref{ass:alternative_space}, we can always find a maximin protocol that guarantees experimentation for each value of $\theta$ in the positive orthant. As a result, no matter how much weight the planner puts on her subjective utility, at the optimum, any optimal protocol will maximize separately (and therefore jointly) maximin welfare and subjective utility from experimentation.} While $t$-tests with all thresholds larger than $\Phi^{-1}(1 - \frac{C(J, \Sigma)}{b\bar{\omega}(J)})$ are maximin optimal, such tests are not uniformly globally optimal. Only $t$-tests with threshold $t=\Phi^{-1}(1 - \frac{C(J, \Sigma)}{b\bar{\omega}(J)})$ for at least some $(J,\Sigma)$ are uniformly globally optimal, and only $t$-tests with threshold $t=\Phi^{-1}(1 - \frac{C(J, \Sigma)}{b\bar{\omega}(J)})$ for all experiments $(J,\Sigma)$ are design-robust globally optimal. This demonstrates that uniform global optimality is a refinement of maximin optimality, restricting attention to protocols with sufficient power for at least some experiment, and that design-robust global optimality is in turn a refinement of uniform global optimality, restricting attention to protocols with sufficient power for all experiments. 

Proposition \ref{cor:threshold_main1} also shows that whether and to what extent the level of these separate tests should depend on the number of hypotheses being tested depends on the structure of the research production function $C(J, \Sigma)$ and on $\bar{\omega}(J)$, and in particular on how they vary with $J$. For example, suppose that $\omega_j = 1$ for all $j$ (all treatments are equally important) so that $\bar{\omega}(J)=|J|$. If $C(J, \Sigma) = \alpha$ for some constant $\alpha$ then a Bonferroni correction is optimal. This corresponds to a stylized case in which the costs of experimentation are fixed regardless of the number of treatments tested ($|J|$) or the precision of the estimates ($\Sigma$).  If, on the other hand, $C(J, \Sigma) = \alpha |J|$ then the optimal level of the test is $\alpha$, irrespective of $|J|$. This might correspond, for example, to a case in which there are no fixed costs and testing each additional treatment requires the same increment to the sample size. The former case arguably captures the lay intuition that if the researcher can test many hypotheses in the hopes of securing some private benefit, then the planner should require a hypothesis testing protocol that discourages this. In the latter case it is still true that the researcher obtains a higher expected reward from taking on projects that test more hypotheses, ceteris paribus, but the appropriate correction for this is already ``built in'' to the costs of conducting research, so that no further correction is required. 

While our original motivation for obtaining the result in Proposition \ref{cor:threshold_main1} was to study the consequences of multiplicity ($|J|$), it follows immediately that, because the costs $C(J, \Sigma)$ may depend on the design $\Sigma$ in addition to $J$, the optimal critical values may as well. For example, if the researcher can choose the number of treatments to test and also the sample size $\bar{n}$ to use per treatment, then her cost structure might take the form $C(J, \Sigma) = c_f +  c_{|J|} |J| + c_{\bar{n}} |J| \bar{n}$, where $c_f$ is a fixed cost (e.g., the costs of staff scientists), $c_{|J|}$ a cost that varies with the number of treatments tested (e.g., the cost of training clinical staff on various treatment protocols), and $c_{\bar{n}}$ a cost per experimental subject (e.g., recruitment costs). In this case (normalizing $b = 1$ and again assuming for simplicity that $\bar{\omega}(J)=|J|$) we obtain optimal thresholds $t=\Phi^{-1}(1 - c_f/|J| - c_{|J|} - c_{\bar{n}} \bar{n})$ which are decreasing in the (per-treatment) sample size $\bar{n}$ as well as in the cost per treatment. Intuitively, to the extent that large-sample experiments are more costly to the researcher to run, the planner need worry less about discouraging the researcher from running such experiments when doing so would not be socially optimal. This point follows from exactly the same economic logic that rationalizes adjusting testing thresholds with respect to $J$, but has not (to our knowledge) come up in past discussions of MHT adjustment. In that sense, it illustrates the value of working out the economic logic underlying MHT adjustment carefully, to make sure we have fully grasped the consequences of any implicit assumptions.

The practical value of Proposition \ref{cor:threshold_main1} lies in the fact that it connects optimal testing protocols to measurable properties of the cost function $C(J,\Sigma)$. We illustrate this in Section \ref{sec:implications_practice} where we develop the application to clinical trials, using publicly available data on moments of their cost structure to derive specific testing thresholds. Appendix \ref{app:jpal} provides a second illustration, applying the framework to experimental program evaluation research in economics using unique data from the Jameel Poverty Action Lab (J-PAL). Readers primarily interested in implications for practice may wish to skip ahead to these exercises.

\begin{rem}[One-sided vs.\ two-sided tests]\label{rem:two-sided tests} Under the assumptions in this section, separate one-sided $t$-tests are uniformly globally optimal. This is because the status quo remains in place if the researcher does not report any findings. This may be one reason that one-sided tests have been seen as appropriate in the drug approval context.\footnote{For example, former FDA advisor Lloyd Fisher writes that ``For drugs that may be tested against placebos, with two positive trials required (as in the United States), it is argued that from both a regulatory and pharmaceutical industry perspective, one-sided tests at the 0.05 significance level are appropriate. In situations where only one trial against a placebo may be done (for example, survival trials), one-sided tests at the 0.025 level are appropriate in many cases.''\citep{fisher1991use}} When there is uncertainty about the planner's action when no findings are reported, on the other hand, our model can justify two-sided hypothesis testing. We report this result in Appendix \ref{app: two-sided tests}.  \qed 
\end{rem}

\begin{rem}[Average size control and FWER control] \label{rem:average}
When the researcher's payoff is linear (Assumption \ref{ass:linear_rule}) and $\omega_j = 1$ for all $j$, Proposition \ref{cor:threshold_main1} implies that we can find uniformly globally optimal protocols that impose average size control, $b/|J|\sum_{j \in J}  P(r_j^*(X;J, \Sigma) = 1 | \theta = 0) \le C(J, \Sigma)/|J|$. Many popular MHT corrections do not directly target average size control and, thus, will generally not be optimal in our model. For example, if $C(J, \Sigma)$ is constant then classical Bonferroni correction is optimal in our model, since it satisfies average size control, but common refinements of Bonferroni such as \citet{holm1979simple}'s method are not. This result is driven by the linearity of the researcher's payoff function; Bonferroni corrections may not be optimal with nonlinear payoff functions (but are maximin optimal for all payoff functions dominated by a linear payoff function; see Proposition \ref{cor:maximin_optimal}). We  discuss the impact of non-linearities for the design of optimal protocols in Appendix \ref{sec:additional_interactions_main}. \qed
\end{rem}

\section{Main extensions}

Here we present two extensions of our main results: a variant of our model where the researcher knows $\theta$ only imperfectly (Section \ref{sec:uncertain}) and a relaxation of the variance homogeneity requirement in Assumption \ref{ass:partially_contractable} (Section \ref{sec:unknown_variances}). Appendix \ref{app:extensions} presents additional extensions.

\subsection{Imperfectly informed researchers} \label{sec:uncertain}

So far, we have assumed that the researcher is perfectly informed and knows $\theta$. Here, we show that our main results continue to hold in settings where the researcher has imperfect information in the form of a prior about $\theta$.\footnote{In the single-hypothesis testing case, \citet{tetenov2016economic} gives results under imperfect information. However, these results rely on the Neyman-Pearson lemma, which is not applicable to multiple tests.} Denote this prior by $\pi' \in \Pi'$, where $\Pi'$ is the class of all distributions over $\Theta$.\footnote{The assumption that $\Pi'$ is unrestricted is made for simplicity. For our theoretical results, we only need that the class of priors $\Pi'$ contains at least one element that is supported on the null space $\bigcap_{J \in \mathcal{J} \setminus \emptyset} \Theta_0(J)$.} The prior $\pi'$ captures knowledge about $\theta$ that is available to the researcher but not to the planner.

We assume that the vector of statistics $X$ is drawn from a normal distribution conditional on $\theta$, where $\theta$ itself is drawn from the prior $\pi'$ with $\int_\Theta \pi'(\theta) d\theta = 1$:
$$
X \mid \theta \sim \mathcal{N}(\theta, \Sigma), \quad \theta \sim \pi', \quad \pi' \in \Pi',
$$  
where $\Sigma$ is positive definite and assumed to be known after being chosen by the researcher.

The researcher acts as a Bayesian decision-maker and chooses 
\begin{equation*} 
\left(J_{r,\pi'}^*, \Sigma_{r,\pi'}^* \right)\in \mathrm{arg} \max_{J \in \mathcal{J}, \Sigma \in \mathcal{S}(J)} \int \beta_r(\theta, J, \Sigma) d\pi'(\theta).
\end{equation*} 
The researcher's prior is correctly specified, and welfare is given by
\begin{equation*} 
\bar{v}_r(\pi', J, \Sigma) =  \int v_r(\theta, J, \Sigma) d\pi'(\theta). 
\end{equation*} 

Under imperfect information, we define maximin protocols with respect to the prior $\pi'$. 
\begin{defn}[$\Pi'$-maximin optimal] \label{defn:delta} We say that $r^*$ is $\Pi'$-maximin optimal if and only if
$$
r^* \in \mathrm{arg} \max_{r\in \mathcal{R}} \inf_{\pi' \in \Pi'} \bar{v}_r(\pi', J_{r,\pi'}^*, \Sigma_{r,\pi'}^*). 
$$ 
\end{defn} 

Definition \ref{defn:delta} generalizes the notion of maximin optimality in Section \ref{sec:multiple_treatments}, which is stated in terms of the parameter $\theta$. When $\Pi'$ contains only point mass distributions, the two notions of maximin optimality are equivalent.

The next proposition shows that one-sided $t$-tests with appropriately chosen critical values are maximin optimal under imperfect information. 
\begin{prop}[Maximin optimality] \label{prop:uninformed} Let Assumptions \ref{ass:linear_rule}, \ref{ass:additive2}, and \ref{ass:partially_contractable} hold.  
Then the protocol 
$$
r_j^t(X; J,\Sigma) = 1\left\{\frac{X_j}{\sqrt{\Sigma_{j,j}}} \ge \Phi^{-1}\left(1 - \frac{C(J, \Sigma)}{b \bar{\omega}(J)} \right)\right\}, \quad \forall j \in J, 
$$ 
is $\Pi'$-maximin optimal. 
\end{prop} 
\begin{proof}
See Appendix \ref{app:sec:prop_uninformed}. 
\end{proof}

Proposition \ref{prop:uninformed} shows that the maximin optimality of separate $t$-tests continues to hold under imperfect information. Intuitively, maximin optimality is preserved here because the worst-case prior is a point mass prior over $\theta$, so that the same reasoning as under perfect information applies. This follows from classical results on linear programs (see Appendix \ref{app:sec:prop_uninformed}). Uniform global optimality of $r^{t}$ follows as a corollary of Proposition \ref{cor:threshold_main1}.

\subsection{Heterogeneous variances} \label{sec:unknown_variances}

Assumption \ref{ass:partially_contractable} restricts the class of designs the researcher can choose from to designs where all $X_j$ have the same variance, $\Sigma_{i,i}=\Sigma_{j,j}$. This assumption may seem strong since one might expect the researcher to choose a design with unequal variances, especially if she believes that the outcomes are more variable under some treatments than under others. Here, we consider settings where the researcher can choose designs with heterogeneous variances. Specifically, we study a version of our model in which the researcher can choose the sample size for each treatment arm, and thus the variances of the test statistics, but has only imperfect knowledge of the underlying heterogeneous outcome variances in the treatment and control groups. See Remark \ref{rem:variance_equalizing_allocation} for a discussion of settings with heterogeneous variances where the researcher has full information and can thus choose the sample size as a function of such variances. 

We assume that for a given $(J,\Sigma)$,  
$X\sim \mathcal{N}(\theta_J,\Sigma)$,
where for each $j$, we interpret $\theta_j$ as a ratio between the treatment effect $\tau_j$ and $\sigma_j := \sqrt{\sigma_{1j}^2/2 + \sigma_{0j}^2/2}$, where $\sigma_{1j}^2$ and $\sigma_{0j}^2$ are the variances of the treated and the control outcomes in the experiment, respectively. 
Define the vector of sample sizes in an experiment with treatments $J$ as $n(J) = \{n_{1j}, n_{0j}\}_{j \in J}$. Here, $n_{1j}$ and $n_{0j}$ are the number of treated and control units in arm $j$ (with $n_{0j}$ potentially constant across $j$ if all arms share the same control group). In this model, $\Sigma_{j,j} = \frac{\sigma_{1j}^2}{\sigma_j^2 n_{1j}} + \frac{\sigma_{0j}^2}{\sigma_j^2 n_{0j}}$. 

Suppose that the researcher only knows $\theta$, the effect measured in standard deviations $\tau_j/\sigma_{j}$ for each $j$, but not $(\tau_j,\sigma_{1j}^2,\sigma_{0j}^2)$ separately. To encode such uncertainty, we assume that the researcher has a prior $(\tau_j,\sigma_{1j}^2,\sigma_{0j}^2) \sim \mathcal{P}_{\theta,j}$, which depends on $\theta$. Here $u(\theta, \{j\}) = \mathbb{E}[\tau_j | \theta]$ is the expected treatment effect $\tau_j$ given $\theta$ under the prior $\mathcal{P}_{\theta,j}$. We assume that (with a slight abuse of notation) $C(J,\Sigma)=C(J,n(J))$ (where $c_\theta(J,\Sigma) = C(J,\Sigma)$ is constant in $\theta$), so that the costs are known to the researcher. That is, the costs can depend on the sample sizes in the experiment, $n(J)$, but not on the unknown to the researcher variances $\{\sigma_{1j}^2,\sigma_{0j}^2\}_{j\in J}$.

The following assumption summarizes the model with unknown heterogeneous variances.
\begin{ass}[Model with unknown heterogeneous variances] \label{ass:t_tests2} Suppose that for all $j$, $\theta_j = \tau_j/\sigma_j$ for some $\sigma_j > 0$, with $\theta_j \in [-M,M]$, for a positive constant $M > 0$. Let $\sigma_j^2 = \sigma_{1j}^2/2 + \sigma_{0j}^2/2$, where $\sigma_{0j}^2$ and $\sigma_{1j}^2$ are bounded away from zero almost surely.  Let 
$(\tau_j,\sigma_{1j}^2,\sigma_{0j}^2) \sim \mathcal{P}_{\theta,j}$ and $u(\theta, \{j\}) = \mathbb{E}[\tau_j| \theta]$. For an experiment with treatments $J$, let $X | (\tau,\sigma) \sim \mathcal{N}(\theta_J,\Sigma)$, with $\Sigma_{j,j} = \frac{\sigma_{1j}^2}{\sigma_j^2 n_{1j}} + \frac{\sigma_{0j}^2}{\sigma_j^2 n_{0j}}$ for all $j \in J$. The class of designs $\mathcal{S}(J)$ is such that $n_{1j}, n_{0j} \le n$ for a finite constant $n < \infty$ and the researcher can choose $\{n_{1j}, n_{0j}\}_{j\in J}$. Finally, assume that $C(J,\Sigma)=C(J,n(J))$.
\end{ass} 

We consider a setting in which the researcher has limited knowledge, captured by the assumption that $(\sigma_{1j}, \sigma_{0j})$  have a common expectation across $j$, conditional on $\theta$. 

\begin{ass}[Common expectation] \label{ass:exchangeable} For some $\bar{\sigma} > 0$,  $\mathbb{E}[\sigma_j|\theta] = \bar{\sigma}$ for all $j$. 
\end{ass}

Under Assumption \ref{ass:exchangeable}, the researcher expects the standard deviations to be the same before running the experiment. Importantly, however, Assumption \ref{ass:exchangeable} allows the realized variances to be heterogeneous.

Under Assumption \ref{ass:t_tests2},  $\{n_{1j},n_{0j}\}_{j\in J}$ are chosen by the researcher before running the experiment and observed by the planner. As a result, the planner can de-facto mandate any choice of sample sizes by only rewarding designs that maximize her utility.

We say that the design $\Sigma$ has a \emph{sample-equalizing allocation} if $n_{1j} = n_{0j} = \bar{m}$ for all $j \in \{1, \dots, |J|\}$ and for some constant $\bar{m} \in (0,n]$ (that can be chosen by the researcher). The next proposition shows that designs with sample-equalizing allocations are optimal.
 
\begin{prop}[Optimality of separate $t$-tests] \label{cor:threshold_aa}  Let Assumptions \ref{ass:linear_rule}, \ref{ass:additive2}, \ref{ass:t_tests2}, and \ref{ass:exchangeable} hold. Then, the testing protocol $r^{\mathrm{N}}(X)$, where
\begin{equation}  \label{eq:optimal_t_test2}
\small 
\begin{aligned}
r_j^{\mathrm{N}}(X;J, \Sigma) = \begin{cases} 
1\left\{\frac{X_j}{\sqrt{\Sigma_{j,j}}} \ge t^*(J,n(J)) \right\}, \quad \forall j \in J & \text{ if } \Sigma \text{ has a sample-equalizing allocation} \\ 
0 & \text{ otherwise}
\end{cases} 
\end{aligned}
\end{equation}  
is maximin and unbiased if $t^*(J,n(J)) = \Phi^{-1}\left(1 - \frac{C(J, n(J))}{b \bar{\omega}(J)} \right)$. 
\end{prop}  

\begin{proof} See Appendix \ref{proof:cor:threshold_aa}. 
\end{proof}

Proposition \ref{cor:threshold_aa} shows that separate one-sided $t$-tests with critical values $t = \Phi^{-1}\left(1 - \frac{C(J, n(J))}{b \bar{\omega}(J)} \right)$ based on sample-equalizing allocations are maximin and unbiased. The planner ``forces'' the researcher to choose designs with sample-equalizing allocations by not rewarding any results from experiments with other designs. Intuitively, since the researcher expects $\sigma_j$ to be the same for all $j$, the optimal protocol equalizes the expected standard errors by requiring a sample-equalizing allocation. Proposition \ref{cor:threshold_aa} provides guidance both on which testing protocol to implement and which design to incentivize. 

 \begin{rem}[Researcher knows and can choose $\Sigma$]\label{rem:variance_equalizing_allocation}
If the researcher knows the outcomes' variances and can choose $\Sigma$ strategically (i.e., by choosing sample sizes as a function of the outcome variances), separate one-sided $t$-tests with critical values $t = \Phi^{-1}\left(1 - \frac{C(J, \Sigma)}{b \bar{\omega}(J)} \right)$ based on variance-equalizing allocations are uniformly globally optimal (by Proposition \ref{cor:threshold_main1}). A variance-equalizing allocation is an allocation such that $\sigma_{0j}^2/n_{0j} + \sigma_{1j}^2/n_{1j} = \sigma^2$ for some constant $\sigma^2 > 0$. This result follows because if the researcher knows and can choose $\Sigma$, then forcing her to impose common variances preserves maximin and uniformly global optimality by arguments in the proof of Proposition \ref{cor:threshold_main1}. 
  \qed     
 \end{rem}

\section{Empirical illustration and broader applicability}
\label{sec:applications}

This section discusses the scope for applying and implementing the framework's implications. Section \ref{sec:implications_practice} considers our running example, the regulatory approval process, while  Section \ref{sec:program_evaluation} comments on potential applications to program evaluation in economics.

\subsection{Empirical illustration}
\label{sec:implications_practice}

Proposition \ref{cor:threshold_main1} showed that the policymaker's preferred MHT adjustments hinge on how research costs $C(J,\Sigma)$ vary with the set of chosen treatments $J$ and the experimental design  $\Sigma$. In particular, denote the level of the separate $t$-tests in Proposition \ref{cor:threshold_main1} as
$$
\alpha(J, \Sigma) := \frac{C(J, \Sigma)}{b \bar{\omega}(J)}.
$$ 
The planner will therefore want to obtain information about $C(J,\Sigma)$, $b$, and $\bar{\omega}(J)$ to compute this level. In the FDA approval context with multiple subgroups, we can think of $b$ as the expected profit per customer, and $\bar{\omega}(J)$ as the total number of customers that would buy the drug if it were approved for every subgroup. To build intuition, it will be helpful to impose the simplifying assumption $\bar{\omega}(J) = |J|$, which holds for example if each subgroup $j$ receives equal weight $\omega_j=1$. 

\smallskip \noindent \textbf{Adjustment factor in general form}. Let $\bar{C}$ denote the cost of a benchmark experiment with a single treatment.\footnote{This benchmark experiment could, for instance, be an experiment with the minimum sample size for a Phase III trial according to FDA guidance (see \citet{fda_sample_size}).} Without loss of generality we can write 
\begin{equation} 
\label{eqn:size_control_alpha}
\alpha(J, \Sigma) = \bar{\alpha} \times \frac{C(J, \Sigma)}{|J|\bar{C}},
\end{equation} 
where $\bar{\alpha} = \bar{C} / b$ denotes the size of the hypothesis test in the benchmark  experiment. This formulation shows that the appropriate size for tests in a study with $|J|$ hypotheses can be calculated as the product of two quantities. The first is the size of the optimal test in the benchmark, single-hypothesis case. The second is the MHT \emph{correction factor} $[ C(J,\Sigma)/\bar{C} \times 1 / |J| ]$, which captures how the cost per test varies as the number of hypotheses tested grows (keeping in mind that this may affect the design $\Sigma$ as well as $J$). Unless all costs are fixed ($C(J, \Sigma) = \bar{C}$) this correction factor will differ from the standard Bonferroni correction factor $1/|J|$. Notice also that if costs are strictly proportional to the number of hypotheses ($C(J, \Sigma) = \bar{C} \times |J|$) then standard inference without adjustment for MHT is optimal. 

\smallskip
\noindent \textbf{Choice of $\bar{\alpha}$}. There are competing benchmarks one might consider for $\bar{\alpha}$, the test size for a benchmark study with a single treatment. FDA guidelines currently recommend a size of 2.5\% for one-sided single hypothesis tests \citep{fda2022guidance}, but \citet{tetenov2016economic}, using data on the costs and expected profits from Phase III trials, proposes a value of 15\%. Given this, and the dispersion in the cost of trials for different drugs \citep{grabowski2002returns}, we provide results for a range of values between 2.5\% and 15\%.

\smallskip 
\noindent \textbf{Modeling costs}. In principle the regulator could evaluate the MHT adjustment term in (\ref{eqn:size_control_alpha}) separately for different categories (e.g., therapeutic classes) or even using data on each study individually. They might require pharmaceutical companies to declare the fixed and variable costs of a study (information about which is often contained in contracts with the hospital or contract research organization organizing a trial) when pre-registering it. Here we wish to illustrate the potential consequences of doing so using existing, published estimates of moments of the cost structure of clinical trials. This requires that we model the cost function. We consider a simple formulation with both fixed and variable costs:
\begin{equation} \label{eqn:cost_examples}
C(J,\Sigma) = c_f + c_v \sum_{j \in J} n_j, 
\end{equation} 
where $c_f$ is a fixed cost invariant to $|J|$ and $c_v$ is a variable cost. This is a special case of the specification in Section \ref{sec: most powerful tests}, where we assume here for simplicity that variable costs vary in proportion to the number of subjects $n_j$. Additional costs that vary with $|J|$, independent of $n_j$, could be accommodated with the appropriate data, and would imply adjustments less conservative than those reported below.

It will be convenient to rewrite this expression (without loss of generality) as 
$$
C(J,\Sigma) = c_f + c_v |J| \bar{m} \frac{\bar{n}}{\bar{m}}, \quad \bar{n}  = \frac{1}{|J|} \sum_{j \in J} n_j,
$$ 
where $\bar{n}$ is the average sample size across subgroups in the trial in question and $\bar{m}$ is the average overall sample size of the single arm benchmark experiment (with size $\bar{\alpha}$). With an abuse of notation, we can then write the optimal level as a function $\alpha\left(\bar{\alpha}, |J|, \frac{\bar{n}}{\bar{m}}\right)$ of the size for the benchmark single-hypothesis experiment $\bar{\alpha}$, the number of treatment arms $|J|$, and the ratio of the average subgroup sample size to the benchmark experiment size $\bar{n}/\bar{m}$.

\medskip \noindent \textbf{Cost calibration}. \citet{SertkayaEtAl2016drivers}, using data on the costs of 31,000 pharmaceutical clinical trials conducted in the United States between 2004 and 2012, estimate that the average fixed costs of a Phase 3 trial were 46\% of the average total cost, with the rest varying either directly with the number of subjects enrolled or with the number of sites at which they were enrolled.\footnote{According to \citet[Table 2]{SertkayaEtAl2016drivers}, average variable costs (i.e., the per-patient and per-site costs) were USD 10,826,880, and average total costs were USD 19,890,000, so that the fraction of fixed costs is $(19,890,000-10,826,880)/19,890,000\approx 0.46$.} 
As an approximation, we set $\bar{m} \bar{J}$ equal to the average historical overall sample size across clinical trials. It follows that $c_f / (c_f + c_v \bar{m} \bar{J}) = 0.46$. Using $\bar{J} = 3$ based on the tabulations in \citet{pocock2002subgroup} yields an MHT correction factor of $(\frac{\bar{n}}{\bar{m}} + 2.56 / |J|) / 3.56$.\footnote{We use the median estimate multiplied by the probability of reporting more than one subgroup. The critical values are not particularly sensitive to $\bar{J}$; if for example we fix $\bar{\alpha}(1) = 0.025$ and double $\bar{J}$ from 3 to 6 this decreases $\alpha(2)$ from $0.016$ to $0.015$, $\alpha(3)$ from $0.013$ to $0.011$, and $\alpha(\infty)$ from $0.007$ to $0.004$.}

Inserting this into (\ref{eqn:size_control_alpha}), we thus arrive at  
\begin{equation}
\alpha\left(\bar{\alpha}, |J|, \frac{\bar{n}}{\bar{m}}\right) = \bar{\alpha} \times  \left[  \underbrace{\frac{1 + 2.56 / |J|}{3.56}}_{\text{Multiplicity adjustment}}  + \underbrace{\frac{1}{3.56} \times \Big(\frac{\bar{n}}{\bar{m}} - 1\Big)}_{\text{Sample size per arm}}\right].
\label{eq:adjustment_alpha_numerical}
\end{equation}
The correction factor here has two terms. The first is a ``pure'' correction for multiple hypothesis testing, accounting for the influence of the number of treatment arms $|J|$. The second corrects for the effects of sample size on study cost: studies with sample sizes larger than $\bar{m}$ ($\bar{n} > \bar{m}$) are more expensive to run and thus require less strict testing thresholds. 

Table \ref{tab:critical_values} illustrates the implications quantitatively. It tabulates the test level implied by (\ref{eq:adjustment_alpha_numerical}) for a range of values of $|J|$ (rows), $\bar{\alpha}$ (columns), and $\bar{n} / \bar{m}$ (column groups). For example, for studies with a benchmark sample size ($\bar{n} = \bar{m}$) and assuming $\bar{\alpha}=0.15$ as in \citet{tetenov2016economic}, those with $|J|=2$ would use size $0.096$, those with $|J|=3$ would use $0.078$, and so on, asymptoting to $0.042$ at $|J| = \infty$. If instead we set $\bar{\alpha} = 0.025$, consistent with FDA guidance, then studies with $|J| = 2$ would use $0.016$, those with $|J| = 3$ would use $0.013$, and so on, asymptoting to $0.007$ at $|J| = \infty$. These thresholds are more conservative than unadjusted ones, but less conservative than those implied by Bonferroni corrections ($\bar\alpha/|J|$).

For further comparison, the last two columns of Table \ref{tab:critical_values} report the adjustment corresponding to FWER control based on the Sidak correction \citep{vsidak1968multivariate}. The Sidak correction, which sets the level of each test to $1 - (1 - \bar{\alpha})^{1/|J|}$, is a useful benchmark because it is exact with independent tests (as for the case of subgroup analyses). It implies more conservative inferences than our tabulated values. For instance, with $\bar{\alpha} = 0.025$ and $|J|=9$, the optimal level is $0.009$ while that under the Sidak correction is $0.003$. 

The table also illustrates how the adjustment factor depends on the (relative) average per-treatment sample size  $\bar{n}/\bar{m}$. The fifth, sixth and seventh columns vary this while holding $\bar{\alpha}$ fixed at 0.025. When $\bar{n}$ is smaller than $\bar{m}$ we require \textit{more} stringent size control, while for larger $\bar{n}$ we require less stringent size control. For example, with $|J|=2$, studies with half the benchmark sample size would use a $0.012$ threshold, while studies with double the benchmark sample size would use $0.023$.

The results in Table \ref{tab:critical_values} should be read as illustrative of what might happen if the FDA were to require cost disclosure and base testing thresholds on the disclosed trial-specific costs, which could be obtained, for example, from privately-owned contract data \citep{SertkayaEtAl2016drivers}. If instead it were to base testing thresholds directly on the number of treatments $|J|$ and samples sizes $\bar{n}$, using the values indicated in the table, this would create incentives for gaming---e.g., reporting results obtained from a single experiment (in the sense that $c_f$ was incurred only once) as if they came from multiple distinct experiments (implying that $c_f$ was incurred more than once). Detecting and deterring such gaming might be easier in some cases than in others. Tests of the same drug in different populations, for example, could be matched based on the chemical formula of the compound in question; tests of different compounds on the same population might be harder to match.

\begin{table}[!t]
{\centering 
\caption{Critical values as functions of hypothesis count and sample size}
\label{tab:critical_values}
\scriptsize
\begin{tabularx}{\textwidth}{c>{\raggedleft\arraybackslash}X
                                   >{\centering\arraybackslash}X
                                   >{\centering\arraybackslash}X
                                   >{\centering\arraybackslash}X
                                   >{\centering\arraybackslash}X
                                   >{\centering\arraybackslash}X
                                   >{\centering\arraybackslash}X
                                   >{\centering\arraybackslash}X
                                   >{\centering\arraybackslash}X
                                   >{\centering\arraybackslash}X
                                   >{\centering\arraybackslash}X}
\toprule
        & $\bar{n} / \bar{m} = $ & \multicolumn{4}{c}{$100\%$} & $50\%$ &  $150\%$ &  $200\%$ & $\alpha_{\text{Šidák}}^{0.025}$ & $\alpha_{\text{Šidák}}^{0.05}$ \\
\cmidrule(lr){3-6}\cmidrule(lr){7-9}\cmidrule(lr){10-11}
$|J|$ & $\bar{\alpha} =$ & $0.025$ & $0.05$ & $0.10$ & $0.15$ & \multicolumn{3}{c}{$0.025$} & $0.025$ & $0.050$ \\
\midrule
1        && 0.025 & 0.050 & 0.100 & 0.150 & 0.021 & 0.029 & 0.032 & 0.025 & 0.050 \\
2        && 0.016 & 0.032 & 0.064 & 0.096 & 0.012 & 0.019 & 0.023 & 0.013 & 0.025 \\
3        && 0.013 & 0.026 & 0.052 & 0.078 & 0.009 & 0.017 & 0.020 & 0.008 & 0.017 \\
4        && 0.012 & 0.023 & 0.046 & 0.069 & 0.008 & 0.016 & 0.019 & 0.006 & 0.013 \\
5        && 0.011 & 0.021 & 0.042 & 0.064 & 0.007 & 0.015 & 0.018 & 0.005 & 0.010 \\
6        && 0.010 & 0.020 & 0.040 & 0.060 & 0.006 & 0.014 & 0.017 & 0.004 & 0.009 \\
7        && 0.010 & 0.019 & 0.038 & 0.058 & 0.006 & 0.014 & 0.017 & 0.004 & 0.007 \\
8        && 0.009 & 0.019 & 0.037 & 0.056 & 0.006 & 0.013 & 0.016 & 0.003 & 0.006 \\
9        && 0.009 & 0.018 & 0.036 & 0.054 & 0.006 & 0.013 & 0.016 & 0.003 & 0.006 \\
$\infty$ && 0.007 & 0.014 & 0.028 & 0.042 & 0.004 & 0.013 & 0.014 & 0.000 & 0.000 \\
\bottomrule
\end{tabularx}
}

{\footnotesize \emph{Notes:} This table tabulates optimal critical values obtained from (\ref{eq:adjustment_alpha_numerical}) for different numbers of hypotheses ($|J|$) and (relative) sample sizes ($\bar{n}/\bar{m}$), all given an assumed critical value for the benchmark case of a single-hypothesis experiment ($\bar{\alpha}$). $\alpha_{\text{Šidák}}$ is the optimal level implied by the Sidak correction \citep{vsidak1968multivariate}, $1 - (1 - \bar{\alpha})^{1/|J|}$. The Sidak correction is exact for controlling the FWER with independent tests.} 
    \normalsize
    \end{table}

\subsection{MHT within economics}
\label{sec:program_evaluation}

Hypothesis testing norms are also a salient issue within economics, given the publication trends noted in Figure \ref{fig:publishing_trends}. Our framework's assumptions arguably correspond most closely to economics papers that report the results of experimental program evaluations, as in that case there are clear policy decisions that the research is explicitly designed to inform (i.e., whether or not to implement or scale the programs being evaluated). Indeed, researchers often conduct studies like these in collaboration with implementation partners, such as governments or NGOs, precisely in order to evaluate the impact of treatments the partners are considering. The paper’s findings may thus affect social welfare because, in addition to potentially being published in an academic journal, they can influence those decisions. Pre-specification of the analysis to be conducted in a pre-analysis plan (corresponding to our assumption that researchers pre-specify their tests) is now common in this genre of work \citep{miguel2021evidence}. And it is also common for such ``policy experiments'' to test more than one treatment as part of the same study. \citet{muralidharan2025factorial} document at least 27 such experiments published in top-5 journals alone between 2007 and 2017.\footnote{Their list includes only studies with interaction arms, so provides a lower bound on the total number of multi-armed evaluations.}

With these points in mind, we also conducted a second quantitative application to program evaluation experiments in development economics, using unique data on their costs, sample sizes, and treatment arm counts which we obtained from the universe of funding proposals submitted to the Abdul Latif Jameel Poverty Action Lab (J-PAL) from 2009 to 2021. In the interests of brevity, we describe the data and analysis in depth in Appendix \ref{app:jpal} and briefly restate the main findings here. We estimate that research costs are significantly and substantially less than proportional to the number of treatments tested, with elasticities ranging from 0.13 to 0.22.\footnote{We obtain these estimates from descriptive regressions; they need not be causal to characterize the cost function $C(J,\Sigma)$ in our model, provided that function is invariant to $r$.
} But they are also not invariant to scale: projects with more arms cost significantly more ($p < 0.05$). As a result the appropriate testing thresholds vary with the number of treatment arms. They are similar to but slightly less conservative than those that result from a Bonferroni correction and those implied by Sidak's correction (which is exact for controlling the FWER for independent tests). Finally, the testing thresholds also vary moderately with the sample size, with larger samples implying (ceteris paribus) less conservative procedures.

This analysis focuses on a specific type of multiplicity, namely multiplicity of treatments. Economists also often deal with multiple outcomes. These do not necessitate multiple tests; indeed, researchers often aggregate the outcomes into summary statistics instead. An earlier version of this paper \citep{VWN2025MHTv8} studied this problem within our framework, showing that optimal rules $r^*$ test for effects on an index formed using statistical weights \citep[similar to][]{anderson2008multiple} when the  outcomes are noisy proxies for some common underlying measure, but using economic weights \citep[as for example in][]{bhatt2024predicting} when they capture distinct components of the planner's utility. The fact that multiple outcomes justify different techniques than do multiple treatments (or subgroups) is noteworthy in the context of historical narratives about MHT practices: the multiple-treatment case---genetic association testing in particular---has often been cited to motivate new MHT procedures \citep[see][for reviews]{dudoit2003multiple, efron2008microarrays}, while the multiple-outcomes case seems to have been bundled with it subsequently, and less intentionally.

One could also move away from the frequentist paradigm (which we have presumed) entirely, towards a Bayesian alternative. Proposals to control the FDR are interesting in this regard. Several papers have pointed out a Bayesian rationale for doing so: controlling the (positive) FDR can be interpreted as rejecting hypotheses with a sufficiently low posterior probability \citep[e.g.,][]{storey2003positive,gukoenker2020invidious,kline2022systemic}. In fact, these arguments apply even in the case of a \emph{single} hypothesis. The essential idea is to balance the costs of false positives and false negatives, rather than prioritize size control at any (power) cost. We thus interpret these arguments less as support for a particular solution to the MHT problem per se, and more as a reminder of the merits of Bayesian approaches generally.

\section*{Acknowledgments}
We are grateful to the Editor and anonymous referees for constructive comments that helped improve the paper. We are also grateful to Nageeb Ali, Isaiah Andrews, Tim Armstrong, Oriana Bandiera, Sylvain Chassang, Kevin Chen, Tim Christensen, Graham Elliott, Stefan Faridani, Will Fithian, Paul Glewwe, Peter Hull, Guido Imbens, Lawrence Katz, Toru Kitagawa, Pat Kline, Michael Kremer, Michal Kolesar, Ivana Komunjer, Damian Kozbur, Lihua Lei, Adam McCloskey, Konrad Menzel, Francesca Molinari, Jose Montiel-Olea, Ulrich Mueller, Mikkel Plagborg-M\o{}ller, David Ritzwoller, Joe Romano, Adam Rosen, Jonathan Roth, Andres Santos, Azeem Shaikh, Jesse Shapiro, Joel Sobel, Sandip Sukhtankar, Yixiao Sun, Elie Tamer, Aleksey Tetenov, Winnie van Dijk, Tom Vogl, Quang Vuong, Michael Wolf, and seminar participants for valuable comments, and to staff at J-PAL, Sarah Kopper and Sabhya Gupta in particular, for their help accessing data. Aakash Bhalothia and Muhammad Karim provided excellent research assistance. Viviano is also affiliated with Y-RISE, and W\"uthrich is also affiliated with CESifo. All errors are our own.

\section*{Funding}
Niehaus and W\"uthrich gratefully acknowledge funding from the UC San Diego Academic Senate. Viviano gratefully acknowledges support from the Griffin Fund at Harvard University and from NSF Grant SES 2447088.

Non-financial support was provided by J-PAL in the form of the data used in Appendix \ref{app:jpal}, which we used under the provisions of a Data Use Agreement (DUA) between J-PAL and UC San Diego. Key provisions are that we are allowed to publish data sufficient to meet the requirements of scholarly journals, subject to various provisions, that J-PAL may curate these data using its established data publication processes to remove any Confidential Data, and that J-PAL may review drafts of the paper prior to publication to ensure that no Confidential Data are disclosed.

\section*{Conflict of Interest} 

Niehaus is co-chair of the Science for Progress Initiative at J-PAL, which provided the data used in Appendix \ref{app:jpal}. The role is uncompensated. He is not an officer, director or board member of J-PAL.

\spacingset{1.25}

\bibliography{my_bib2}
\bibliographystyle{chicago}

\spacingset{1.5}

\appendix

\newpage

\setcounter{page}{1}

\begin{center}
    \LARGE Online Appendix
\end{center}

\startcontents[sections]

\section{Application to economic experiments}
\label{app:jpal}

In this appendix we consider the application of the framework to academic economic research. Not all academic research fits within our framework, tailored as it is to the regulatory approval process. Research intended primarily to increase generalizable, conceptual knowledge, for example, without reference to any particular policy decision, does not map naturally. And data on costs are not available systematically for all economic research. We focus on program evaluation experiments. These (arguably) map well to our framework, in that they are typically designed at least in part to inform an immediate policy decision about whether to retain or scale up a policy intervention, and we have access to unique data on the costs of conducting them that we obtained for this purpose.\footnote{To map this setting into the formalisms of our model, one might interpret the researcher's payoff (\ref{eqn:B_R_general}) as capturing the expected value of academic publications (which depend on hypothesis rejections), net of costs. The policy-makers decision to implement the papers findings need not be contingent on their publication, which one can think of as reflecting the fact that policymakers do not discriminate between papers based on the academic prestige of the outlet.}

\smallskip \noindent \textbf{Data.} The data cover (essentially) all funding proposals for evaluations submitted to the Abdul Latif Jameel Poverty Action Lab (J-PAL) from 2009 to 2021. J-PAL is the leading funder and facilitator of experimental economic research in low-income countries, and funds projects that are typically designed to inform policy in those countries. In this sense the characteristics of these projects align fairly closely with the assumptions in our framework. The data contain the reported total financial cost of each project (including the amount requested from sources other than J-PAL, and with personnel costs broken out separately) as well as the overall sample size and the number of experimental arms in the study. This allows us to examine how costs vary with the design of the experiment and with hypothesis multiplicity with respect to the number of treatments. Financial costs are not the only costs incurred, of course, but are likely to be highly correlated with other relevant ones that a researcher deciding whether to undertake a project would consider: ceteris paribus, larger budgets will tend to mean more researcher effort raising funds and managing teams of research assistants, for example. 

\smallskip \noindent \textbf{Modeling costs.} To see how testing adjustments would vary as a function of the average sample size $\bar{n}$ and the number of treatment arms $|J|$ we follow steps (and use notation) similar to those in Section \ref{sec:implications_practice}. Specifically, we again assume equal weights $\omega_j = 1$ and let $\overline{C}$ and $\overline{\alpha}$ denote the cost and the testing threshold, respectively, for a benchmark experiment with average sample size $\bar{m}$, so that \eqref{eqn:size_control_alpha} again gives the optimal size of a test. For this application we posit a Cobb-Douglas form
\begin{equation}
\small 
\begin{aligned}
    \label{eqn:cost_function_jpal}
    C(J, \Sigma) = k |J|^{\beta} \left( \bar{n} \right)^{\iota}
    \end{aligned} 
\end{equation}
for the cost function, which will fit the cost data reasonably well, and yields
\begin{equation}
\small 
\begin{aligned}
\label{eqn:optimal_alpha_jpal}
\alpha(|J|, \bar{n}) = \bar{\alpha} \times |J|^{\beta - 1} \times \left(\frac{\bar{n}}{\bar{m}}\right)^{\iota} 
\end{aligned}
\end{equation}
where the later two factors adjust for hypothesis count and sample size, respectively.

\smallskip \noindent \textbf{Empirical analysis.}
We take this to the data by taking logarithms of \eqref{eqn:cost_function_jpal} and estimating the resulting linear model via OLS. An open question is how to model potential variation in $k$, the fixed cost factor. The correct interpretation of the regression results is that they identify values of the parameters $\beta$ and $\iota$ that would be appropriate for a planner to substitute into (\ref{eqn:optimal_alpha_jpal}) while setting $\overline{\alpha}$ to the appropriate test size for an experiment testing one hypothesis, with sample size $\bar{m}$, \emph{and with the same fixed cost factor} $k$. It is therefore useful to consider specifications that include further controls that may capture variation in $k$ across classes of projects. We include (incrementally) indicators for project type (pilot, full study, or add-on funding), indicators for the J-PAL initiative in question, and the (log of) personnel costs as potential proxies for $k$. Across all specifications we report results from tests of the hypotheses that $\beta$ is zero and one, respectively. The first condition holds if costs are invariant with respect to the number of arms, in which case average size control is indicated (Proposition \ref{cor:threshold_main1}). The second holds if costs are proportionate to the number of arms, in which case no MHT adjustment is indicated. Tests of the nulls that $\iota$ is zero and one assess corresponding conditions with respect to sample size.

The data reject both of these hypotheses (see Table \ref{tab:main_economics}). Regardless of the specification, costs are significantly and substantially less than proportional to the number of treatments tested, with estimated elasticities ranging from 0.13 to 0.22. But they are also not invariant to scale: projects with more arms cost significantly more, as these estimates are all significantly different from zero at the 5\% level.\footnote{To avoid a reductio ad absurdum we do not apply MHT adjustments to our tests of the hypothesis that MHT adjustments are required.} Interpreted through the lens of our model, these patterns provide both a prima facie justification for applying MHT adjustment to studies of this sort, and also imply that simply controlling the average size of tests in these studies (e.g., via a Bonferroni correction) would be too conservative. 

In Table \ref{tab:critical_values_economics} we make this point concrete, calculating the specific adjustments implied (and, for comparability, using the same format as Table \ref{tab:critical_values}). Adjustments vary significantly as the number of treatment arms increases, but are less conservative than Bonferroni corrections. They are also less conservative to those implied by FWER control using the Sidak correction, which is exact when the tests are independent, although the differences are smaller than in Section \ref{sec:implications_practice}. Interestingly, the adjustments vary only moderately as a function of the sample size, as illustrated in columns 6 - 9 of Table \ref{tab:critical_values_economics}.

\begin{table}[!ht] 
  \begin{center}
  \caption{Multiple hypothesis testing adjustment in economics} 
  \scriptsize
  \label{tab:main_economics} 
\begin{tabular}{@{\extracolsep{5pt}}lcccc} 
\toprule
\midrule
 & \multicolumn{4}{c}{\textit{Dependent variable: log(total project cost)}} \\ 
\cline{2-5} 
\\[-1.8ex] & (1) & (2) & (3) & (4)\\ 
\midrule
 $\log($Treatment Arms$)[\beta]$ & 0.216$^{***}$ & 0.140$^{**}$ & 0.222$^{***}$ & 0.132$^{**}$ \\ 
  & (0.078) & (0.070) & (0.072) & (0.054) \\ 
  & & & & \\ 
 $\log($Number of surveys per arm$)[\iota]$ & 0.253$^{***}$ & 0.107$^{***}$ &  0.136$^{***}$ & 0.075$^{***}$ \\ 
  & (0.024) & (0.023) & (0.024) & (0.020)\\ 
  & & & & \\ 
$\log($Personnel Costs$)$ &  &  &  & 0.531$^{***}$ \\ 
  &  &  &  & (0.030) \\ 
  & & & & \\ 
 Proposal Type FEs & No  & Yes  & Yes & Yes   \\ 
 Initiative FEs &  No & No  & Yes  & Yes   \\ 
  \midrule
 $p$-value, $H_0:\beta = 0$ & 0.005 & 0.045 & 0.002 & 0.014 \\ 
$p$-value, $H_0:\beta = 1$ & 0.000 & 0.000 & 0.000 & 0.000  \\
 \midrule
 $p$-value, $H_0:\iota = 0$ & 0.000 & 0.000 & 0.000 & 0.000 \\ 
$p$-value, $H_0:\iota = 1$ & 0.000 & 0.000 & 0.000 & 0.000  \\
  \midrule
Observations & 730 & 728 & 692 & 617  \\ 
Adjusted R$^{2}$ & 0.129 & 0.334 & 0.410 & 0.656 \\ 
\midrule
\bottomrule
\end{tabular} 

\end{center}
\footnotesize{\emph{Note:} The dependent variable in all specifications is the (log of) the total cost of the proposed project. Proposal type fixed effects include indicators for full projects, pilot projects, and add-on funding to existing projects. Initiative fixed effects include an indicator for each J-PAL initiative that received funding applications. Heteroskedasticity-robust standard errors in parenthesis.}
\end{table}

\begin{table}[!ht]
\begin{center} 
\caption{Critical values as functions of hypothesis count and sample size in economics}
\label{tab:critical_values_economics}
\scriptsize
\begin{tabularx}{\textwidth}{c>{\raggedleft\arraybackslash}X
                                   >{\centering\arraybackslash}X
                                   >{\centering\arraybackslash}X
                                   >{\centering\arraybackslash}X
                                   >{\centering\arraybackslash}X
                                   >{\centering\arraybackslash}X
                                   >{\centering\arraybackslash}X
                                   >{\centering\arraybackslash}X
                                   >{\centering\arraybackslash}X
                                   >{\centering\arraybackslash}X
                                   >{\centering\arraybackslash}X}
\toprule
        & $\bar{n} / \bar{m} = $ & \multicolumn{4}{c}{$100\%$} & $50\%$ &  $150\%$ &  $200\%$ & $\alpha_{\text{Šidák}}^{0.025}$ & $\alpha_{\text{Šidák}}^{0.050}$\\
\cmidrule(lr){3-6}\cmidrule(lr){7-9}\cmidrule(lr){10-11}
$|J|$ & $\bar{\alpha} =$ & $0.025$ & $0.050$ & $0.100$ & $0.150$ & \multicolumn{3}{c}{$0.050$} & $0.025$ & $0.050$\\
\midrule
1        && 0.025 & 0.050 & 0.100 & 0.150 & 0.047 & 0.051 & 0.052 & 0.025 & 0.050\\
2        && 0.013 & 0.027 & 0.054 & 0.081 & 0.026 & 0.028 & 0.028 & 0.013 & 0.025\\
3        && 0.009 & 0.019 & 0.038 & 0.057 & 0.018 & 0.019 & 0.020 & 0.008 & 0.017\\
4        && 0.007 & 0.015 & 0.030 & 0.045 & 0.014 & 0.015 & 0.015 & 0.006 & 0.013\\
5        && 0.006 & 0.012 & 0.024 & 0.037 & 0.011 & 0.012 & 0.013 & 0.005 & 0.010\\
6        && 0.005 & 0.010 & 0.021 & 0.031 & 0.010 & 0.010 & 0.011 & 0.004 & 0.009\\
7        && 0.004 & 0.009 & 0.018 & 0.027 & 0.008 & 0.009 & 0.009 & 0.004 & 0.007\\
8        && 0.004 & 0.008 & 0.016 & 0.024 & 0.007 & 0.008 & 0.008 & 0.003 & 0.006\\
9        && 0.003 & 0.007 & 0.014 & 0.022 & 0.007 & 0.007 & 0.007 & 0.003 & 0.006\\
\bottomrule
\end{tabularx}

\end{center}

{\footnotesize \emph{Notes:} This table tabulates optimal critical values obtained from the last column in Table \ref{tab:main_economics} (using J-PAL data) for different numbers of hypotheses ($|J|$) and (relative) sample sizes ($\bar{n}/\bar{m}$), all given an assumed critical value for the benchmark case of a single-hypothesis experiment ($\bar{\alpha}$). $\alpha_{\text{Šidák}}$ is the optimal level implied by the Sidak correction \citep{vsidak1968multivariate}, $1 - (1 - \bar{\alpha})^{1/|J|}$. The Sidak correction is exact for controlling the FWER with independent tests.} 
    \normalsize
    \end{table}

\smallskip \noindent \textbf{Heterogeneity analysis.} The specification underlying Table \ref{tab:main_economics} allows for variation in the fixed costs, but there may also be variation across categories of experiments in the slope coefficients $\beta$ and $\iota$. As a final exercise we explore this possibility. We focus in particular on the distinction between experiments in high- as opposed to low-income countries. While the latter are J-PAL's main focus and make up 80\% of the data, the sample also includes a smaller and more recent subset of projects set in high-income countries (primarily the United States) which may differ from those in low income countries. Table \ref{tab:tab1} presents results estimated for these two subsamples separately.

The results for low income countries (Columns 1--4) are similar to those in Table \ref{tab:main_economics}, as one would expect. In high income countries, however, studies with more arms cost \emph{less} on average in several specifications (Columns 5--7), presumably reflecting heterogeneity in project types. As we control for more proxies for fixed costs $k$ this relationship becomes smaller and statistically insignificant. When in particular we control for personnel costs the estimated relationship between total costs and experimental arms is---unlike that in low-income countries---very close to zero. This is consistent with the idea that projects in low-income countries tend to incur substantial variable costs---collecting original survey data in rural areas, for example---while those in high-income countries tend to incur primarily fixed costs (as for example if they are able to take advantage of pre-existing administrative data sources to measure impacts). While only suggestive, this result illustrates how the cost structures of experimental research may vary by context, implying in turn differences in the appropriate MHT adjustments.\footnote{Our data do not let us observe and draw out sharper distinctions between different projects (e.g., whether an experiment is a field or online experiment). A planner who could observe such information would generally wish to condition testing protocols on it. In some cases researchers might be able to credibly disclose the cost structure of studies in a similar class to theirs, and pre-specify test procedures based on these. If so one would expect this to lead to widespread disclosure via the unraveling effects that are standard in disclosure games.}

\begin{table}[!ht]
\scriptsize
\begin{center}
  \caption{Experimental research costs: heterogeneity analysis} 
  \label{tab:tab1} 
 \begin{tabular*}{\textwidth}{@{\extracolsep{\fill}}lcccc|cccc}
 \toprule
\midrule
& \multicolumn{4}{c}{Main sample (low income countries) } & \multicolumn{4}{c}{High income countries} \\ \cline{2-5}  \cline{6-9}   \\
& (1) & (2) & (3) & (4) & (5) & (6) & (7) & (8)\\ 
\hline
 $\log($Treatment Arms$)[\beta]$ & 0.380$^{***}$ & 0.280$^{***}$ & 0.312$^{***}$ & 0.163$^{***}$ & $-$0.309$^{*}$ & $-$0.260$^{*}$ & $-$0.209 & 0.003 \\ 
  & (0.084) & (0.077) & (0.071) & (0.056) & (0.165) & (0.142) & (0.146) & (0.115) \\ 
  & & & & \\ 
 $\log($Number of surveys per arm$)[\iota]$ & 0.295$^{***}$ & 0.153$^{***}$ & 0.155$^{***}$ & 0.081$^{***}$ & 0.193$^{***}$ & 0.037 & 0.083 & 0.068 \\  
  & (0.027) & (0.027) & (0.026) & (0.020) & (0.053) & (0.045) & (0.054) & (0.041) \\ 
  & & & & \\ 
$\log($Personnel Costs$)$ &  &  &  & 0.531$^{***}$ &  &  &  & 0.530$^{***}$ \\ 
  &  &  &  & (0.027) &  &  &  & (0.089) \\ 
  & & & & \\ 
 Proposal Type FEs & No  & Yes  & Yes & Yes & No & Yes & Yes & Yes   \\ 
 Initiative FEs &  No & No  & Yes  & Yes & No & No & Yes & Yes   \\ 
\hline 
$p$-value, $H_0:\beta = 0$ & 0.000 & 0.000 & 0.000 & 0.012 &  0.060 & 0.067 & 0.151 & 0.977 \\ 
$p$-value, $H_0:\beta = 1$ & 0.000 & 0.000 & 0.000 & 0.000 & 0.000 & 0.000 & 0.000 & 0.000 \\ 
\hline
Observations & 610 & 608 & 577 & 518 & 117 & 115 & 101 & 85 \\ 
Adjusted R$^{2}$ & 0.173 & 0.362 & 0.400 & 0.678 & 0.104 & 0.332 &  0.323 & 0.530 \\ 
\midrule
\bottomrule
  \end{tabular*}
  
\end{center}
\footnotesize{ \emph{Note:} The dependent variable in all specifications is the (log of) the total cost of the proposed project. Proposal type fixed effects include indicators for full projects, pilot projects, and add-on funding to existing projects. Initiative fixed effects include an indicator for each J-PAL initiative that received funding applications. ``High income countries'' are as according to the World Bank classification. Heteroskedasticity-robust standard errors in parenthesis.}
\end{table}

\section{Additional extensions}
\label{app:extensions}
Throughout the remainder of the appendix, we will often suppress the dependence of the testing protocols on $J$ and $\Sigma$ and simply write $r(X)$.
All proofs are in Appendix \ref{app:omitted_proofs}.

\subsection{Two-sided tests}
\label{app: two-sided tests}

The model in the main text naturally justifies one-sided hypothesis testing. To move away from one-sided hypothesis testing we need to allow policymakers to take actions different from the baseline treatment of ``do nothing'' when no recommendations are made. We therefore consider a model where a policymaker, who does not coincide with the planner, may implement treatments if no recommendation is made. This justifies two-sided testing. 

Consider a model where the researcher reports a vector of recommendations, $r(X)$, and the sign of each statistic $j$, $\sgn(X_j)$, for which $r_j(X)=1$. That is, for each treatment, the researcher reports either a positive recommendation ($r_j(X)=1$ and $\sgn(X_j)= 1$), a negative recommendation ($r_j(X)=1$ and $\sgn(X_j)\in\{-1,0\}$), or no recommendation ($r_j(X)=0$).   

The key feature of the model is that if $r_j(X) = 0$, the policymaker may or may not implement treatment $j$. Define the indicator $p$, where $p = 1$ if the policymaker implements treatments with $r_j(X) = 0$ and $p = 0$ otherwise.
The planner does not know $p$, and we consider a worst case approach with respect to $p$, consistent with the maximin approach we consider in the main text.
If $r_j(X) = 1$, the policymaker implements treatment $j$ if $\sgn(X_j) = 1$ and does not implement the treatment if $\sgn(X_j)\in\{-1,0\}$.

Welfare conditional on experimentation is
$
\sum_{j \in J}  r_j(X)  1\{\sgn(X_j) = 1\}\theta_j + p \sum_{j \in J}  (1 - r_j(X)) \theta_j,  
$
which can be rewritten (up to constant terms) as 
\begin{eqnarray*}
\small 
\begin{aligned}
v^{\mathrm{two}}(\theta, J, \Sigma, p) &= & (1 - p) \sum_{j \in J}  \omega_j r_j(X, J, \Sigma) \theta_j 1\{\sgn(X_j) = 1\} - p \sum_{j \in J} \omega_j   r_j(X, J, \Sigma) \theta_j 1\{\sgn(X_j) \in\{-1,0\} \}
\end{aligned} 
\end{eqnarray*}
The expected utility of the researcher is (setting $b = 1$ without loss of generality)
\begin{eqnarray*}
\small 
\beta_r^{\mathrm{two}}(\theta, J, \Sigma, p)&=&\sum_{j \in J}  \omega_j \mathbb{E}[r_j(X, J, \Sigma) 1\{\sgn(X_j) = 1\} | \theta] (1 - p)\\
&&+\sum_{j \in J} \omega_j \mathbb{E}[r_j(X, J, \Sigma) 1\{\sgn(X_j) \in \{-1,0\}\} | \theta] p - C(J, \Sigma)
\end{eqnarray*}


We consider maximin testing protocols with respect to $\theta$ and $p$. The following proposition shows that standard separate two-sided $t$-tests are maximin optimal in this modified model. 
\begin{prop}[Maximin optimality of two-sided tests] \label{prop:two_sided}  Consider the model described in this section and the testing protocol $\tilde{r}^t(X)=(\tilde{r}_1^t(X),\dots,\tilde{r}_{|J|}^t(X))^\top$, where
$$
\small
\begin{aligned} 
\tilde{r}_j^t(X) = 1\left\{\frac{|X_j|}{\sqrt{\Sigma_{j,j}}} \ge \Phi^{-1}\left(1 - \frac{C(J,\Sigma)}{b \bar{\omega}(J)}\right)\right\},\quad \forall j\in J. 
\end{aligned} 
$$
Let Assumptions    \ref{ass:linear_rule}, \ref{ass:additive2}, and \ref{ass:partially_contractable} hold.  Then,  $\tilde{r}^t$ is maximin optimal, i.e.,
$$
\small 
\tilde{r}^t \in \arg \max_{r\in \mathcal{R}}\min_{\theta \in \Theta} \min_{p \in \{0,1\}} v_r^{\mathrm{two}}(\theta, J_{r,\theta, p}^*,\Sigma_{r,\theta, p}^*, p), \text{ where } (J_{r,\theta, p}^*,\Sigma_{r,\theta, p}^*) \in \mathrm{arg} \max_{J \in \mathcal{J}, \Sigma \in \mathcal{S}(J)} \beta_r^{\mathrm{two}}(\theta, J, \Sigma, p).
$$
\end{prop} 

The critical value of the optimal two-sided $t$-tests is the same as for the one-sided $t$-tests in Proposition \ref{cor:threshold_main1}. This is because the planner is maximin with respect to $p$.
Using similar arguments as in the main text, one can show that two-sided $t$-tests are also uniformly globally optimal. 

\subsection{Interactions between treatments}
\label{sec:additional_interactions_main}

The results in Section \ref{sec:multiple_treatments} apply to a broad class of researcher benefit and welfare functions, including settings with interactions between treatments. The presence of interactions (in particular in the researcher's benefits) may justify notions of size control that are different from average size control. In this section, we outline the conclusions that one can draw when there are additional interactions.

Following Example \ref{exmp:fwer}, we start by replacing the linear benefits function in Assumption \ref{ass:linear_rule} with a threshold function, where studies yield a (constant) positive payoff if and only if they produce at least one finding,
\begin{equation}  \label{eqn:researcher_payoff_interaction}
\small 
\begin{aligned}
\beta_r(\theta, J, \Sigma) = b \int 1\Big\{\sum_{j \in J}  r_j(x) \ge 1\Big\} dF_\theta(x) - C(J, \Sigma).
\end{aligned} 
\end{equation} 
With a threshold crossing payoff function, the incremental value of a finding to the researcher depends on the number of other findings. For this class of protocols, the first condition in Equation \eqref{cor:sufficient_maximin_eq1} for maximin optimality reads as
\begin{equation} \label{eqn:restriction_fwer}
\small 
\begin{aligned} 
1 - P\Big(\sum_{j \in J} r_j(X) = 0 | \theta\Big) \le C(J, \Sigma)/b, \quad \forall \theta \in \Theta_0(J). 
\end{aligned} 
\end{equation} 
This leads to more complicated optimal hypothesis testing protocols that depend on the joint distribution of $X$.

Suppose first that $r_j(X) \perp r_{j'}(X)$ with $j \neq j'$, which is satisfied in settings where $X$ has independent entries and $r_j$ only depends on $X_j$. An empirical example is when studying treatment effects on independent subgroups of individuals, and considering testing protocols that only depend on the effect on a given subgroup. Under independence, optimal protocols satisfy $P(r_j(X) = 1) = p^*$ for some $p^*$ and impose that $1 - (1 - p^*)^{|J|} \le C(J, \Sigma)/b$ so that $p^* \le 1 - (1 - C(J,\Sigma)/b)^{1/|J|}$.  When $C(J, \Sigma)/b = \alpha$, which is equivalent to assuming that the costs are constant in the number of findings, we can show that the right-hand side is of order $1/|J|$ as $|J| \rightarrow \infty$. Thus, asymptotically, fixed-cost research production functions again rationalize Bonferroni-type corrections. 

Absent restrictions on the correlation structure, one can construct maximin optimal protocols by invoking Proposition \ref{cor:maximin_optimal}, bounding $1 - P\left(\sum_{j \in J} r_j(X) = 0 | \theta\right)$ as 
$$
\small 
\begin{aligned} 
1 - P\Big(\sum_{j \in J} r_j(X) = 0 | \theta\Big)\le  \sum_{j \in J} P(r_j(X) = 1| \theta),
\end{aligned} 
$$ and imposing maximin optimality with respect to this upper bound. This type of correction again amounts to a Bonferroni-type correction with a constant cost function. 

Finally, suppose in addition that (combinations of) treatments are exclusive so that $\sum_{j \in J} r_j(X) \le 1$ and $X$ satisfies Assumption \ref{ass:partially_contractable} with $\Sigma$ being a diagonal matrix (e.g.,\ $X_j$ are effects estimated on independent subgroups). This introduces an additional form of interaction because only one entry of $r$ can be positive (e.g., encoding any form of interaction effects in the welfare function).  
An example of a maximin optimal protocol is $r_j(X) = 1\{j \in \mathrm{arg} \max_{j'} X_{j'}\} \times 1\{\max_{j'} X_{j'} \ge t\}$, where the threshold $t$ must satisfy the restriction in  Equation \eqref{eqn:restriction_fwer}, such that $P(\max_{j'} X_{j'} \ge t|\theta) \le C(J,\Sigma)/b$ for all $\theta \in \Theta_0(J)$. 
This directly relates to weak FWER control.

\subsection{Comparisons with other notions of optimality} \label{sec:discussion_objective}

In this section, we compare the proposed notion of optimality to other notions of optimality. We  consider a simplified version of the model in the main text where the researcher only decides whether or to not run an experiment with a given set of treatments $\bar{J}$ and design $\Sigma$.

\begin{ass} \label{ass:two_decisions} Suppose that $J \in \mathcal{J} = \Big\{\emptyset, \bar{J}\Big\}$ for some $\bar{J}$, with $|\bar{J}| > 1$, with $\mathcal{S}(\bar{J}) = \{\Sigma\}$ for some positive definite $\Sigma$.  
\end{ass} 

Assumption \ref{ass:two_decisions} simplifies the subsequent analysis, while allowing us to highlight the main insights. We keep the notation $J_{r,\theta}^*$ and $\Sigma_{r,\theta}^*$, as in Equation \eqref{eqn:optimality_sigma}, where $J_{r,\theta}^*$ and $\Sigma_{r,\theta}^*$ now reflect the researcher's optimal choices over the restricted sets $\mathcal{J}$ and $\mathcal{S}(\bar{J})$ in Assumption \ref{ass:two_decisions}.

\subsubsection{Uniformly most powerful testing protocols} A natural approach for choosing optimal hypothesis testing protocols would be to set $\lambda = 0$ and choose maximin protocols that uniformly dominate all other maximin protocols, as \cite{tetenov2016economic} does in the $|\bar{J}|=1$ case. This corresponds to looking for uniformly most powerful (UMP) tests in the terminology of classical hypothesis testing. Unfortunately, this approach is not applicable in our context since there is no dominant maximin protocol. 
\begin{prop}[No dominant maximin protocol]  \label{prop:no_admissible} Let Assumptions \ref{ass:linear_rule}, \ref{ass:additive2}, and \ref{ass:two_decisions} hold. Suppose that $0 < C(\bar{J}, \Sigma) < \bar{\omega}(\bar{J})$. Then there exists a compact parameter space $\Theta$, a distribution $\{F_\theta, \theta \in \Theta\}$, and weights $\omega > 0$ such that no maximin protocol  (weakly) dominates all other maximin protocols. Moreover, there exists such a distribution $F_\theta$ satisfying Assumption \ref{ass:partially_contractable}.  
\end{prop} 

In the terminology of classical hypothesis testing, Proposition \ref{prop:no_admissible} states that there are settings in which no UMP tests exist with multiple treatments. This is an important difference between the MHT case we study and the single hypothesis case considered in \citet{tetenov2016economic}, where UMP protocols do exist. 

Note that Proposition \ref{prop:no_admissible} does not imply, however, that there are no maximin protocols that are strictly dominated. For example, separate $t$-tests with $t=\infty$ are maximin optimal (Proposition \ref{cor:threshold_main1}) but dominated by protocol \eqref{eq:optimal_t_test} with $t = \Phi^{-1}(1- C(J,\Sigma)/b \bar{\omega}(J))$.

\subsubsection{Local power} 

Here we compare maximin protocols in terms of their power over a local (to $\theta=0$) alternative space. We consider a notion of local power with the property that locally most powerful protocols are also admissible when $\lambda=0$. This notion of local power is inspired by the corresponding notions in Section 4 of \citet{romano2011consonance} and Chapter 9.2 of \citet{lehmann2005testing}. We show that any uniformly globally optimal protocol is also locally most powerful (and thus admissible if $\lambda=0$) under linearity and normality.

We start by defining a suitable local alternative space.

\begin{defn}[$\epsilon$-alternatives] \label{defn:loc_power} For $\epsilon > 0$, define the local alternative space (with an abuse of notation) as 
$
\Theta_1(\epsilon) :=  \Big\{\theta:  \theta_j =  \epsilon \text{ for some } j , \theta_{j'} \in [0, \epsilon]  \text{ for all other } j' \Big\}. 
$
\end{defn} 

The set of $\epsilon$-alternatives $\Theta_1(\epsilon)$ is the set of parameters for which, for some policy decision, welfare is strictly positive by $\epsilon$. 

Based on Definition \ref{defn:loc_power}, we introduce the following notion of local power.
\begin{defn}[Locally more powerful] \label{defn:most_favorable} A testing protocol $r$ is locally more powerful than $r'$ if
\begin{equation} \label{eqn:most_pow}
\small 
\begin{aligned}
\lim \inf_{\epsilon \downarrow 0}  \left\{\frac{1}{\epsilon}\inf_{\theta \in \Theta_1(\epsilon)} v_{r}(\theta, J_{r,\theta}^*, \Sigma_{r,\theta}^*) - 
\frac{1}{\epsilon}\inf_{\theta' \in \Theta_1(\epsilon)} v_{r'}(\theta', J_{r',\theta'}^*, \Sigma_{r',\theta'}^*)\right\} \ge 0.   
\end{aligned}
\end{equation}
\end{defn} 

Definition \ref{defn:most_favorable} introduces a partial ordering of protocols based on their worst-case performance under $\epsilon$-alternatives. It considers parameter values in an alternative space that contains the origin as $\epsilon \rightarrow 0$. The rescaling by the location parameter $\epsilon$ avoids trivial solutions. 
Under the notion of local power in Definition \ref{defn:most_favorable}, the planner prioritizes power for detecting small effects.

The next proposition characterizes locally most powerful protocols. For simplicity, we focus on settings with equal weights $\omega_j = 1$ for all $j$. The results extend straightforwardly to settings with general weights. 

\begin{prop}[Separate size control is locally most powerful] \label{prop:locally_most_powerful} Let Assumptions \ref{ass:linear_rule}, \ref{ass:additive2}, \ref{ass:partially_contractable}, and \ref{ass:two_decisions} hold with $\omega_j = 1$ for all $j$. Then $r^* \in \mathcal{R}$ is maximin optimal and locally most powerful if it satisfies Equations \eqref{cor:sufficient_maximin_eq1}, \eqref{cor:sufficient_maximin_eq2}, and 
\begin{equation} \label{eqn:separate} 
\small 
\begin{aligned} 
P(r_j^*(X;J,\Sigma) = 1 | \theta = 0) = \frac{C(J,\Sigma)}{b \bar{\omega}(J)}  \quad \forall j \in \{1, \dots, J\}.  
\end{aligned} 
\end{equation}
\end{prop} 
 

The following corollary shows that separate $t$-tests are also locally most powerful. 

\begin{cor}[Separate $t$-tests are locally most powerful] \label{cor:any_global}  Let Assumption  \ref{ass:alternative_space} and the conditions in Proposition \ref{prop:locally_most_powerful} hold. Then separate $t$-testing with $t = \Phi^{-1}\Big(1 - \frac{C(\bar{J},\Sigma)}{b \bar{\omega}(\bar{J})}\Big)$ in Proposition \ref{cor:threshold_main1} is also locally most powerful. 
\end{cor} 

The proof follows directly from Propositions \ref{cor:threshold_main1} and \ref{prop:locally_most_powerful}. 
 More generally, any uniformly globally most powerful test is locally most powerful when at $\theta = 0$ all rejection probabilities are the same.

\subsubsection{Alternative planner objectives}
\label{app: globally optimal protocols}

In the main text, we consider the planner utility  $U$ defined in \eqref{eqn:plannerobjective}. Here we formally compare $U$ to alterative planner objective functions.

\smallskip

\noindent \textbf{Comparison to $U'$ defined in \eqref{eqn:plannerobjective_prime}.}
It is natural to ask whether there are maximin protocols that maximize the second component of $U'$ for \textit{any} $w$. Such a result would guarantee that for any choice of $w$ (and $\lambda$), we can always find a maximizer of $U'$, as was the case with the planner utility $U$ in the main text. 
We show that such a protocol does not exist. For simplicity, we state the result for $\omega_j = 1$ for all $j$.

\begin{prop}[Optimal protocols depend on weights] \label{prop:weights_power} Suppose that Assumptions  \ref{ass:linear_rule}, \ref{ass:additive2}, \ref{ass:partially_contractable}, and \ref{ass:two_decisions} hold with $\omega_j = 1$ for all $j$. For any maximin protocols $r^*$, there exists another maximin protocol $r' \neq r^*$ and a set of weights $w(\theta): \int_{\bar{\Theta}_1} w(\theta) = \int_{\Theta} w(\theta)$ such that
$$
\small 
\begin{aligned}
\int_{\bar{\Theta}_1}  (v_{r^*}(\theta, J_{r^*,\theta}^*, \Sigma_{r^*,\theta}^*) - v_{r'}(\theta, J_{r',\theta}^*, \Sigma_{r',\theta}^*)) w(\theta) d \theta < 0.  
\end{aligned}
$$ 
\end{prop}

Proposition \ref{prop:weights_power} highlights an important limitation of working with $U'$: optimal hypothesis protocols depend on the weights $w$. This sensitivity to the choice of weights is undesirable in practice because choosing suitable weights over the high-dimensional alternative space $\bar{\Theta}_1$ is typically very difficult and somewhat arbitrary. By contrast, the results in Section \ref{sec: most powerful tests} show that the proposed model yields optimal protocols that satisfy unbiasedness---a standard requirement for statistical tests---and do not depend on $\pi$ or $\lambda$.

Nevertheless, it is possible to show that $U$ is \textit{approximately} equal to $U'$ for a suitable choice of the parameter spaces $\bar{\Theta}_1' \subset \bar{\Theta}_1$. The key insight here is that the weight $\pi$ can take arbitrary positive values on $\bar{\Theta}_1$ and therefore can approximately match the welfare effects multiplied by the weights $w$ (over the positive orthant $\bar{\Theta}_1$). This implies that any protocol that maximizes $U$ for all $(\lambda,\pi)$, such as the one in Proposition \ref{cor:threshold_main1}, also maximizes $U'$ for any $w$ supported on $\bar{\Theta}_1' \subset \bar{\Theta}_1$ up-to a possibly small error. We formalize this intuition in the following proposition. To state the result, let $v_{r = 1}(\theta) = v_{r = 1}(\theta,\bar{J}, \Sigma)$, where $r = 1$ is a protocol with $r_j(X) = 1$ almost surely for all $j$.

\begin{prop}[Approximate optimality] \label{prop:apprx2} Let Assumption \ref{ass:two_decisions} hold. Let $\pi(\theta) = v_{r = 1}(\theta) w(\theta)$ and $w(\theta)$ be a weight (i.e., $w(\theta) \ge 0$) with $\int_{\bar{\Theta}_1'} w(\theta) d\theta = \int_{\Theta} w(\theta) d\theta = 1$ for some $\bar{\Theta}_1' \subset \bar{\Theta}_1$.  Then
$$
\small 
\begin{aligned}
\Big|U'(r;\lambda, w)  - U(r;\lambda, \pi)\Big| \le \lambda \max_{\theta \in \bar{\Theta}_1'} |v_{r = 1}(\theta) - v_{r}(\theta, \bar{J}, \Sigma)|  
\end{aligned} 
$$
\end{prop}

Proposition \ref{prop:apprx2} shows that we can bound the difference between $U$ and $U'$ by the difference in social welfare when $r = 1$. To gain further insights, suppose that $v_r(\theta, \bar{J}, \Sigma)$ is linear (Assumption \ref{ass:additive2}) and $\omega_j=1$ for all $j$, so that we can write
$$
\small 
\begin{aligned}
v_{r = 1}(\theta) - v_{r}(\theta, \bar{J}, \Sigma) = \sum_{j \in \bar{J}} \theta_j\Big(1 - \mathbb{E}[r_j(X) | \theta]\Big). 
\end{aligned} 
$$
It follows that for any testing protocol for which $\mathbb{E}[r_j(X)|\theta]$ is decreasing in $\Sigma_{j,j}$ for $\theta$ in the alternative space $\bar{\Theta}_1'$ (e.g., separate $t$-tests), the difference between $U'$ and $U$ is small for $n$ sufficiently large. As a result, the uniformly globally optimal separate $t$-tests in Proposition \ref{cor:threshold_main1} are also approximately optimal under $U'$.  We illustrate this in the following corollary. 

\begin{cor} \label{cor:final}  Suppose that Assumptions  \ref{ass:linear_rule}, \ref{ass:additive2}, \ref{ass:partially_contractable}, and \ref{ass:two_decisions} hold and let $\theta \in [-M, M]$ for a finite constant $M$, and $\omega_j = 1$ for all $j$. Consider $r^t$ defined in Proposition \ref{cor:threshold_main1} with $t = \Phi^{-1}\Big(1 - \frac{C(\bar{J},\Sigma)}{\bar{\omega}(\bar{J})}\Big)$. Then for any $w(\theta)$ satisfying the conditions in Proposition \ref{prop:apprx2} with $\bar{\Theta}_1' = \Big\{\theta: u(\theta, j)> \mu, \forall j \in \bar{J}\Big\}$ for a positive constant $\mu$ and any $\lambda \ge 0$, 
$$
\small 
\begin{aligned} 
\max_r U'(r;\lambda, w) - U'(r^t;\lambda, w) \le  \lambda \cdot M \sum_{j \in \bar{J}} \Phi\Big(t - \frac{\mu}{\sqrt{\Sigma_{j,j}}}\Big) 
\end{aligned} 
$$
\end{cor} 

To interpret Corollary \ref{cor:final}, note that the standard error $\sqrt{\Sigma_{j,j}}$ is typically of order $1/\sqrt{n}$, where $n$ denotes the sample size of the experiment, so that $\max_r U'(r;\lambda, w) - U'(r^t;\lambda, w)$ decreases exponentially fast in $\sqrt{n}$.

\smallskip

\noindent \textbf{An objective function internalizing the researcher's utility.} Consider the following planner's objective that in addition to the maximin component incorporates the researcher's utility,
 $$
 U''(r;\lambda, w) = \min_{\theta \in \Theta} v_r(\theta, J_{r,\theta}^*, \Sigma_{r,\theta}^*) + \lambda \int \beta_r(\theta, J_{r,\theta}^*,\Sigma_{r,\theta}^*) w(\theta) d\theta,
 $$ 
 for some given weights $w(\theta)$. The following proposition shows that separate $t$-tests, which are optimal under $U$, are also approximately optimal under $U''$.

\begin{prop}[Approximate optimality of $t$-tests] \label{cor:final2}  Suppose that Assumptions \ref{ass:linear_rule}, \ref{ass:additive2}, \ref{ass:partially_contractable}, and \ref{ass:two_decisions} hold with $\omega_j = 1$ for all $j$. Consider $r^t$ defined in Proposition \ref{cor:threshold_main1} with $t = \Phi^{-1}\Big(1 - \frac{C(\bar{J},\Sigma)}{\bar{\omega}(\bar{J})}\Big)$. Then for any $w(\theta)$ satisfying the conditions in Proposition \ref{prop:apprx2} with $\bar{\Theta}_1' = \Big\{\theta: u(\theta, j)> \mu, \forall j \in \bar{J}\Big\}$ for a positive constant $\mu$ and any $\lambda \ge 0$, 
$$
\small 
\begin{aligned} 
\max_r U''(r;\lambda, w) - U''(r^t;\lambda, w) \le  \lambda \sum_{j \in \bar{J}} \Phi\Big(t - \frac{\mu}{\sqrt{\Sigma_{j,j}}}\Big). 
\end{aligned} 
$$
\end{prop}

\section{Review of single hypothesis case in \citet{tetenov2016economic}} \label{sec:review_single_hypothesis}
 
\citet{tetenov2016economic} considers a game between an informed agent and a regulator, where the agent decides whether to run a pre-specified experiment. We explain his results using the terminology of our framework. Define the \textit{null space} of parameter values as the set of parameters such that implementing the (single) treatment being studied would reduce welfare, $\Theta_0 := \left\{\theta: u(\theta) < 0\right\}$. Similarly, define the \emph{alternative space} of parameter values as the set of parameters such that the treatment increases welfare $\Theta_1 := \left\{\theta: u(\theta) \ge  0\right\}.$ Welfare is
$v_r(\theta)=u(\theta)$ if $\int r(x) dF_{\theta}(x) \ge C$, where $C$ is the cost of running the experiment, and zero otherwise. That is, welfare is non-zero if the expected payoff from experimenting, $\int r(x) dF_{\theta}(x)$, is larger than the cost of experimentation.

To justify single hypothesis testing, \citet[][]{tetenov2016economic} focuses on maximin optimal testing protocols, i.e., testing protocols that maximize worst-case welfare, 
$
r^* \in \arg \max_{r\in \mathcal{R}}  \min_{\theta \in \Theta} v_r(\theta).
$
Proposition 1 in \citet[][]{tetenov2016economic} shows that a testing protocol is maximin optimal
if and only if 
$
   \int r^*(x) dF_{\theta}(x) \le C  \text{ for all }\theta \in \Theta_0. 
 $ 
The model thus rationalizes standard size control. 

To select among the many alternative maximin testing protocols, \citet{tetenov2016economic} provides admissibility results under an additional monotone likelihood ratio property. He shows that admissible testing protocols satisfy the following condition
$ 
    \int r^*(x) dF_0(x) = C, 
  $ 
with the testing protocol taking the form of threshold crossing protocols, $r(X)=1\{X\ge t^\ast\}$. This result provides a formal justification for standard (one-sided) tests.

\section{Proofs of results in main text}

Throughout the remainder of the paper, we will write the researcher's cost function as $C(J,\Sigma)$ whenever the costs do not depend on $\theta$.  To simplify the exposition, we will sometimes write $C$ instead of $C(J,\Sigma)$ and $P(r_j(X) | \theta)$ instead of $P(r_j(X) = 1|\theta)$ whenever it does not cause any confusion. Without loss of generality, unless otherwise specified, we normalize $b$ to $b = 1$.  We denote by $\mathcal{M}$ the set of maximin protocols. 
 Because $\Theta$ is a compact space, without loss of generality, we will assume that $|\theta_j| \le M$ for some finite constant $M < \infty$ and all $j$. We will use that $\Theta_0(J) \neq \emptyset$ for all $J \neq \emptyset$ and $\bar{\Theta}_1 \neq \emptyset$ (as discussed in the main text). Finally, we will write $t$ in lieu of $t(J,\Sigma)$ to alleviate the notation.

\subsection{Proof of Proposition \ref{cor:unbiasedness} } \label{proof:cor:unbiasedness}

Before providing the results in the main text, we introduce the following lemma. 

  \begin{lem}[Globally optimal protocols for given $\pi, \lambda$] \label{prop:global_optimal} Let Assumption \ref{ass:alternative_space} hold. Define 
 $$
 \mathcal{R}^\ast(\pi, \lambda) = \Big\{r: r \text{ is maximin optimal and } \lambda \int_{\Theta} 1\{ \beta_r^*(\theta) \ge 0\} \pi(\theta) d\theta = \lambda \int_{\bar{\Theta}_1} \pi(\theta) d\theta\Big\}, 
 $$
 with $\beta_r^*(\theta)$ defined in Equation \eqref{eqn:unbiased}. 
Assume that $\mathcal{R}^\ast(\pi, \lambda) \neq \emptyset$. Then  
 $$
\mathcal{R}^\ast(\pi, \lambda) = \arg \max_{r\in \mathcal{R}} \left\{\min_{\theta \in \Theta} v_r(\theta, J_{r,\theta}^*,\Sigma_{r,\theta}^*) + \lambda \int_{\Theta} e_r^\ast(\theta) \pi(\theta) d\theta\right\},   
$$
where the set $\mathcal{R}^*(\pi,\lambda)$ is a function of $(\pi, \lambda)$. 
 \end{lem} 

 \begin{proof} 
Recall that $\int_\Theta \pi(\theta) d\theta = \int_{\bar{\Theta}_1} \pi(\theta) d\theta$ and $\pi(\theta) \ge 0$ under Assumption \ref{ass:alternative_space}. Notice that any maximin protocol  maximizes $\min_{\theta \in \Theta} v_r(\theta, J_{r,\theta}^*,\Sigma_{r,\theta}^*)$ by definition. Notice further that for any $\theta \in \bar{\Theta}_1$ $e_r^\ast(\theta) = 1\{\beta_r^*(\theta) \ge 0\}$ for $\theta \in \bar{\Theta}_1$, where the weak inequality follows from the tie-breaking rule (since for $\theta \in \bar{\Theta}_1$ welfare effects are always weakly positive). This implies that the set of protocols that maximize $\lambda \int_\Theta e_r^\ast(\theta) \pi(\theta) d\theta$ coincides with the set of protocols satisfying $\lambda \int_\Theta 1\{\beta_r^*(\theta) \ge 0\} \pi(\theta) d\theta = \lambda \int_\Theta \pi(\theta) d\theta= \lambda \int_{\bar{\Theta}_1} \pi(\theta) d\theta$, so that $\lambda \int_\Theta e_r^\ast(\theta) \pi(\theta) d\theta = \lambda  \int_\Theta \pi(\theta) d\theta$, which is the largest value it can achieve (since $\pi(\theta) \ge 0$). This completes the proof.
 \end{proof} 

 We can now prove the main statement. 
The ``if'' direction is a direct corollary of Lemma \ref{prop:global_optimal}, since a maximin and unbiased protocol belongs to $\bigcap_{\pi \in \Pi, \lambda \ge 0} \mathcal{R}^*(\pi, \lambda)$. The ``only if'' direction proceeds by contraposition. Take any protocol $r$ that is not maximin. Then  this protocol does not maximize the planner's utility at $\lambda = 0$. Take any protocol $r$ that is not unbiased (but could be maximin). Then because $r$ is not unbiased we can find a set of weights $\pi \in \Pi$ such that $\int e_r^*(\theta)\pi(\theta) d\theta < \int \pi(\theta) d\theta$, and thus $r$ does not maximize the planner's utility $U(r;\lambda, \pi)$ for any $\lambda > 0$ (and therefore is not uniformly globally optimal). The proof completes because we assumed existence of unbiased maximin protocol.

\subsection{Proof of Proposition \ref{prop:maximin2}}
\label{app:prop:maximin}

The proof proceeds in two steps. We first provide an equivalent condition for maximin optimality and then use this condition to prove the result.

\smallskip

\noindent \textbf{Step 1: Equivalent condition for maximin optimality.} Define the worst-case $\theta$ as a function of the protocol $r$ as $\theta^*(r) = \arg\min_{\theta \in \Theta} v_r(\theta, J_{r,\theta}^*, \Sigma_{r,\theta}^*)$. In this step, we prove the following claim: A protocol $r^*$ is maximin optimal if and only if 
$$
v_{r^*}\left(\theta^*(r^*), J_{r^*,\theta^*(r^*)}^*, \Sigma_{r^*,\theta^*(r^*)}^*\right) = 0.
$$
To prove this claim, note first that for any maximin optimal $r^*$, $v_{r^*}\left(\theta^*(r^*), J_{r^*,\theta^*(r^*)}^*, \Sigma_{r^*,\theta^*(r^*)}^*\right) \ge 0$ since the planner can always choose $\tilde{r}(X;J,\Sigma) = (0, 0, \dots, 0)$ for all $(X,J,\Sigma)$ and obtain $v_{\tilde{r}}(\theta, J_{\tilde{r},\theta}^*, \Sigma_{\tilde{r},\theta}^*)= 0 $ for all $ \theta \in \Theta$. 
Next, we show that $v_{r^*}(\theta^*(r^*), J_{r^*,\theta^*(r^*)}^*, \Sigma_{r^*,\theta^*(r^*)}^*)\le 0$ for any $r^*$ (and therefore also for any maximin optimal $r^*$). Pick any $\theta' \in \bigcap_{J \in \mathcal{J} \setminus \emptyset} \Theta_0(J)$, where $\bigcap_{J \in \mathcal{J} \setminus \emptyset} \Theta_0(J) \neq \emptyset$ by Assumption \ref{ass:non_emptyness}. Then it follows that 
$v_{r^*}(\theta^*(r^*), J_{r^*,\theta^*(r^*)}^*, \Sigma_{r^*,\theta^*(r^*)}^*) = \min_{\theta \in \Theta} v_{r^*}(\theta, J_{r^*,\theta}^*, \Sigma_{r^*,\theta}^*) \le v_{r^*}(\theta', J_{r^*,\theta'}^*, \Sigma_{r^*,\theta'}^*) \le 0$.  This completes the proof of the claim. 

\medskip

\noindent \textbf{Step 2: Proof of main result.} 
To prove the ``if'' direction, we only need to show that under $r^*$, $v_{r^*}(\theta^*(r^*), J_{r^*,\theta^*(r^*)}^*, \Sigma_{r^*,\theta^*(r^*)}^*) \ge 0$ (which, by the argument in Step 1, implies that $v_{r^*}(\theta^*(r^*), J_{r^*,\theta^*(r^*)}^*, \Sigma_{r^*,\theta^*(r^*)}^*) = 0$). Equation \eqref{eqn:constraint} implies that $v_{r^*}(\theta, J_{r^*,\theta}^+, \Sigma_{r^*,\theta}^+) = 0$ for all $\theta \in \Theta_0^*(r^*)$. By definition, $J_{r^*,\theta^*(r^*)}^* \in \{\emptyset, J_{r^*,\theta^*(r^*)}^+\}$. Because $v_{r^*}(\theta^*(r^*), \emptyset, \Sigma_{r^*,\theta^*(r^*)}^*) = 0$ by construction, it must be that 
$v_{r^*}(\theta^*(r^*), J_{r^*,\theta^*(r^*)}^*, \Sigma_{r^*,\theta^*(r^*)}^*) \ge 0$ if $\theta^*(r^*) \in \Theta_0^*(r^*)$ by the tie-breaking rule (since we are assuming $\lambda = 0$). This is because for $\theta^*(r^*) \in \Theta_0^*(r^*)$, $J_{r^*,\theta^*(r^*)}^* = J_{r^*,\theta^*(r^*)}^+$ only when the researcher is indifferent between whether to conduct the experiment or not (i.e., is indifferent between $J_{r^*,\theta^*(r^*)}^* \in \{\emptyset, J_{r^*,\theta^*(r^*)}^+\}$), in which case by the tie-breaking rule, the researcher does what is most convenient for the social planner. If, instead, $\theta^*(r^*) \in \Theta \setminus \Theta_0^*(r^*)$, we have that $v_{r^*}(\theta^*(r^*), J_{r^*,\theta^*(r^*)}^*, \Sigma_{r^*,\theta^*(r^*)}^*) \ge 0$ by construction of $\Theta_0^*(r^*)$. 

We now prove the ``only if'' direction. Consider first the case where the second condition in 
Equation \eqref{eqn:constraint} fails.  Then this implies we can find a protocol $\tilde{r}(X) = (0, \dots, 0)$ so that $\min_{\theta \in \Theta} v_{r^*}(\theta, J_{r^*, \theta}^*, \Sigma_{r^*,\theta}^*) < 0 = \min_{\theta \in \Theta} v_{\tilde{r}}(\theta, J_{\tilde{r}, \theta}^*, \Sigma_{\tilde{r},\theta}^*)$, which implies that $r^*$ is not maximin optimal.  

Consider now the case where the first condition in Equation \eqref{eqn:constraint} fails. 
We use a contradiction argument. Suppose that 
$\beta_{r^*}(\theta, J_{r^*,\theta}^+, \Sigma_{r^*,\theta}^+) > 0$ for some $\theta \in \Theta_0^*(r^*)$. Since $J_{r^*,\theta}^* \in \{\emptyset, J_{r^*,\theta}^+\}$, we can find some $\theta \in \Theta_0^*(r^*)$ so that $J_{r^*,\theta}^* = J_{r^*,\theta}^+$ and $\Sigma_{r^*,\theta}^* = \Sigma_{r^*,\theta}^+$. This follows because the researcher's best action when $\beta_{r^*}(\theta, J_{r^*,\theta}^+, \Sigma_{r^*,\theta}^+) > 0$ is to experiment. 
Because $\beta_{r^*}(\theta, J_{r^*,\theta}^+, \Sigma_{r^*,\theta}^+) > 0$, and $c_\theta(J_{r^*,\theta}^+, \Sigma_{r^*,\theta}^+) \ge 0$ for all $\theta$ (see Equation \eqref{eqn:B_R_general}), it must be under the payoff in Equation \eqref{eqn:B_R_general} that $\int \delta_k(r(x; J_{r^*,\theta}^+, \Sigma_{r^*,\theta}^+), J_{r^*,\theta}^+) dF_{\theta, J_{r^*,\theta}^+, \Sigma_{r^*,\theta}^+}(x) > 0$ for some $k$. 
By definition of $\Theta_0^*(r^*)$, this implies that  
$$
\begin{aligned} 
v_{r^*}(\theta, J_{r^*,\theta}^*, \Sigma_{r^*,\theta}^*) = \sum_k u_k(\theta, J_{r^*,\theta}^+) \int \delta_k(r(x; J_{r^*,\theta}^+, \Sigma_{r^*,\theta}^+), J_{r^*,\theta}^+) dF_{\theta, J_{r^*,\theta}^+, \Sigma_{r^*,\theta}^+}(x)  < 0
\end{aligned} 
$$
for $\theta \in \Theta_0^*(r^*)$. This violates the equivalent condition for maximin optimality in Step 1.

\subsection{Proof of Corollary \ref{cor:sufficient_maximin}} \label{proof:cor:sufficient_maximin}  

By Proposition  \ref{cor:unbiasedness}, it suffices to show that the chosen protocol is both maximin optimal and unbiased.  
Since our chosen tie-breaking rule favors the planner whenever $\beta_{r^*}(\theta, J_{r^*,\theta}^*,\Sigma_{r^*,\theta}^*) = 0$, we assume that the researcher experiments for $\theta \in \bar{\Theta}_1$ (since this leads positive planner's utility) and does not experiment for all $\theta$ such that $v_{r^*}(\theta, J_{r^*,\theta}^*,\Sigma_{r^*,\theta}^*) < 0$ (since this leads to weakly negative planner utility because $\pi(\theta)$ has support over $\theta \in \bar{\Theta}_1$ by Assumption \ref{ass:alternative_space} and $v_r(\theta, J,\Sigma) \ge 0$ for all $\theta \in \bar{\Theta}_1$).\footnote{Note that there could be other possible tie-breaking rules depending on $r$. However, this does not matter for of the argument below, since the planner's utility is equivalent under all tie-breaking rules that maximize her utility when the researcher is indifferent.}  

\smallskip

\noindent \textbf{Maximin optimality.}
Whenever Equation \eqref{cor:sufficient_maximin_eq1} holds, the first condition in Equation \eqref{eqn:constraint} holds. Therefore, it suffices to only verify the second condition in Equation \eqref{eqn:constraint} to prove maximin optimality. To show this, note that for any $\theta$ so that $\beta_{r^*}(\theta, J_{r^*,\theta}^*,\Sigma_{r^*,\theta}^*) < 0$, $v_{r^*}(\theta, J_{r^*,\theta}^*,\Sigma_{r^*,\theta}^*) = 0$  because the researcher does not experiment in this case. In addition, by our tie-breaking rule, when $\beta_{r^*}(\theta, J_{r^*,\theta}^*,\Sigma_{r^*,\theta}^*) = 0$, the researcher does not experiment, unless  $v_{r^*}(\theta, J_{r^*,\theta}^*,\Sigma_{r^*,\theta}^*) \ge 0$. Therefore, we can write $v_{r^*}(\theta, J_{r^*,\theta}^*,\Sigma_{r^*,\theta}^*) \ge v_{r^*}(\theta, J_{r^*,\theta}^*,\Sigma_{r^*,\theta}^*) 1\{\beta_{r^*}(\theta, J_{r^*,\theta}^*,\Sigma_{r^*,\theta}^*) > 0\}$. From this identity, it follows that Equation \eqref{cor:sufficient_maximin_eq2} implies the second condition in Equation \eqref{eqn:constraint}, proving maximin optimality.

\smallskip

\noindent \textbf{Unbiasedness.} Unbiasedness of $r^\ast$ follows from Equation \eqref{cor:sufficient_maximin_eq3}, since $\beta_{r^*}(\theta, J, \Sigma) \ge 0$ for all $J\in \mathcal{J}\setminus \emptyset$ and $ \Sigma\in \mathcal{S}(J)$ implies that $\beta_{r^*}^*(\theta) \ge 0$  for all $\theta \in \bar{\Theta}_1$ as defined in Equation \eqref{eqn:unbiased}. This completes the proof. 

\subsection{Proof of Proposition \ref{cor:iff_robust}} \label{proof:cor:iff_robust}
We start by noting that Proposition \ref{cor:unbiasedness} and Proposition \ref{prop:maximin2} (and thus Corollary \ref{cor:sufficient_maximin}) hold verbatim once we restrict the feasible set to be $\mathcal{D}$ for any $\mathcal{D} \subseteq \{(J,\Sigma): J \in \mathcal{J},\Sigma \in \mathcal{S}(J)\}$. This follows because the arguments for proving these propositions apply verbatim after redefining the set of treatments to only include the relevant combinations of treatments in $\mathcal{D}$.

\smallskip

\noindent \textbf{``If'' direction.} Take a set of feasible designs $(J,\Sigma) \in \mathcal{D}$.
By
Corollary \ref{cor:sufficient_maximin}, a protocol $r^*$ is uniformly globally optimal once with respect to $\mathcal{D}$ if Equations \eqref{cor:sufficient_maximin_eq1}, \eqref{cor:sufficient_maximin_eq2}, and \eqref{cor:sufficient_maximin_eq3} hold for all $(J,\Sigma) \in \mathcal{D}, J \neq \emptyset$. 

Take any subset $\mathcal{D}$. 
Because $\mathcal{D} \subseteq \{(J,\Sigma): J \in \mathcal{J}, \Sigma \in \mathcal{S}(J)\}$, requiring that Equations \eqref{cor:sufficient_maximin_eq1}, \eqref{cor:sufficient_maximin_eq2} and \eqref{cor:sufficient_maximin_eq3} hold for all $J \in \mathcal{J} \setminus \emptyset,\Sigma \in \mathcal{S}(J)$  is a more stringent condition than requiring them to hold for all $(J,\Sigma) \in \mathcal{D}$ (after excluding $J = \emptyset$ for which the equations trivially hold). Clearly this argument applies to all sets $\mathcal{D}$. This implies that any protocol that satisfies Equations \eqref{cor:sufficient_maximin_eq1}, \eqref{cor:sufficient_maximin_eq2} and \eqref{cor:sufficient_maximin_eq3} for all $J \in \mathcal{J} \setminus \emptyset,\Sigma \in \mathcal{S}(J)$ is design-robust globally optimal.

\smallskip

\noindent \textbf{``Only if'' direction.} To prove the ``only if'' direction, it suffices to find a set $\mathcal{D}$ and some $(\bar{J},\bar{\Sigma}) \in \mathcal{D}, \bar{J} \neq \emptyset$, so that if either Equation \eqref{cor:sufficient_maximin_eq1},  \eqref{cor:sufficient_maximin_eq2}, or \eqref{cor:sufficient_maximin_eq3} is violated for some $(\bar{J},\bar{\Sigma})$, then $r^*$ is not uniformly globally optimal under the feasible set $\mathcal{D}$.  

 To do so, suppose that for some $(\bar{J}, \bar{\Sigma})$ either Equation \eqref{cor:sufficient_maximin_eq1} or \eqref{cor:sufficient_maximin_eq2} or \eqref{cor:sufficient_maximin_eq3} are violated.
Take $\mathcal{D} = \{(\emptyset, \emptyset), (\bar{J},\bar{\Sigma})\}$. Specializing our notation to this choice of $\mathcal{D}$, we have $\Theta_0^*(r^*) = \Theta_0(\bar{J})$, $\bar{\Theta}_1 = \Theta_1(\bar{J})$, $J_{r^*,\theta}^* \in \{\emptyset, \bar{J}\}, J_{r^*,\theta}^+ = \bar{J}$, and $\Sigma_{r,\theta}^* = \bar{\Sigma}$ when $J_{r,\theta}^* \neq \emptyset$. 

If Equation \eqref{cor:sufficient_maximin_eq3} is violated, $r^*$ is not unbiased and therefore not uniformly globally optimal by Proposition \ref{cor:unbiasedness}. 

Suppose next that Equation \eqref{cor:sufficient_maximin_eq1} is violated. 
Then,  $J_{r^*,\theta}^* = \bar{J}$, and because $\beta_{r^*}(\theta,\bar{J},\bar{\Sigma}) > 0$, it follows that $r^*$ is not maximin optimal by Proposition \ref{prop:maximin2} and thus not uniformly globally optimal by Proposition \ref{cor:unbiasedness}. 

Finally, suppose that Equation \eqref{cor:sufficient_maximin_eq2} is violated.  Then this implies that for some $\theta \in \Theta \setminus \Theta_0(\bar{J})$, we have $v_{r^*}(\theta, \bar{J},\bar{\Sigma}) < 0$ and $\beta_{r^*}(\theta, \bar{J},\bar{\Sigma}) > 0$. This implies that $J_{r^*,\theta}^* = \bar{J}$. This  violates the second condition in Proposition \ref{prop:maximin2}. Therefore $r^*$ is not maximin optimal and therefore not uniformly globally optimal by Proposition \ref{cor:unbiasedness}. The proof is complete.

\subsection{Proof of Proposition \ref{cor:maximin_optimal}}  \label{proof:maximin_3}

In this proof, we make the dependence on the researcher's payoff function $\beta$ explicit and write $\left(J_{r, \theta}^*(\beta), \Sigma_{r,\theta}^*(\beta)\right) = \mathrm{arg} \max_{J \in \mathcal{J}, \Sigma \in \mathcal{S}(J)} \beta_r(\theta, J, \Sigma)$. 

\medskip

\noindent \textbf{Proof of Part (i).} Following Step 1 of Proposition \ref{prop:maximin2}, a protocol is maximin optimal for given $\beta_r(\theta, J, \Sigma)$ if $v_{r^*}(\theta; J_{r, \theta}^*(\beta), \Sigma_{r,\theta}^*(\beta)) \ge 0$ for all $\theta \in \Theta$. 

Take $\theta \in \Theta_0(J)$. Then, since $\beta_{r^*}'(\theta, J,\Sigma) \le \beta_{r^*}(\theta, J, \Sigma) $ for all $ \theta \in \Theta, J \in \mathcal{J}, \Sigma \in \mathcal{S}(J)$, it follows that $\beta_{r^*}'(\theta, J, \Sigma) \le 0$.   In addition, we have that $1\{\beta_r(\theta, J, \Sigma) \ge 0\} \ge 1\{\beta_r'(\theta, J, \Sigma) \ge 0\}$. Therefore, $v_{r^*}(\theta; J_{r, \theta}^*(\beta'), \Sigma_{r,\theta}^*(\beta')) \ge 0$ for all $\theta \in \Theta_0(J), J \in \mathcal{J}$ under the tie-breaking rule.\footnote{As  we discuss in the proof of Corollary \ref{cor:sufficient_maximin}, it suffices to focus on the tie-breaking rule where the researcher if indifferent of whether to run an experiment or not, does not run an experiment whenever $\theta \in \cup_{J \in \mathcal{J}} \Theta_0(J)$ and runs the experiment when $\theta \in \bar{\Theta}_1$.} 

Take $\theta \not \in \Theta_0(J)$ (the subsequent argument applies to any $\theta \in \Theta_0(J)$ and any $J \in \mathcal{J} \setminus \emptyset$). There are two cases: (i) $\theta$ is such that $J_{r^*,\theta}^*(\beta') \neq \emptyset$ and (ii) $\theta$ is such that $J_{r^*,\theta}^*(\beta') = \emptyset$. Consider first the case where $J_{r^*,\theta}^*(\beta') \neq \emptyset$. Then $r^*$ satisfying the first two condition in Corollary \ref{cor:sufficient_maximin} (Equations \eqref{cor:sufficient_maximin_eq1} and \ref{cor:sufficient_maximin_eq2}) implies that $v_{r^*}(\theta, J, \Sigma) 1\{\beta_r(\theta,J,\Sigma) \ge 0\} \ge 0$ for all $\Sigma \in \mathcal{S}(J)$. Because this argument applies to all $J \in \mathcal{J} \setminus \emptyset$, it must be (since $1\{\beta_{r^*}(\theta,J_{r^*,\theta}^*,\Sigma_{r^*,\theta}^*) \ge 0\} = 1$ if $J_{r^*,\theta}^* \neq \emptyset$)  that $v_{r^*}(\theta, J_{r^*,\theta}(\beta'), \Sigma_{r,\theta}^*(\beta')) \ge 0$ when $J_{r^*,\theta}(\beta') \neq \emptyset$ proving maximin optimality case in this case. Suppose instead that $\theta$ is such that $J_{r^*,\theta}^*(\beta') = \emptyset$. Then we must have $v_{r^*}(\theta, J_{r^*,\theta}^*(\beta'), \Sigma_{r,\theta}^*(\beta')) = 0$ for $\theta$ such that $J_{r^*,\theta}^*(\beta') = \emptyset$. 

Collecting all cases, this proves that for all $\theta \in \Theta$, $v_{r^*}(\theta, J_{r^*,\theta}^*(\beta'), \Sigma_{r,\theta}^*(\beta')) \ge 0$.  

\medskip

\noindent \textbf{Proof of Part (ii).} The proof follows similarly to the proof of Proposition \ref{cor:unbiasedness}. The protocol $r^*$ described in (ii) maximizes the maximin component of the planner's utility by construction. Moreover, it maximizes the second component of the planner's utility because it guarantees that $\int e_{r^*}(\theta) d\pi'(\theta) = \int d\pi'(\theta)$ for all $\pi' \in \tilde{\Pi}$, by construction of $\tilde{\Pi}$. 

\subsection{Proof of Proposition \ref{cor:threshold_main1}} 
\label{app:cor:threshold}

\subsubsection*{Maximin optimality of separate $t$-tests}
We start by proving the ``if'' statement. We will prove that the conditions in Equations \eqref{cor:sufficient_maximin_eq1} and \eqref{cor:sufficient_maximin_eq2} hold for any $(J,\Sigma)$ with $J \neq \emptyset$ and $\Sigma$ satisfying Assumption \ref{ass:partially_contractable}. We will therefore suppress the dependence of $r(\cdot;J,\Sigma)$ on $(J,\Sigma)$ for notational convenience. 
Clearly, for any $J$ such that $|J| = 1$, Equations \eqref{cor:sufficient_maximin_eq1} and \eqref{cor:sufficient_maximin_eq2} trivially hold, since $P(r_j(X)=1 | \theta)$ is monotonically increasing in $\theta_j$ and constant in $\theta_{-j}$, where $\theta_{-j}$ denotes all elements in $\theta$ except for the $j$th element. It therefore suffices to focus on settings where $|J| > 1$. Recall that we write $\bar{\omega}(J) = \sum_{j \in J} \omega_j$. 

 Because $P(r_j(X)=1 | \theta)$ is monotonically increasing in $\theta_j$ and constant in $\theta_{-j}$, Equation \eqref{cor:sufficient_maximin_eq1} holds for any $t \ge \Phi^{-1}(1 - C(J,\Sigma)/\bar{\omega}(J))$. We now show that Equation \eqref{cor:sufficient_maximin_eq2} is satisfied.

To show this, it suffices to show that the worst-case objective function is weakly positive for any $t \ge \Phi^{-1}\left(1 - C(J,\Sigma)/\bar{\omega}(J) \right)$. With an abuse of notation, we denote by $\theta_j$ the treatment effect divided by $\sigma:= \sqrt{\Sigma_{j,j}}$, which is finite and strictly positive by assumption. 
Note further that
$
C(J,\Sigma) = \bar{\omega}(J) (1 - \Phi(t^\ast)),$ $ t^\ast = \Phi^{-1}(1 - C(J,\Sigma)/\bar{\omega}(J)) 
$
and that, for any $t' \ge t^\ast$, we have $C(J,\Sigma) \ge \bar{\omega}(J) (1 - \Phi(t'))$. 

Using such standardization, 
since $u(\theta,j)= \sigma \theta_j$ by Assumption \ref{ass:partially_contractable} (where $\theta_j$ now denotes the treatment effect divided by $\sigma$) and $\Theta$ is compact, $\min_\theta v_{r^t}(\theta, J, \Sigma) 1\{\beta_{r^t}(\theta,J,\Sigma) \ge 0\}$ must be bounded from below  (since $\Sigma_{j,j} = \sigma^2 > \underline{\gamma} > 0$). It follows that for any $t \ge t^\ast$, proving that $\min_\theta v_{r^t}(\theta, J, \Sigma) 1\{\beta_{r^t}(\theta,J,\Sigma) \ge 0\}$ weakly positive is equivalent to proving weak positivity of the value of the objective function (at the optimum) in the following problem: 
\begin{equation} \label{eqn:lll} 
\small 
\begin{aligned}
\min_{\theta \in [-M, M]^{|J|
}} \sigma \sum_{j \in J} \omega_j (1 - \Phi(t- \theta_j)) \theta_j , ~~\text{s.t. } \sum_{j \in J} \omega_j (1 - \Phi(t- \theta_j))  \ge C(J, \Sigma) \ge \bar{\omega}(J) (1 - \Phi(t)). 
\end{aligned}
\end{equation}  
Note that the value of the objective function in Equation \eqref{eqn:lll} is bounded from below by the value of the following objective function (at the optimum)   
\begin{equation} \label{eqn:lll_v2} 
\small 
\begin{aligned}
\min_{\theta \in [-M, M]^{|J|}} \sigma \sum_{j \in J} \omega_j (1 - \Phi(t- \theta_j)) \theta_j , ~~ \text{s.t. } \sum_{j \in J} \omega_j (1 - \Phi(t- \theta_j))  \ge  \bar{\omega}(J) (1 - \Phi(t)),  
\end{aligned}
\end{equation}  
since we relaxed the lower bound on the constraint. 
We focus on the case where $t$ is finite (if $t = \infty$ the planner's utility is always zero and the result trivially holds). 
It therefore suffices to show that the value of the objective function (at the optimum) in Equation \eqref{eqn:lll_v2} is weakly positive to obtain the desired claim of maximin optimality. 

In the following, we prove by contradiction that the objective in Equation \eqref{eqn:lll} is weakly positive for any $t \ge \Phi^{-1}\left(1 - C(J,\Sigma)/\bar{\omega}(J) \right)$ by first assuming that the objective of Equation \eqref{eqn:lll} is strictly negative and then using a contradiction argument. 

\smallskip

\noindent \textbf{Step 1: Preliminary observation.} Observe that welfare is negative only if the minimizer $\theta^*$ is such that for some $j$, $\theta^*_j < 0$, and for some other $j' \neq j$, $\theta_{j'}^* > 0$ because $\omega_j \ge 0$ for all $j$. To see this, note that to have negative welfare, there must exist some $\theta_j^*<0$. Moreover, there must exist some  $\theta_{j'}^*>0$ because if $\theta^*_j< 0$ for all $j$ the constraint is violated. This shows that a necessary condition for welfare to be strictly negative is that for some $j \neq j'$, $\theta_j^*<0$ and $\theta_{j'}^*>0$. 

\smallskip
\noindent \textbf{Step 2: Focus on interior solution.} Note that if we replace the constraint in Equation \eqref{eqn:lll_v2}, $\theta_J \in [-M, M]^{|J|}$, with $\theta \in [-M', M']^{|J|}, M' > M$ for some large enough but finite $M'$, the solution to the corresponding optimization problem is a lower bound for Equation \eqref{eqn:lll_v2}.  

Now we argue that it suffices to focus on solutions $\theta^*$ in the interior of $[-M', M']^{|J|}$ for some (possibly) large but finite $M'$, as we replace $\theta_J \in [-M, M]^{|J|}$ with $\theta_J \in [-M', M']^{|J|}$ to show negativity of the objective function. It suffices to show that $\theta_j \neq -\infty$ for all $j$ to claim that if the worst-case objective is negative, the minimizer must be in the interior of $[-M', M']^{|J|}$. 
To see why, observe that if $\theta_j^*=-\infty$ for at least one $j$, its contribution to the objective function is zero (since $z (1 - \Phi(t - z)) \rightarrow 0$ as $z \rightarrow -\infty$). However, it forces some of the elements of $\theta_{-j}$ to be positive via its impact on the constraint. 
Hence, if the minimum of the objective is strictly smaller than zero, there must exist a minimizer $\theta^*$, which is in the interior of $[-M', M']^{|J|}$ for some (possibly) large but finite $M'$. Therefore, we will replace $\theta_J \in [M, M]^{|J|}$ with $\theta_J \in [-M', M']^{|J|}, M' >M$ for $M'$ finite but large enough such that this constraint is not binding in the following steps.

\smallskip
\noindent \textbf{Step 3: Constraint qualification.} Next, we show that for finite $t$ the KKT conditions are necessary for the optimality of $\theta_J^*$ in the interior of $[-M', M']^{|J|}$. Note that $t > -\infty$ by construction since  $\bar{\omega}(J) > C(J,\Sigma)$ in Assumption \ref{ass:linear_rule}. To show necessity of KKT conditions for any local (and therefore also for the global) optimizer $\theta_J^* \in [-M',M']$ for some finite $M'$, we use the linear independence constraint qualification (LICQ). Observe that the gradient of the constraint function has component $j$ equal to 
$
-\omega_j \phi(t - \theta_j). 
$
Here checking the LICQ conditions amount to check whether $\omega_j \phi(t - \theta_j) \neq 0$ for any finite $t$ and $\theta_J^*$ in the interior of $[-M',M']^{|J|}$ for some finite $M'$. 
Clearly, $\omega_j \phi(t - \theta_j) \neq 0$ for all $j$, 
for finite $t$ and any finite $\theta$ since $\omega_j > 0$ for all $j$.  

\smallskip
\noindent \textbf{Step 4: Lagrangian.} We now study necessary conditions for the optimal solution of the problem in Equation \eqref{eqn:lll_v2}. Consider the Lagrangian function
\begin{equation} \label{eqn:lagrangian1}
\small 
\begin{aligned}
\sigma \sum_{j \in J} \omega_j (1 - \Phi(t- \theta_j)) \theta_j
&& + \nu \Big[ \bar{\omega}(J) (1 - \Phi(t)) - \sum_{j \in J} \omega_j (1 - \Phi(t- \theta_j))\Big] \\
&&+ \mu_{1, j}[\theta - M'] + \mu_{2, j}[-\theta - M']. 
\end{aligned} 
\end{equation} 
Now observe that by the argument in Step 2 and complementary slackness, we can focus on the cases where $\mu_{1, j} = \mu_{2,j} = 0$ for all $j$ (i.e., $\theta_J^*$ is an interior of $[-M', M']^{|J|}$ for some finite $M'$ which we can choose large enough).
Taking first-order conditions of the Lagrangian and using the fact that $\omega_j > 0$ for all $j$, we obtain
$
\omega_j \phi(t - \theta_j) \theta_j + \omega_j (1 - \Phi(t - \theta_j)) = \frac{\nu}{\sigma} \omega_j \phi(t - \theta_j) 
$
or, equivalently, 
\begin{equation}
\small 
\begin{aligned}
\frac{\nu}{\sigma} =   \frac{1}{\phi(t - \theta_j)} \Big[ \phi(t - \theta_j) \theta_j + (1 - \Phi(t - \theta_j))\Big] .\label{eq:optimal_nu}
\end{aligned} 
\end{equation}

\smallskip
\noindent \textbf{Step 5: Contradiction argument.} We conclude the proof of the ``if'' direction using a contradiction argument showing that strict negativity of the objective function violates the necessary condition established as a preliminary observation in Step 1. Suppose that the objective function is strictly negative. Then there must exists a $j$ such that $\theta_j^* <0$ and $j' \neq j$ such that $\theta_{j'}^* > 0$. In addition, observe that using Equation \eqref{eq:optimal_nu}, we can write 
$$
\small 
\begin{aligned}
0 > \theta_j^* = \frac{\nu}{\sigma} - \frac{(1 - \Phi(t - \theta_j^*))}{\phi(t - \theta_j^*)} \quad 0 < \theta_{j'}^* =  \frac{\nu}{\sigma} - \frac{(1 - \Phi(t - \theta_{j'}^*))}{\phi(t - \theta_{j'}^*)}.
\end{aligned} 
$$ 
Using the fact that $t$ is finite, it follows that 
$
\frac{1 - \Phi(t - \theta_{j'}^*)}{\phi(t - \theta_{j'}^*)} < \frac{\nu}{\sigma} < \frac{1 - \Phi(t - \theta_j^*)}{\phi(t - \theta_j^*)}. 
$ 
Observe now that the expression implies
$
\frac{1 - \Phi(z)}{\phi(z)} < \frac{1 - \Phi(z')}{\phi(z')} \text{ for some } z < z'. 
$
However, since $(1 - \Phi(z))/\phi(z)$ is monotonically decreasing in $z$ we have a contradiction.

\subsubsection*{Unbiasedness of separate $t$-tests}
To establish unbiasedness, it suffices to show that the researcher would choose to experiment for all $\theta \in \bar{\Theta}_1$ and at least some $(J, \Sigma)$ (since $(J,\Sigma)$ are chosen by the researcher). By our tie-breaking rule, this follows because $\beta_{r^*}(0, J,\Sigma)= 0$  and $\beta_{r^*}(\theta, J,\Sigma)$ is weakly increasing in each element of $\theta$ under Assumption \ref{ass:linear_rule}.

\subsubsection*{``Only if'' direction for maximin optimality and unbiasedness} 
 
We first prove the ``only if'' direction for maximin optimality. Suppose there exists $r^{t'}$ for $t'(J,\Sigma) < \Phi^{-1}\left(1 - \frac{C(J, \Sigma)}{b \bar{\omega}(J)} \right)$ for some $(J,\Sigma)$, so that $r^{t'}$ is maximin optimal. Take $\theta = (-\epsilon, \dots, -\epsilon)$. By continuity of the Gaussian CDF,  we can find some $\epsilon$ sufficiently small, so that $\beta_{r^{t'}}(\theta, J,\Sigma) > 0$. In turn, this implies that $\beta_{r^{t'}}(\theta, J_{r^{t'},\theta}^*, \Sigma_{r^{t'},\theta}^*) \ge \beta_{r^{t'}}(\theta, J,\Sigma) > 0$ and  therefore at the optimal choice $\left(J_{r^{t'},\theta}^*, \Sigma_{r^{t'},\theta}^*\right)$ the researcher conducts the experiment. Because $\theta = (-\epsilon, \cdots, -\epsilon)$ this violates the maximin optimality conditions in \ref{prop:maximin2}.

To prove the ``only if'' direction for maximin optimality and unbiasedness, suppose there exists a maximin optimal and unbiased $r^{t'}$ with $t'(J,\Sigma) > \Phi^{-1}\left(1 - \frac{C(J, \Sigma)}{b \bar{\omega}(J)} \right)$ for all $(J,\Sigma)$. It follows that at $\theta = 0$, $\beta_{r^{t'}}(\theta, J, \Sigma) < 0$ for all $(J,\Sigma)$. Take $\pi \in \Pi$ corresponding to a point mass distribution at $\theta = (\epsilon, \dots, \epsilon)$ for $\epsilon > 0$. By continuity, we can find some $\epsilon > 0$ sufficiently small so that $\beta_{r^{t'}}(\theta, J, \Sigma) < 0$ for all $(J,\Sigma)$ and therefore also for $(J,\Sigma) = (J_{r^{t'},\theta}^*, \Sigma_{r^{t'},\theta}^*)$. This contradicts the premise that $r^{t'}$ is unbiased, completing the proof. 

\subsubsection*{Proof of design-robust global optimality} 

Take $t^* = \Phi^{-1}\left(1 - \frac{C(J,\Sigma)}{b \underline{\omega}(J)}\right)$.
We prove that a threshold protocol with threshold $t$, $r^{t}$, is design-robust globally optimal if and only if $t = t^*$. 
We start by proving the ``if'' result. 
The first part of our proof above titled ``Maximin optimality of separate $t$-tests'' shows that $r^{t^*}$ satisfies the conditions in Equations \eqref{cor:sufficient_maximin_eq1} and \eqref{cor:sufficient_maximin_eq2}. In addition, because $\beta_{r^{t^*}}(0,J,\Sigma) = 0$ and $\beta_{r^{t^*}}(\theta,J,\Sigma)$ is increasing in each element of $\theta$ by Assumptions \ref{ass:linear_rule} and \ref{ass:partially_contractable}, Equation \eqref{cor:sufficient_maximin_eq3} also holds. Therefore, the ``if'' direction follows directly from Proposition \ref{cor:iff_robust}. 

We now prove the ``only if'' direction. Suppose first that $r^t$ is such that $t(J,\Sigma) < t^*(J,\Sigma)$ for some $(J,\Sigma)$ with $J \neq \emptyset$. Then take $\theta = (-\epsilon, \dots, -\epsilon)$. By continuity of the Gaussian CDF,  we can find some $\epsilon$ sufficiently small, so that $\beta_{r^{t'}}(\theta, J,\Sigma) > 0$. This implies that Equation \eqref{cor:sufficient_maximin_eq1} is violated, and therefore $r^t$ is not design-robust globally optimal. Now suppose that $t(J,\Sigma) > t^*(J,\Sigma)$ for some $(J, \Sigma)$ with $J \neq \emptyset$. Take $\theta = (\epsilon, \dots, \epsilon)$ for $\epsilon > 0$. By continuity, we can find some $\epsilon > 0$ sufficiently small so that $\beta_{r^{t}}(\theta, J, \Sigma) < 0$ for some $(J,\Sigma)$ with  $J \neq \emptyset$ violating Equation \eqref{cor:sufficient_maximin_eq3}. The proof completes by Proposition \ref{cor:iff_robust}.

\subsection{Proof of Proposition \ref{prop:uninformed}} \label{app:sec:prop_uninformed} 

We will prove that maximin optimality holds for all $(J,\Sigma)$ under Assumption \ref{ass:partially_contractable}, therefore implying that it also holds for the researcher's best response $J_{r^\ast,\theta}^*,\Sigma_{r^\ast,\theta}^*$. 

We will use the following lemma in the proof.  To alleviate the notation, we keep the dependence of $v_r(\theta;J,\Sigma)$ and $\bar{v}_r(\pi;J,\Sigma)$ on $(J,\Sigma)$ implicit whenever there is no ambiguity. 
\begin{lem}[Sufficient conditions for maximin optimality] \label{prop:maximin_pi} $r^*$ is $\Pi'$-maximin optimal (Definition \ref{defn:delta}) 
if $ \inf_{\pi' \in \Pi'} \bar{v}_{r^*}(\pi', J, \Sigma) \ge 0$, for all $(J,\Sigma)$.   
\end{lem} 
\begin{proof}
Clearly, for $J = \emptyset$ the result trivially holds. Therefore, it suffices to show that this holds also for $J \neq \emptyset$. By construction, $v_r(\theta) \le 0$ for all $\theta \in \Theta_0(J)$ and all $r\in \mathcal{R}$. Because $\Theta_0(J) \neq \emptyset$ for all $r\in \mathcal{R}$, we have that $\int_{\Theta} v_r(\theta)d \pi'(\theta) \le 0$ for all $\pi' \in \Pi'$ such that $\int_{\Theta_0(J)} \pi'(\theta) d\theta = 1$. Because $\Pi'$ is unrestricted, there exists at least one $\pi'$ such that $\int_{\Theta_0(J)} \pi'(\theta) d\theta=1$. It follows that $\inf_{\pi' \in \Pi'} \widetilde{v}_r(\pi') \le 0$ for all $r\in \mathcal{R}$. Therefore, $r$ is maximin optimal if $\inf_{\pi' \in \Pi'} \widetilde{v}_r(\pi') \ge 0$ for all $(J,\Sigma)$.
\end{proof}

In view of Lemma \ref{prop:maximin_pi}, it suffices to show that the worst-case objective function is weakly positive  for any $(J,\Sigma)$. Without loss of generality (since $\Sigma$ has non-zero and finite homogeneous component variances under Assumption \ref{ass:partially_contractable}), we normalize $\Sigma_{j, j} = 1$ for all $j$ and define $\theta_j$ as the treatment effect divided by $\sqrt{\Sigma_{j,j}}$ (which are homogeneous by assumption) so that $\theta_j \in [-M', M']^J$ for some finite $M'$ (because $\sqrt{\Sigma_{j,j}} > 0$ by assumption). 

\smallskip
\noindent \textbf{Step 1: Preliminaries for maximin optimality.} 
By Lemma \ref{prop:maximin_pi}, to establish maximin optimality, it suffices to show that the solution to the following optimization problem is weakly positive
\begin{equation} \label{eqn:jjh2} 
\small 
\begin{aligned}
\min_{\pi' \in \Pi'}  \int \sum_{j \in J} \omega_j (1 - \Phi(t- \theta_j)) \theta_j d\pi'(\theta) , ~\text{ s.t. } \int \sum_{j \in J} \omega_j (1 - \Phi(t- \theta_j)) d\pi'(\theta)  \ge \bar{\omega}(J) (1 - \Phi(t)). 
\end{aligned} 
\end{equation}  
Observe that since $C(J,\Sigma) > 0$, $t$ is finite. 

\smallskip
\noindent \textbf{Step 2: Finite dimensional optimization program.} 
Because the optimization problem only depends on the marginal distributions of $\theta_j$, the minimization over $\pi' \in \Pi'$ can be equivalently rewritten as a minimization over marginal distributions $\pi_1', \cdots, \pi_J'$, $ \theta_j \sim \pi_j'$. 
From standard duality theory (see e.g., \citet{gaivoronski1986linearization}), 
\begin{equation} \label{eqn:dual_pi}
\small 
\begin{aligned}
\eqref{eqn:jjh2} \ge \min_{\pi' \in \Pi'}  \left\{\int \sum_{j \in J} \omega_j (1 - \Phi(t- \theta_j)) \theta_j d\pi'(\theta) + \nu \Big(\bar{\omega}(J) (1 - \Phi(t)) - \int \sum_{j \in J} \omega_j (1 - \Phi(t- \theta_j)) d\pi'(\theta) \Big)\right\}
\end{aligned} 
\end{equation} 
for $\nu \ge 0$ denoting the Lagrangian multiplier. Applying 
Theorem 1, result 1 in \citet{gaivoronski1986linearization} (which here implies that we can find an optimizer that is a point mass distribution), the value of the objective function in Equation \eqref{eqn:jjh2} at its minimum is equal to the solution of the following program
\begin{equation} \label{eqn:jjh2b} 
\small 
\begin{aligned}
\eqref{eqn:dual_pi} = \min_{\theta \in \Theta} \left\{\sum_{j \in J} \omega_j (1 - \Phi(t- \theta_j)) \theta_j  + \nu \Big(\bar{\omega}(J) (1 - \Phi(t)) - \sum_{j \in J} \omega_j (1 - \Phi(t- \theta_j)) \Big)\right\}
\end{aligned} 
\end{equation}  
At this point, the Lagrangian in Equation \eqref{eqn:jjh2b} corresponds to the Lagrangian of the optimization program in Equation \eqref{eqn:lll_v2} (where we normalized $\Sigma_{j,j} = \sigma = 1$). We can follow verbatim the proof of Proposition \ref{cor:threshold_main1} to show that Equation \eqref{eqn:jjh2b} (and therefore Equation \eqref{eqn:jjh2} by duality) is weakly positive. The proof completes by Lemma \ref{prop:maximin_pi}.

\subsection{Proof of Proposition \ref{cor:threshold_aa}} \label{proof:cor:threshold_aa} 

First, note that under $r^{N}$, the researcher will always choose $\Sigma_{j,j}$ under a sample-equalizing allocation, since otherwise her utility is zero provided she experiments. Therefore, by the tie-breaking rule, it suffices to show that $r^N$ is maximin and unbiased under a sample equalizing allocation.  Note that $\Sigma_{j,j} = \frac{\sigma_{j1}^2}{\sigma_j^2 n_{1j}} + \frac{\sigma_{0j}^2}{\sigma_j^2 n_{0j}} = \frac{2}{\bar{m}}$ for some constant $\bar{m}$ under a sample equalizing allocation. Therefore, $\Sigma_{j,j}$ is only a function of $\bar{m}$ but not of $\sigma_j^2$.

\smallskip

\noindent \textbf{Maximin optimality.} 
To show maximin optimality, we show how to modify the proof of Proposition \ref{cor:threshold_main1} so that it applies directly to this case. Note that in light of the sufficient conditions in Corollary \ref{cor:sufficient_maximin}, it suffices to show that maximin optimality holds for any $(\Sigma, J)$ with  a sample-equalizing allocation. As it will become clear below, this amounts to showing that maximin optimality holds for any $\bar{m} \in (0, n]$, $n< \infty$.
Under Assumption \ref{ass:t_tests2}, we have $X | (\tau,\sigma) \sim \mathcal{N}(\theta,\Sigma)$ and $u(\theta,j) = \mathbb{E}[\tau_j | \theta]$.  
Therefore, we can write by  Assumption \ref{ass:additive2},  
$$
\small 
\begin{aligned} 
v_r(\theta, J, \Sigma)  & = \sum_{j \in J} \omega_j \mathbb{E}[r_j(X, J, \Sigma) | \Sigma, \theta] u(\theta,j)  = \sum_{j \in J} \omega_j \mathbb{E}[r_j(X, J, \Sigma) | \Sigma, \theta] \mathbb{E}[\tau_j | \theta]   \\ 
&= \sum_{j \in J} \omega_j \mathbb{E}[r_j(X, J, \Sigma) | \Sigma, \theta] \mathbb{E}[\frac{\tau_j}{\sigma_j} \sigma_j | \theta] 
= \sum_{j \in J} \omega_j \mathbb{E}[r_j(X, J, \Sigma) | \Sigma, \theta] \theta_j \mathbb{E}[\sigma_j | \theta]. 
\end{aligned} 
$$ 
Note that under Assumption \ref{ass:exchangeable}, $\mathbb{E}[\sigma_j|\theta] = \bar{\sigma}$ for some positive constant $\bar{\sigma} > 0$. Therefore, to prove maximin optimality it suffices to show  that $\min_\theta \sum_{j \in J} \omega_j \mathbb{E}[r_j^N(X, J, \Sigma) | \Sigma, \theta] \theta_j \ge 0$ whenever the researcher experiments. 
Let $t^* := t^*(J,n(J))$, which is known to the researcher since $C(J,n(J))$ is known. 
We can write the researcher's payoff as 
$
\sum_{j \in J} \omega_j \mathbb{E}[r_j(X, J, \Sigma) | \Sigma, \theta] = \sum_{j \in J} \omega_j \Big(1 - \Phi(t^* - \theta \sqrt{\frac{\bar{m}}{2}})\Big).
$
Note that the researcher payoff's is known to the researcher ex-ante because it only depends on $t^*$, $\bar{m}$ and $\theta$ all assumed to be known to the researcher. 

Therefore, following the argument in the proof of Proposition \ref{cor:threshold_main1}, proving maximin optimality only requires  proving that 
\begin{equation} 
\small 
\begin{aligned} 
\label{eqn:variance_balance} 
\min_\theta \sum_{j \in J} \omega_j \Big(1 - \Phi(t^* -  \theta  \sqrt{\frac{\bar{m}}{2}})\Big)  \sqrt{\frac{\bar{m}}{2}} \theta_j, \quad \text{ such that } \sum_{j \in J} \omega_j \Big(1 - \Phi(t^* -  \theta  \sqrt{\frac{\bar{m}}{2}})\Big) \ge C(J,n(J))
\end{aligned} 
\end{equation} 
is weakly positive for any finite $t^*$, $\bar{m} \in (0, n]$ for finite $n <\infty$ (which is equivalent to proving it for any $\Sigma$ with a sample-equalizing allocation) where without loss we multiplied welfare by $ \sqrt{\frac{\bar{m}}{2}}$ (which does not depend on $j$). Here, the constraint is the (known to the researcher) individual rationality constraint (i.e., the constraint that the researcher's payoff needs to be weakly positive for the experiment to be conducted). The argument that shows weak positivity of Equation \eqref{eqn:variance_balance} then follows verbatim the proof of Proposition \ref{cor:threshold_main1}, with $\theta  \sqrt{\frac{\bar{m}}{2}}$ in lieu of $\theta$ (see Equation \eqref{eqn:lll_v2} and discussion below), since $ \theta  \sqrt{\frac{\bar{m}}{2}}$ lies in a compact space given that $\bar{m} \le n < \infty$. 

\smallskip

\noindent \textbf{Unbiasedness.} To establish unbiasedness, it suffices to show that the researcher's payoff is weakly positive for all $\theta \ge 0$ (since $u(\theta,j) = \theta \mathbb{E}[\sigma_j|\theta] = \theta \bar{\sigma}$ where $\bar{\sigma} > 0$). This follows directly from the same argument in the proof of Proposition \ref{cor:threshold_main1}. 
 
\section{Proofs of results in Online Appendix} 
\label{app:omitted_proofs}

\subsection{Proof of Proposition \ref{prop:two_sided}} \label{app:twosided_proof} To prove the claim, following verbatim the argument in the proof of Corollary \ref{cor:sufficient_maximin},  it suffices to show that $v_{\tilde{r}}^{\mathrm{two}}(\theta, J, \Sigma, p) 1\{\beta_{\tilde{r}}(\theta, J, \Sigma,p) \ge 0\} \ge 0$ for all $\theta \in \Theta, p \in \{0,1\}, J \in \mathcal{J}, \Sigma \in \mathcal{S}(J)$. 
Define $t^\ast = \Phi^{-1}(1 - \frac{C(J,\Sigma)}{b \bar{\omega}(J)})$ and, for notational convenience, let $\Sigma_{i,i} = 1$ without loss of generality (under Assumption \ref{ass:partially_contractable}). We write 
$$
\small 
\begin{aligned} 
\mathbb{E}[\tilde{r}_j^{t^*}(X, J, \Sigma) 1\{\sgn(X_j)= 1\} \theta | \theta_j] & = \Big[1 - \frac{\Phi(t^\ast - \theta_j) - \Phi(-\theta_j)}{1 - \Phi(-\theta_j)}\Big] (1 - \Phi(-\theta_j)) =  (1 - \Phi(t^\ast - \theta_j)).
\end{aligned}
$$
Similarly, we can write 
$
\mathbb{E}[\tilde{r}_j^{t^*}(X, J, \Sigma) 1\{s_j(X) = -1\} \theta_j | \theta_j]  = \Phi(-t^\ast - \theta_j).
$
Collecting the terms, upon  experimentation (occurring only if $\beta_{\tilde{r}_j^{t^*}}(\theta, J, \Sigma,p) \ge 0$),
$$
\small 
\begin{aligned}
v_r^{\mathrm{two}}(\theta, J, \Sigma, p) = \sum_{j \in J} \omega_j \theta_j \Big[(1 - \Phi(t^\ast - \theta_j)) (1 - p) - p\Phi(-t^* - \theta_j^\ast)\Big]. 
\end{aligned} 
$$
To show maximin optimality, it suffices to show (by the same argument in Proposition \ref{cor:threshold_main1}) 
\begin{equation} \label{eqn:two_sided}
\small 
\begin{aligned} 
& \min_{\theta, p}\sum_{j \in J} \omega_j \theta_j \Big[(1 - \Phi(t^\ast - \theta_j)) (1 - p) - p\Phi(-t^* - \theta_j)\Big] \\
& \text{ such that } \sum_{j \in J} \omega_j \Big[(1 - \Phi(t^\ast - \theta_j))(1 - p) + p \Phi(-t^\ast - \theta_j)\Big] \ge C(J, \Sigma)
\end{aligned} 
\end{equation}
is weakly positive. 

Consider first the case where $p = 0$. In this case, we can show that Equation \eqref{eqn:two_sided} is weakly positive using the arguments in the proof of Proposition \ref{cor:threshold_main1}. 

Consider now the case where $p = 1$. Define $\tilde{\theta}_j = -\theta_j$, and note that  $\Phi(-t^\ast + \tilde{\theta}_j) = 1 - \Phi(t^\ast - \tilde{\theta}_j)$. Then, we can use the arguments in the proof of Proposition \ref{cor:threshold_main1} with $\tilde{\theta}_j$ in lieu of $\theta_j$ to show that \eqref{eqn:two_sided} is weakly positive. This completes the proof.

\subsection{Proof of Proposition \ref{prop:no_admissible}}
\label{app:prop:no_admissible}
  The researcher's net benefit is
$
 \sum_{j \in J}   P(r_j(X) | \theta)  - C, 
$
 where we suppress the dependence of $r(X,J,\Sigma)$ and $C(J,\Sigma)$ on $J$ and $\Sigma$ to alleviate the notation and, as before, normalize $b = 1$ without loss of generality. In addition, we set $\omega_j = 1$ for all $j$, although one could directly extend our reasoning to more general $\omega$.

 Recall that under Assumption \ref{ass:two_decisions}, $\mathcal{J} = \{\emptyset, \bar{J}\}$. Therefore we can write the social planner's utility accounting for the researcher's best response as 
\begin{equation} \label{eqn:two_cases_v}
\small 
\begin{aligned}
v_r(\theta) = \begin{cases} 
 v_r(\theta, \bar{J},\Sigma), \quad & \text{ if } \beta_r(\theta, \bar{J},\Sigma) > 0 \\ 
 \max\{v_r(\theta, \bar{J},\Sigma), 0\}, \quad & \text{ if } \beta_r(\theta, \bar{J},\Sigma) = 0 \\ 
 0 & \text{ otherwise}, 
\end{cases} 
\end{aligned} 
\end{equation} 
where the second case follows from the tie-breaking rule. We will refer to $\Theta_0$ as $\Theta_0(\bar{J})$ suppressing its dependence on $\bar{J}$ whenever clear from the context.

 Take the parameter space (suppressing its dependence on $\bar{J}$ for expositional convenience)
$
\Theta = \Big\{\theta \in [-M,M]^{\bar{J}} \text{ such that } \text{sign}(\theta_1) = \text{sign}(\theta_2) = \cdots = \text{sign}(\theta_{\bar{J} - 1})\Big\}.
$ 
To prove the statement we show that there exists a (set of) maximin protocol that strictly dominates all others over an arbitrary set $\Theta' \subseteq \Theta$, and a different (set of) maximin testing protocol that it strictly dominates all others over some arbitrary set $\Theta'' \cap \Theta' = \emptyset, ~ \Theta '' \subseteq \Theta$. 
We choose $\Theta' = (0, \cdots, 0, t)$ for a small $t$ and $\Theta'' = (t, \cdots,t, 0)$ for a small $t$. We choose 
$
X \sim \mathcal{N}(\theta, I),
$ 
which satisfies Assumption \ref{ass:partially_contractable}. 
Observe that by definition of $\Theta_0$, 
$
\Theta_0  \subseteq \tilde{\Theta}_0 = \Big\{\theta:  \theta_j \le 0 \text{ for all } j\Big\}, 
$ 
where $\tilde{\Theta}_0$ also contains those elements that lead to weakly negative welfare. 

\smallskip
\noindent \textbf{Step 1: Construction of the function class.} 
Define
$
\mathcal{R}^1 = \Big\{r \in \mathcal{R}: C \ge  P\Big(r_{|\bar{J}|}^1(X) | \theta =  0\Big) >  \frac{C}{ \bar{J}}\text{ and } r \text{ is  maximin}\Big\}.
$
We claim that $\mathcal{R}^1 \neq \emptyset$, i.e., there exists a function $r \in \mathcal{R}^1$. An example is a threshold crossing testing protocol of the form 
\begin{equation} \label{eqn:block_quote_exmp}
\small 
\begin{aligned} 
r_j(X) = 1\{X_j > t_j\}, t_{j < \bar{J}} = \infty, t_{\bar{J}} = \Phi^{-1}(1 - \min\{C, 1\}).
\end{aligned} 
\end{equation}
In addition, we observe that 
$
\sup_{r \in \mathcal{R}^1} P\Big(r_{|\bar{J}|}(X) | \theta_{\bar{J}} = 0, \theta_{j < \bar{J} } = 0 \Big) \ge \min\{C, 1\}
$
from the example in Equation \eqref{eqn:block_quote_exmp}. Define $\tau =  \min\{C, 1\}$ for the rest of the proof.

\medskip
\noindent \textbf{Step 2: Comparisons with maximin protocols.} 
We now claim that for $\theta = (0, 0, \cdots, 0, t)$, for $t$ approaching zero, there exists a maximin testing protocol $r^1 \in \mathcal{R}^1$ which leads to strictly larger welfare than any maximin decisions $r^2 \in \mathcal{M} \setminus \mathcal{R}^1$, where $\mathcal{M}$ denotes the set of maximin protocols. To show our claim, it suffices to compare $r^1$ to any maximin testing protocol $r^2 \not \in \mathcal{R}^1$ with 
$
P\Big(r_{|\bar{J}|}^2(X) | \theta = 0  \Big) \le \frac{C}{\bar{J}}.
$
To see why, observe that whenever the above probability is between $(\frac{C}{\bar{J}}, \tau]$, we contradict the statement that $r^2 \not \in \mathcal{R}^1$. When instead
$
P\Big(r_{|\bar{J}|}^2(X) | \theta = 0  \Big) > \tau,
$
$r^2$ is not maximin optimal, since this implies that $C < 1$, which in turn implies that by Assumption \ref{ass:partially_contractable}, the researcher would experiment under $(\theta_{\bar{J}} = -t, \theta_{j < \bar{J}} = -t)$, for some small positive $t$, leading to strictly negative welfare. 

\medskip
\noindent \textbf{Step 3: Comparisons of welfare.} For $\theta = (0, 0, \cdots, 0, t)$ write  
$
v_{r^1}(0, 0, \cdots, 0, t) = t \times P\Big(r_{|\bar{J}|}^1(X_1, X_2, \cdots, X_{\bar{J}})\Big| \theta_{-\bar{J}} = 0, \theta_{\bar{J}} = t\Big).
$
By continuity of $P(\cdot)$ in $t$, it follows that we can choose $t > 0$ sufficiently small so that $v_{r^1}(\theta = (0,0, \cdots, t)) - v_{r^2}(\theta = (0,0, \cdots, t)) > 0$.

\smallskip
\noindent \textbf{Step 4: Testing protocol $r^1$ is not dominant.} 
We are left to show that there exists a function $r^3 \in \mathcal{M} \setminus \mathcal{R}^1$ that leads to strictly larger welfare than any $r^1 \in \mathcal{R}^1$ for some different $\theta$. Choose $\theta = (t, t, \cdots, 0)$. Let 
$
\mathcal{R}^2 = \mathcal{M} \setminus \mathcal{R}^1.
$
We claim that $\mathcal{R}^2$ is non-empty. An example is
$
r^3_j(X) = 1\{X_j > t_j\},~  t_j  = \Phi^{-1}\left(1 - \min\left\{\frac{C}{\bar{J} - 1}, 1\right\}\right), ~ j < \bar{J}, ~ t_{\bar{J}} = \infty. 
$ 
Consider the alternative $\check{\theta} = (t, \cdots, t, 0)$. Observe now that we have 
$
 v_{r^3}(\check{\theta}) - 
\sup_{r^1 \in \mathcal{R}^1} v_{r^1}(\check{\theta})  = t \times  \Big[\sum_{j < \bar{J}} P(r_j^3(X)  | \theta = \check{\theta}) - \sup_{r^1 \in \mathcal{R}^1} \sum_{j < \bar{J}}  P(r_j^1(X)  | \theta = \check{\theta})  \Big]. 
$
Next, we claim that 
\begin{equation} \label{eqn:statementaa}
\small 
\begin{aligned} 
\sum_{j < \bar{J}} P(r_j^1(X) | \theta = 0) < (\bar{J} - 1)(\min\{C/(\bar{J}-1), 1\}). 
\end{aligned} 
\end{equation} 
We prove the claim by contradiction. Suppose that the above equation does not hold. Then it must be that (since $P(r_{|\bar{J}|}^1(X) |\theta = 0) > C/\bar{J}$)
\begin{equation} \label{eqn:contr}
\small 
\begin{aligned} 
\sum_{j < \bar{J}} P(r_j^1(X)  | \theta = 0) + P(r_{|\bar{J}|}^1(X)  | \theta = 0) > (\bar{J} - 1)(\min\{C/(\bar{J}-1), 1\}) +  \frac{C}{\bar{J}}. 
\end{aligned} 
\end{equation}  
Clearly if $C/(\bar{J} - 1) \le 1$, Equation \eqref{eqn:statementaa} is true since otherwise by Equation \eqref{eqn:contr} we would contradict maximin optimality of $r^1$. Suppose that $C/(\bar{J}-1) > 1$. Then for $r^1$ to be maximin optimal we must have that $(\bar{J} - 1) + C/\bar{J} \le C$. However, it is easy to show that this implies that $C/\bar{J} \ge 1$ which leads to a contradiction.   

Finally, using continuity, we obtain that for $t$ small enough any $r^1 \in \mathcal{R}^1$ is dominated by $r^3$, completing the proof.

\subsection{Proof of Proposition \ref{prop:locally_most_powerful}} \label{app:prop:locally_most_powerful}

As in the proof of Proposition \ref{prop:no_admissible}, we write the researcher's net benefit as 
$
 \sum_{j \in J}  P(r_j(X) | \theta)  - C. 
$
Maximin optimality directly follows from Corollary \ref{cor:sufficient_maximin}. Thus, we focus on local power.
The proof proceeds as follows. We first find a lower bound on the worst-case power of $r^*$. We then argue that any other maximin testing protocol attains power below this lower bound (and therefore has lower power than $r^*$). 

Define $v_r(\theta)$ as in Equation \eqref{eqn:two_cases_v} under Assumption \ref{ass:two_decisions}. As in the statement of the proposition, we will be assuming that $\omega_j = 1$ for all $j$ in Assumption \ref{ass:linear_rule}, although our reasoning can extend to general $\omega > 0$.

\smallskip
\noindent \textbf{Step 1: Lower bound on worst-case power.} We claim that
$
\lim \inf_{\epsilon \downarrow 0} \frac{1}{\epsilon} \inf_{\theta \in \Theta_1(\epsilon)} v_{r^*}(\theta) \ge \frac{C}{\bar{J}}. 
$
We now show why. Denote by $\theta(\epsilon) \in \Theta_1(\epsilon)$ the parameter under the local alternative. Observe that the welfare under the local alternative reads as follows
$$
(A) = \inf_{\theta(\epsilon)} \sum_{j \in \bar{J}}  \theta(\epsilon) P(r_j^*(X) | \theta = \theta(\epsilon)), \text{ such that } \theta_j(\epsilon) = \epsilon \text{ for some } j, \quad \theta_j(\epsilon) \in [0,\epsilon] \quad \forall j. 
$$
We write
$$
\small 
\begin{aligned} 
 (A) &\ge  \inf_{w \in [0,\epsilon]^{\bar{J}}: \sum_j w_j \ge \epsilon, \theta(\epsilon) \in \Theta_1(\epsilon)} \sum_{j \in \bar{J}}  w_j P(r_j^*(X) | \theta = \theta(\epsilon)) \\ &\ge \inf_{w \in [0,\epsilon]^{\bar{J}}: \sum_j w_j \ge \epsilon, \theta' \in [0,\epsilon]^{\bar{J}}: \sum_j \theta_j \ge \epsilon}  \sum_j w_j P(r_j^*(X) | \theta = \theta') := g(\epsilon).
\end{aligned} 
$$
Define $\mathcal{W}(\epsilon_1, \epsilon_2) = \Big\{(w,\theta) \in [0,\epsilon_1]^{\bar{J}} \times [0,\epsilon_2]^{\bar{J}}: \sum_j w_j \ge \epsilon_1, \sum_j \theta_j \ge \epsilon_2\Big\}$ and write 
$$
\small 
\begin{aligned} 
\frac{1}{\epsilon} g(\epsilon) = \inf_{(w,\theta') \in \mathcal{W}(1, \epsilon) }  \sum_{j \in \bar{J}}  w_j P(r_j^*(X) | \theta = \theta') = \inf_{(w,\theta') \in \mathcal{W}(1, 1) }  \sum_{j \in \bar{J}}  w_j P(r_j^*(X) | \theta = \epsilon \theta'). 
\end{aligned} 
$$ 

Observe that $\mathcal{W}(1, 1)$ is a compact space. In addition $P(r_j^*(X) | \theta = \epsilon \theta')$ is continuous in $\epsilon$ for any $\theta' \in \Theta$ by Assumption \ref{ass:partially_contractable}. As a result, $g(\epsilon)/\epsilon$ is a continuous function in $\epsilon$. Therefore, 
$
\lim_{\epsilon \rightarrow 0} \frac{g(\epsilon)}{\epsilon} = \inf_{(w,\theta') \in \mathcal{W}(1, 1) }  \sum_{j \in \bar{J}}  w_j P(r_j^*(X) | \theta = \theta' \times 0) = \inf_{(w,\theta') \in \mathcal{W}(1, 1) }  \sum_{j \in \bar{J}}  w_j \frac{C}{\bar{J}} =  \frac{C}{\bar{J}}. 
$
This completes the proof of our claim. 

\smallskip
\noindent \textbf{Step 2: Alternative set of maximin protocols.} We now claim that any maximin protocol $r'$, which does not satisfy Equation \eqref{eqn:separate}, must satisfy for some $j \in \{1, \cdots, \bar{J}\}$, 
\begin{equation} \label{eqn:mot1} 
\small 
\begin{aligned} 
P\left(r'_j(X) = 1 | \theta = 0\right) < \frac{C}{\bar{J}}. 
\end{aligned} 
\end{equation} 
We prove the claim by contradiction. 
Consider a maximin protocol $r'$ such that for all $j$ Equation \eqref{eqn:mot1} is violated. Then
if $r'$ is maximin optimal and satisfies Equation \eqref{eqn:mot1} with equality for all $j$, there must be an $r^*$ defined as in the proposition statement equal to $r'$, which leads to a contradiction. Therefore it must be that if $r'$ does not satisfy Equation \eqref{eqn:mot1} for some $j$, $r'$ is such that for some $j$ Equation \eqref{eqn:mot1} is satisfied with reversed \textit{strict} inequality and for all $j$ is satisfied with reversed weak inequality. In such a case, it follows that 
$
\sum_j P\Big(r_j'(X) = 1 | \theta = 0\Big) > C. 
$
This violates maximin optimality, as we can take $\theta = (-t, -t, \cdots, -t) \in \Theta_0$ for some small $t$. By Assumption \ref{ass:partially_contractable} (namely, by continuity of $F_\theta$), we have for $t$ small enough,
$
\sum_j  P\Big(r_j'(X) = 1 | \theta_k = -t, ~ \forall k\Big) > C. 
$
Therefore, for $t$ small enough, the protocol $r'$ induces the researcher to experiment leading to negative welfare. This violates maximin optimality by the argument in Step 1 of the proof of Proposition \ref{prop:maximin2}. 

\medskip
\noindent \textbf{Step 3: Power comparison.} Observe now that 
$
 \inf_{\theta \in \Theta_1(\epsilon)} \sum_{j \in \bar{J}}  \theta P(r_j(X) | \theta) \le \epsilon P(r_j(X) = 1 | \theta_j = \epsilon, \theta_{-j} = 0),  
$ 
since the vector $(\theta_j = \epsilon, \theta_{-j} = 0) \in \Theta_1(\epsilon)$. Using Assumption \ref{ass:partially_contractable} we have  
$
\lim_{\epsilon \rightarrow 0} P(r_j(X) = 1 | \theta_j = \epsilon, \theta_{-j} = 0) = P(r_j(X) = 1 | \theta = 0)<  \frac{C}{\bar{J}}. 
$
This completes the proof of the if statement.


\subsection{Proof of Proposition \ref{prop:weights_power}} \label{app:sec:global}

Under Proposition \ref{prop:maximin2} and continuity of $X$ (Assumption \ref{ass:partially_contractable}) every maximin protocol must satisfy (recalling the normalization $b = 1$)
$
\sum_{j \in J}  P(r_j(X) = 1 | \theta = 0) \le C(J) . 
$
We can write the weighted welfare under $r$ as 
$
\int_{\bar{\Theta}_1} w(\theta) \sum_{j \in J}  P(r_j(X) = 1| \theta) \theta_j d\theta. 
$ 
We now discuss two cases: (i) $P(r_{|\bar{J}|}(X) = 1 | \theta = 0) = C(\bar{J},\Sigma)$, and (ii) $P(r_{|\bar{J}|}(X) = 1 | \theta = 0) < C(\bar{J}, \Sigma)$. 
Define $v_r(\theta)$ as in Equation \eqref{eqn:two_cases_v}. 

\medskip
\noindent \textbf{Case (i):} 
Suppose first that $P(r_{|\bar{J}|}(X) = 1 | \theta = 0) = C(\bar{J}, \Sigma)$, which implies that $P(r_1(X) = 1| \theta = 0) = 0$. Then choose $w(\theta) = 1\{(\theta_1, \cdots, \theta_{\bar{J}}) = (\epsilon, 0, \dots, 0)\}$ for some small $\epsilon > 0$. Take $r'$ such that 
$
r_1'(X) = 1\Big\{X_1/\sqrt{\Sigma_{1,1}} \ge \Phi^{-1}(1 - C(\bar{J}, \Sigma)) 
\Big\}, ~ r_j'(X) = 0, ~ \forall j > 1. 
$
It is easy to show that $r'(X)$ is maximin under Assumptions \ref{ass:linear_rule} and \ref{ass:partially_contractable}. Then it follows that 
$$
\small 
\begin{aligned} 
\int_{\theta \in \bar{\Theta}_1}  \Big(v_r(\theta) - v_{r'}(\theta)\Big) w(\theta) d \theta= \epsilon \Big(P(r_1(X) = 1| \theta = (\epsilon, 0, \dots,0) ) - P(r_1'(X) = 1| \theta = (\epsilon, 0, \dots,0))\Big). 
\end{aligned} 
$$ 
By continuity, it follows that, as $\epsilon \downarrow 0$,
$$
\small 
\begin{aligned} 
P(r_1(X) = 1| \theta = (\epsilon, 0, \dots,0))  \rightarrow 0, \quad  P(r_1'(X) = 1| \theta = (\epsilon, 0, \dots,0))  \rightarrow C(\bar{J}, \Sigma) > 0. 
\end{aligned} 
$$ Hence, by continuity, we can take $\epsilon > 0$ small enough so that 
$
\int_{\theta \in \bar{\Theta}_1} w(\theta) \Big(v_r(\theta) - v_{r'}(\theta)\Big) < 0. 
$

\medskip
\noindent \textbf{Case (ii):}
Suppose now that $P(r_{|\bar{J}|}(X) = 1 | \theta = 0) < C(\bar{J}, \Sigma)$. Then, we can take 
$$
\small 
\begin{aligned} 
r_{|\bar{J}|}'(X) = 1\Big\{X_{|\bar{J}|}/\sqrt{\Sigma_{|\bar{J}|,|\bar{J}|}} \ge \Phi^{-1}(1 - C(\bar{J}, \Sigma)) 
\Big\}, \quad r_j'(X) = 0, \quad \forall j < \bar{J},  
\end{aligned} 
$$ 
and $w(\theta) = 1\{\theta = (0, \dots, 0, \epsilon)\}$. The argument now follows verbatim as in the previous case, with the first entry replacing the last entry.

\subsection{Proof of Proposition \ref{prop:apprx2}} \label{proof:prop:apprx2}
Under Assumption \ref{ass:two_decisions}, we can write $e_r^*(\theta) = 1\{\beta_r(\theta, \bar{J}, \Sigma) \ge 0\}$. 
We can write 
$$
\small 
\begin{aligned} 
& \Big|U'(r;\lambda, w)  - U(r;\lambda, \pi)\Big|  = \lambda \Big|\int v_r(\theta, \bar{J}, \Sigma) e_r^*(\theta)  w(\theta) d\theta  - \int e_r^*(\theta)  \pi(\theta) d\theta \Big| \\ & = 
\lambda \Big|\int v_r(\theta, \bar{J}, \Sigma) e_r^*(\theta)  w(\theta) d\theta  - \int e_r^*(\theta)  v_{r=1}(\theta) w(\theta) d\theta \Big| = \lambda \Big|\int \Big(v_r(\theta, \bar{J}, \Sigma) - v_{r=1}(\theta)\Big) e_r^*(\theta)  w(\theta) d\theta \Big|. \\ 
\end{aligned} 
$$
Note that $\int |e_r^*(\theta)  w(\theta)| d\theta \le \int | w(\theta)| d\theta = 1$. Therefore, using Holder's inequality, 
$
\Big|\int \Big(v_r(\theta, \bar{J}, \Sigma) - v_{r=1}(\theta)\Big) e_r^*(\theta)  w(\theta) d\theta \Big| \le \max_{\theta \in \bar{\Theta}_1'} |v_{r = 1}(\theta) - v_{r}(\theta, \bar{J}, \Sigma)|,
$
where the maximum is over $\bar{\Theta}_1'$ since $w(\theta)$ has support only over $\bar{\Theta}_1'$. The proof completes.

\subsection{Proof of Corollary \ref{cor:final}} \label{proof:cor:final}
  Define 
$r^* = \mathrm{arg} \max_r U'(r;\lambda, w)$. Because $r^t$ is maximin and unbiased and by definition of $\bar{\Theta}_1$, since the maximin component is always weakly negative (see proof of Proposition \ref{prop:maximin2}), 
we have
$
U'(r^*;\lambda, w)   \le \lambda \int v_{r=1}(\theta) w(\theta)d\theta \quad  \text{and} \quad
 U(r^t;\lambda, \pi)  = \lambda \int \pi(\theta)d\theta.
 $
 for any $\pi$ satisfying Assumption \ref{ass:alternative_space}. Therefore, for any $\pi$ defined in Proposition \ref{prop:apprx2}, we have  
$
U'(r^*;\lambda, w) - U'(r^t;\lambda, w)  =  U'(r^*;\lambda, w) - U(r^t;\lambda, \pi) + U(r^t;\lambda, \pi) - U'(r^t;\lambda, w) \le U(r^t;\lambda, \pi) - U'(r^t;\lambda, w).
$
The proof now follows directly from Proposition \ref{prop:apprx2}.

\subsection{Proof of Proposition \ref{cor:final2}} \label{proof:cor:final_2}
The proof mimics the proof of Corollary \ref{cor:final} with appropriate changes. 
  Define $\beta_{r=1}(\theta) = \beta_{r=1}(\theta, \bar{J}, \Sigma)$. (Recall here we assume that $J \in \{\emptyset, \bar{J}\}$ and $\mathcal{S}(\bar{J}) = \{\Sigma\}$.) Let 
$r^* = \mathrm{arg} \max_r U''(r;\lambda, w)$. Because $r^t$ is maximin and unbiased and by definition of $\bar{\Theta}_1$, since the maximin component is always weakly negative (see proof of Proposition \ref{prop:maximin2}), 
it follows that
$ 
U''(r^*;\lambda, w)   \le \lambda \int \beta_{r=1}(\theta) w(\theta)d\theta \quad  \text{and} \quad
 U(r^t;\lambda, \pi)  = \lambda \int \pi(\theta)d\theta,
 $
 for any $\pi$ satisfying Assumption \ref{ass:alternative_space}. Therefore, take $\pi(\theta) = w(\theta) \beta_{r=1}(\theta)$. We have  
$$ 
\small 
\begin{aligned} 
U''(r^*;\lambda, w) - U''(r^t;\lambda, w) & =  U''(r^*;\lambda, w) - U(r^t;\lambda, \pi) + U(r^t;\lambda, \pi) - U''(r^t;\lambda, w) \le U(r^t;\lambda, \pi) - U''(r^t;\lambda, w).
\end{aligned} 
$$
We can now follow the proof of Proposition \ref{prop:apprx2} with appropriate changes. We can write 
$$
\small 
\begin{aligned} 
& \Big|U''(r^t;\lambda, \pi) - U(r^t;\lambda, w)\Big|  = \lambda \Big|\int \beta_r(\theta, \bar{J},\Sigma)e_r^*(\theta) w(\theta)d\theta - \int e_r^*(\theta) \pi(\theta) d\theta  \Big|  \\ 
&= \lambda \Big|\int \beta_r(\theta, \bar{J},\Sigma)e_r^*(\theta) w(\theta)d\theta - \int e_r^*(\theta) \beta_{r=1}(\theta) w(\theta) d\theta  \Big| \le \max_{\theta \in \bar{\Theta}_1'}|\beta_{r = 1}(\theta) - \beta_r(\theta, \bar{J},\Sigma)|,
\end{aligned} 
$$
where the argument follows similarly to the one in Proposition \ref{prop:apprx2} invoking Holder's inequality. The conclusion follows from the expression for $\beta_r$ under Assumption \ref{ass:linear_rule}.

\end{document}